\documentclass[12pt,letter,openany,oneside%,draft
]{book}
\usepackage{epsfig}
\usepackage{amsmath}
\usepackage{amsfonts}
\usepackage{titlesec}
\usepackage[nottoc,notlot,notlof]{tocbibind}
\usepackage{caption}
\usepackage[left=3cm, right=2.3cm, top=3cm, bottom=2cm]{geometry}
\usepackage{color}
\usepackage{calc}
\usepackage{titletoc}
\usepackage{pifont}
\usepackage{graphicx}
\usepackage{relsize}

\newcommand{\bigrule}{\titlerule[1mm]}
\definecolor{c1}{rgb}{0,0.5,0}
\definecolor{c2}{rgb}{0.9,.0,0}
\definecolor{c3}{rgb}{.465,.535,.605}%color del panel
\definecolor{c4}{rgb}{.6,.6,.6}%color de los botones del panel

\titleformat{\chapter}[display]
{\bfseries\Large}{
\titlerule
\vspace{1mm} \filleft \Large\chaptertitlename \resizebox{!}{2cm}{\bf\textup{{\thechapter}}}}
{15mm} \filcenter [\vspace{0.5mm} \titlerule \vskip 3pt \bigrule]

\titleformat{\section}[frame]
{\bfseries\large} { \filright {\colorbox{c4}{ \footnotesize \enspace SECTION \thesection }}\enspace} {6mm} \filright

\titleformat{\subsection}[hang]
{\bfseries\large} { \filright \colorbox{c4}\thesubsection} {6mm} \filright

\widenhead*{0pt}{\marginparsep}
\renewpagestyle{plain}{}
\newpagestyle{latex}[\sl\scriptsize]{
\headrule
\sethead[\thepage][\sectiontitle][]{\sectiontitle}{}{\small\thepage}}
\newcommand{\be}{\begin{equation}}
\newcommand{\ee}{\end{equation}}
\newcommand{\bea}{\begin{eqnarray}}
\newcommand{\eea}{\end{eqnarray}}
\newcommand{\beal}{\begin{align}}
\newcommand{\eeal}{\end{align}}

\renewcommand{\thefootnote}{\#\arabic{footnote}}

\newcommand{\gapp}{\mathrel{\raise.3ex\hbox{$>$}\mkern-14mu
              \lower0.6ex\hbox{$\sim$}}}
\newcommand{\lapp}{\mathrel{\raise.3ex\hbox{$<$}\mkern-14mu
              \lower0.6ex\hbox{$\sim$}}}

\newcommand{\lsim}{\mbox{\raisebox{-.9ex}{~$\stackrel{\mbox{$<$}}{\sim}$~}}}
\newcommand{\gsim}{\mbox{\raisebox{-.9ex}{~$\stackrel{\mbox{$>$}}{\sim}$~}}}
\newcommand\vev[1]{{\langle {#1} \rangle}}
\renewcommand\({\left(}
\renewcommand\){\right)}
\renewcommand\[{\left[}
\renewcommand\]{\right]}

\newcommand\eq[1]{Eq.~(\ref{#1})}

\newcommand\eqreff[1]{(\ref{#1})}

\newcommand\pa{\partial}

\def\calb{{\cal B}}

\def\call{{\cal L}}

\def\calp{{\cal P}}

\def\calt{{\cal T}}
\def\calv{{\cal V}}
\def\calpz{{\calp_\zeta}}
\def\calbz{{\calb_\zeta}}
\def\caltz{{\calt_\zeta}}
\newcommand\bfA{{\mathbf A}}

\newcommand\bfd{{\mathbf d}}

\newcommand\bfk{{\mathbf k}}

\newcommand\bfn{{\mathbf N}}

\newcommand\bfp{{\mathbf p}}
\newcommand\bfq{{\mathbf q}}

\newcommand\bfx{{\mathbf x}}

\newcommand\MeV{\,\mbox{MeV}}

\newcommand\sub[1]{_{\rm #1}}
\newcommand\su[1]{^{\rm #1}}
\newcommand\mone{^{-1}}

\newcommand\half{^{1/2}}

\newcommand\threehalf{^{3/2}}

\newcommand\mn{{\mu\nu}}
\newcommand{\fnl}{f\sub{NL}}

\newcommand{\tnl}{\tau\sub{NL}}
\newcommand{\gnl}{g\sub{NL}}

\newcommand\pz{P_\zeta}
\newcommand\bz{B_\zeta}
\newcommand\tz{T_\zeta}
\newcommand\nz{n_\zeta}
\newcommand\gz{g_\zeta}
\newcommand\dn{\delta N}
\newcommand\no{\nonumber}
\begin{document}
\frontmatter
\renewcommand\bibname{REFERENCES}
\renewcommand\contentsname{CONTENTS}
\renewcommand{\listfigurename}{LIST OF FIGURES}
\renewcommand\appendixname{}
\chaptermark{\thechapter}

\renewcommand{\footnoterule}{\vspace*{-3pt}
  \noindent\rule{5cm}{1pt}\vspace*{2.6pt}}

%%%%%%%%%%%%%%%%%%%%%%%%%%%%%%%%%%%%%%%%%%%%%%%%%%%%%%%%%%%%%%%%%%%%%%%%%%%%%%%%%%%%%%%%%%%%%%%%%%%%%%%%%
%%%%%%%%%%%%%%%%%%%%%%%%%%%%%%%%%%%%%%%%%%%%%%%%%%%%%%%%%%%%%%%%%%%%%%%%%%%%%%%%%%%%%%%%%%%%%%%%%%%%%%%%%
\begin{titlepage}
\begin{center} $$ \left . \right.$$
{\LARGE \bf  NON-GAUSSIANITY AND STATISTICAL\\ ANISOTROPY IN COSMOLOGICAL \\ INFLATIONARY MODELS\\}

\vfill by

\vfill{{\Large \bf C\'ESAR ALONSO VALENZUELA TOLEDO}\\{\large Physicist, MSc}}

\vfill{\centering\includegraphics[width=0.45\textwidth]{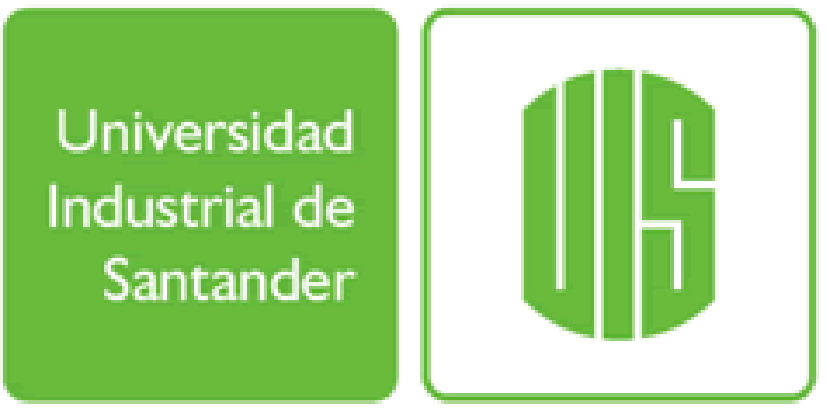}}
\vfill{
\sc Grupo de Investigaci\'on en Relatividad y Gravitaci\'on \\
\sc Escuela de F\'isica, Universidad Industrial de Santander \\
\sc Ciudad Universitaria, Bucaramanga, Colombia\\
\sc Grupo de f\'isica-fenomenolog\'ia  de part\'iculas elementales y cosmolog\'ia
\sc Centro de Investigaciones, Universidad Antonio Nari\~no\\
\sc  Cra 3 Este \# 47A-15, Bogot\'a D.C., Colombia.}
\vspace{\stretch{0.3}}
\rule{\textwidth}{3pt}
\end{center}
\end{titlepage}
%%%%%%%%%%%%%%%%%%%%%%%%%%%%%%%%%%%%%%%%%%%%%%%%%%%%%%%%%%%%%%%%%%%%%%%%%%%%%%%%%%%%%%%%%%%%%%%%%%%%%%%%%
%%%%%%%%%%%%%%%%%%%%%%%%%%%%%%%%%%%%%%%%%%%%%%%%%%%%%%%%%%%%%%%%%%%%%%%%%%%%%%%%%%%%%%%%%%%%%%%%%%%%%%%%%
\begin{titlepage} \begin{center} $$ \left . \right.$$
{\LARGE \bf  NON-GAUSSIANITY AND STATISTICAL\\ ANISOTROPY IN COSMOLOGICAL \\ INFLATIONARY MODELS\\}
\vfill{{\Large \bf C\'ESAR ALONSO VALENZUELA TOLEDO}\\{\large Physicist, MSc}}
\vfill { \large \sc Thesis directed by
\\ \large \bf Dr. YEINZON RODR\'IGUEZ GARC\'IA}
\vfill{\centering\includegraphics[width=0.45\textwidth]{UIS.eps}
\vfill \large\sc A thesis submmited in partial fulfillment of the requirements for the degree of
Doctor in Sciences -- Physics\\
\vspace{0.8cm}
 \sc \today}
\vspace{\stretch{0.3}}
\rule{\textwidth}{3pt}
\end{center}
\end{titlepage}

%%%%%%%%%%%%%%%%%%%%%%%%%%%%%%%%%%%%%%%%%%%%%%%%%%%%%%%%%%%%%%%%%%%%%%%%%%%%%%%%%%%%%%%%%%%%%%%%%%%%%%%%%
%%%%%%%%%%%%%%%%%%%%%%%%%%%%%%%%%%%%%%%%%%%%%%%%%%%%%%%%%%%%%%%%%%%%%%%%%%%%%%%%%%%%%%%%%%%%%%%%%%%%%%%%%

\begin{titlepage}
\begin{flushleft}
\vspace*{0.25\textheight} \hspace*{0.45\textwidth}
\begin{minipage}{0.5\textwidth}
\begin{flushright}
{\fontfamily{pzc}\selectfont To Juan Camilo, C\'esar David and Mar\'ia Sof\'ia.}
\end{flushright}
\end{minipage}
\end{flushleft}
\end{titlepage}

%To the memory of my grandfather Juan Evangelista.
%%%%%%%%%%%%%%%%%%%%%%%%%%%%%%%%%%%%%%%%%%%%%%%%%%%%%%%%%%%%%%%%%%%%%%%%%%%%%%%%%%%%%%%%%%%%%%%%%%%%%%%%%
%%%%%%%%%%%%%%%%%%%%%%%%%%%%%%%%%%%%%%%%%%%%%%%%%%%%%%%%%%%%%%%%%%%%%%%%%%%%%%%%%%%%%%%%%%%%%%%%%%%%%%%%%

\chapter*{ACKNOWLEDGMENTS}

The time has come to thank each one of the people who made this Thesis possible. I would like to begin with my parents and 
siblings, because they have been a constant source of support since I left my home some 15 years ago. Without them, there 
would not either be BSc, nor MSc, nor PhD thesis. So, thanks a lot.

I want to thank my supervisor Yeinzon Rodr\'iguez Garc\'ia for trusting me and for making possible the realization of this 
thesis. It was a real pleasure to be able to work with him. He started as my advisor but now he is one of my best friends.

I must also want to thank the members of the Grupo de Investigaci\'on en Relatividad y Gravitaci\'on (GIRG).

I want to thank the members of the Cosmology and Astroparticle Physics Group at  the University of Lancaster specially to David Lyth for their hospitality during my visit in 2008. 

Finally, it is time for me to apologize with my son Juan Camilo. Many times I changed his company for academic things, related to this 
thesis. Many times we stopped going to the park, going walking, going playing, ... please forgive me.

\vspace{2cm}
{\it Bucaramanga COL, \today}

\makeatletter
\def\@fnsymbol#1{\ensuremath{\ifcase#1\or *\or \dagger\or \ddagger\or
\mathsection\or \mathparagraph\or \|\or **\or \dagger\dagger \or
\ddagger\ddagger\or \mathsection\mathsection \or
\mathparagraph\mathparagraph \or \|\|\else\@ctrerr\fi}}
\renewcommand{\thefootnote}{\arabic{footnote}}
\makeatother

%%%%%%%%%%%%%%%%%%%%%%%%%%%%%%%%%%%%%%%%%%%%%%%%%%%%%%%%%%%%%%%%%%%%%%%%%%%%%%%%%%%%%%%%%%%%%%%%%%%%%%%%%
%%%%%%%%%%%%%%%%%%%%%%%%%%%%%%%%%%%%%%%%%%%%%%%%%%%%%%%%%%%%%%%%%%%%%%%%%%%%%%%%%%%%%%%%%%%%%%%%%%%%%%%%%
\begin{titlepage}

{\bf T\'ITULO :} NO GUSSIANIDAD Y ANISOTROP\'IA ESTAD\'ISTICA EN MODELOS COSMOL\'OGICOS INFLACIONARIOS
\footnote{T\'esis de Doctorado.}.

{\bf AUTOR :} VALENZUELA TOLEDO, C\'esar
Alonso \footnote{Facultad de Ciencias, Escuela de F\'isica,
Yeinzon Rodr\'iguez Garc\'ia (Director).}.

{\bf PALABRAS CLAVES:} Cosmolog\'ia, No gaussianidad, Inflaci\'on, Anisotrop\'ia estadist\'ica, Teor\'ia de perturbaciones cosmol\'ogicas, Perturbaci\'on primordial en la curvatura.

{\bf DESCRIPCI\'ON:}
Se estudian los descriptores estad\'isticos para algunos modelos cosmol\'ogicos inflacionarios que permiten
obtener altos niveles de no gaussianidad y violacion de la isotrop\'ia estad\'istica. B\'asicamente, se estudian
dos tipos de modelos: modelos que involucran s\'olo campos escalares, particularmente un modelo inflacionario de
rodadura lenta con potencial escalar cuadr\'atico de dos componentes con t\'erminos cin\'eticos can\'onicos, y modelos que 
incluyen campos escalares y vectoriales.

Se muestra que para el modelo de rodadura lenta con potencial escalar cuadr\'atico de dos componentes,
es possible obtener valores altos y observables para los niveles de no gaussianidad $f_{NL}$ y $\tnl$ en el 
bi-espectro $B_\zeta$ y en el tri-espectro $\tz$, respectivamente, de la perturbaci\'on primordial en la curvatura $\zeta$. 
Se consideran contribuciones a nivel \'arbol y a un lazo en el espectro $\pz$, en el bi-espectro $B_\zeta$ y en el tri-espectro 
$\tz$. Se muestra que valores considerables se pueden obtener aun cuando $\zeta$ es generada durante inflaci\'on. Cinco aspectos 
son considerados cuando se extrae el espacio disponible de par\'ametros: 1. El asegurar la existencia de un r\'egimen
perturbativo de tal manera que la expansi\'on en serie de $\zeta$, y su truncamiento, sean v\'alidas. 2. El determinar las condiciones correctas que 
determinan el peso relativo de las correcciones a nivel \'arbol y a un lazo. 3. El satisfacer la condici\'on de normalizaci\'on de
espectro. 4. El cumplir la restricci\'on observacional del \'indice espectral. 5. El asegurar un monto de inflaci\'on
m\'inimo necesario para resolver el problema de horizonte.

Para los modelos que incluyen campos escalares y vectoriales, nuevamente se estudia  el espectro $\pz$, el bi-
espectro $B_\zeta$ y el tri-espectro $\tz$ de la perturbaci\'on primordial en la curvatura, cuando $B_\zeta$ y
$\tz$ son generados por perturbaciones escalares y vectoriales. Se estudian las contribuciones a nivel \'arbol y a
un lazo, considerando que las ultimas puedan dominar sobre las primeras. Se calculan los niveles de no
gaussianidad  $f_{NL}$ y $\tnl$, y se encuentran relaciones de consistencia entre \'estos y el nivel de anisotrop\'ia 
estad\'istica $\gz$ en el espectro $\pz$, concluyendo que para valores peque\~nos de $\gz$ los niveles de no-gaussianidad 
pueden ser altos, en algunos casos excediendo las cotas observacionales actuales.

\end{titlepage}

%%%%%%%%%%%%%%%%%%%%%%%%%%%%%%%%%%%%%%%%%%%%%%%%%%%%%%%%%%%%%%%%%%%%%%%%%%%%%%%%%%%%%%%%%%%%%%%%%%%%%%%%%
%%%%%%%%%%%%%%%%%%%%%%%%%%%%%%%%%%%%%%%%%%%%%%%%%%%%%%%%%%%%%%%%%%%%%%%%%%%%%%%%%%%%%%%%%%%%%%%%%%%%%%%%%

\begin{titlepage}

{\bf TITLE}: NON-GAUSSIANITY AND STATISTICAL ANISOTROPY IN COSMOLOGICAL INFLATIONARY MODELS
\footnote{PhD Thesis.}.

{\bf AUTHOR:} VALENZUELA TOLEDO,
C\'esar Alonso \footnote{Facultad de Ciencias, Escuela de F\'isica,
Yeinzon Rodr\'iguez Garc\'ia (Supervisor).}.

{\bf KEY WORDS:} Cosmology, Non-gaussianity, Inflation, Statistical anisotropy, Cosmological perturbation theory,
Primordial curvature perturbation.

{\bf DESCRIPTION:}
We study the statistical descriptors for some cosmological inflationary models that allow us to get large levels of 
non-gaussianity and violations of statistical isotropy. Basically, we study two different class of models: a model
that include only scalar field perturbations, specifically a subclass of small-field {\it slow-roll} models of
inflation with canonical kinetic terms, and models that admit both vector and scalar field perturbations.

We study the former to show that it is possible to attain very high, {\it including observable}, values for
the levels of non-gaussianity $f_{NL}$ and $\tnl$ in the bispectrum $B_\zeta$ and trispectrum $\tz$ of the
primordial curvature perturbation $\zeta$ respectively. Such a result is obtained by taking care of loop
corrections in the spectrum $P_\zeta$, the bispectrum $B_\zeta$ and the trispectrum $\tz$. Sizeable values for
$f_{NL}$ and $\tnl$ arise even if $\zeta$ is generated during inflation. Five issues are considered when
constraining the available parameter space:  1. we must ensure that we are in a perturbative regime so that the
$\zeta$ series expansion, and its truncation, are valid. 2. we must apply the correct condition for the (possible)
loop dominance in $B_\zeta$ and/or $P_\zeta$. 3. we must satisfy the spectrum normalisation condition.  4. we must
satisfy the spectral tilt constraint. 5. we must have enough inflation to solve the horizon problem.

For the latter we study the spectrum $\calp_\zeta$, bispectrum
$B_\zeta$ and trispectrum of the primordial curvature perturbation  when $\zeta$ is generated by
scalar and vector field perturbations. The tree-level and one-loop contributions from vector field perturbations
are worked out considering the possibility that the one-loop contributions may be dominant over the tree level
terms. The levels of non-gaussianity  $f_{NL}$ and $\tnl$, are calculated and related to  the level of statistical
anisotropy in the power spectrum, $g_\zeta$. For very small amounts of statistical anisotropy in the power
spectrum, the levels of non-gaussianity may be very high, in some cases exceeding the current observational limit.

\end{titlepage}

%%%%%%%%%%%%%%%%%%%%%%%%%%%%%%%%%%%%%%%%%%%%%%%%%%%%%%%%%%%%%%%%%%%%%%%%%%%%%%%%%%%%%%%%%%%%%%%%%%%%%%%%%
%%%%%%%%%%%%%%%%%%%%%%%%%%%%%%%%%%%%%%%%%%%%%%%%%%%%%%%%%%%%%%%%%%%%%%%%%%%%%%%%%%%%%%%%%%%%%%%%%%%%%%%%%
\tableofcontents
\listoffigures

\mainmatter

%%%%%%%%%%%%%%%%%%%%%%%%%%%%%%%%%%%%%%%%%%%%%%%%%
%%%%%%%%%%%%%%%%%%%%%%%%%%%%%%%%%%%%%%%%%%%%%%%%%
\chapter{INTRODUCTION}                        %%%
%%%%%%%%%%%%%%%%%%%%%%%%%%%%%%%%%%%%%%%%%%%%%%%%%
%%%%%%%%%%%%%%%%%%%%%%%%%%%%%%%%%%%%%%%%%%%%%%%%%

\begin{figure}[!h]
\begin{center}
\includegraphics[scale=0.55]{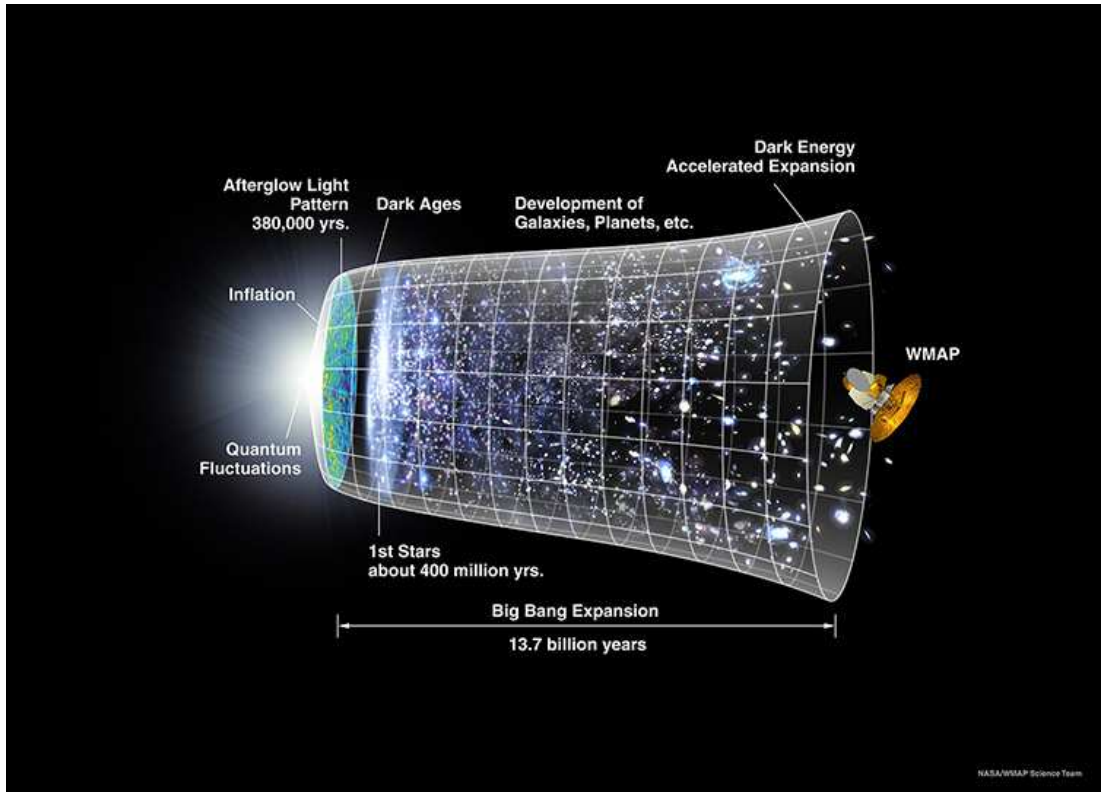}
\end{center}
\caption[A representation of the evolution of the universe over 13.7 billion years.]{A representation of the
evolution of the universe over 13.7 billion years. The far left depicts the earliest moment we can now probe, when
a period of `` inflation " produced a burst of exponential growth in the universe. (Size is depicted by the
vertical extent of the grid in this graphic.) For the next several billion years, the expansion of the universe
gradually slowed down as the matter in the universe pulled on itself via gravity. More recently, the expansion has
begun to speed up again as the repulsive effects of dark energy have come to dominate the expansion of the
universe. The afterglow light seen by WMAP was emitted about 380,000 years after inflation and has traversed the
universe largely unimpeded since then. The conditions of earlier times are imprinted on this light; it also forms
a backlight for later developments of the universe (Courtesy of the NASA/WMAP Science Team \cite{wmap}).}
\label{cmbe}
\end{figure}

The corner-stone of modern cosmology is that, at least on large scales, the visible universe seems to be the same in all
directions around us and around all points, i.e. the Universe is almost homogeneous and isotropic. This is borne out by a 
variety of observations, particulary observations of cosmic microwave background (CMB); this radiation has been traveling to us 
for about 14000 million years (see \hbox{Fig. \ref{cmbe}}), supporting the conclusion that the Universe at sufficiently large 
distances is nearly the same. On the other hand, it is apparent that nearby regions of the observable Universe are at present 
highly inhomogeneous, with material clumped into stars, galaxies and galaxy clusters. It is believed that these structures have 
formed over the time via gravitational attraction, from a distribution that was more homogeneous in the past.

The large-scale behavior of the Universe can be described by assuming a homogeneous background. On this background, we can 
superimpose the short scale irregularities. For much of the evolution of the observable Universe, these irregularities can be 
considered to be small perturbations on the evolution of the background (unperturbed) Universe. The metric of unperturbed 
Universe is called the Friedman-Leimatre-Roberson-Walker metric, and its line element can be to written as:
\be
ds^2=-dt^2+a^2(t)\(dr^2+r^2(d\theta^2+\sin\phi^2d\phi^2\),
\ee
where $a(t)$ is the scale factor and $r,\theta,\phi$ are the spherical comoving coordinates\footnote{A particle in this metric
have fixed-coordiantes.}

The model described by the above metric is known as the standard cosmological model (known also as Big-Bang cosmological 
model) \cite{friedman1,robertson1,robertson2,robertson3,walker} and is the successful
framework that describes the observed properties of the Universe: homogeneity and isotropy at large scales, Hubble
expansion, almost 14 billion years of evolution in agreement with globular clusters and radioactive isotopes
dating, cosmic microwave background radiation (CMB) confirmed by Penzias and Wilson's discovery in 1965
\cite{dicke,penziaswilson}, and the relative abundances of light elements
\cite{alpher1,alpher2,gamow,hoyle,olive,wagoner,walkeretal} in full agreement with observation.

\begin{figure*}[!h]
\begin{center}
\includegraphics[scale=1.2]{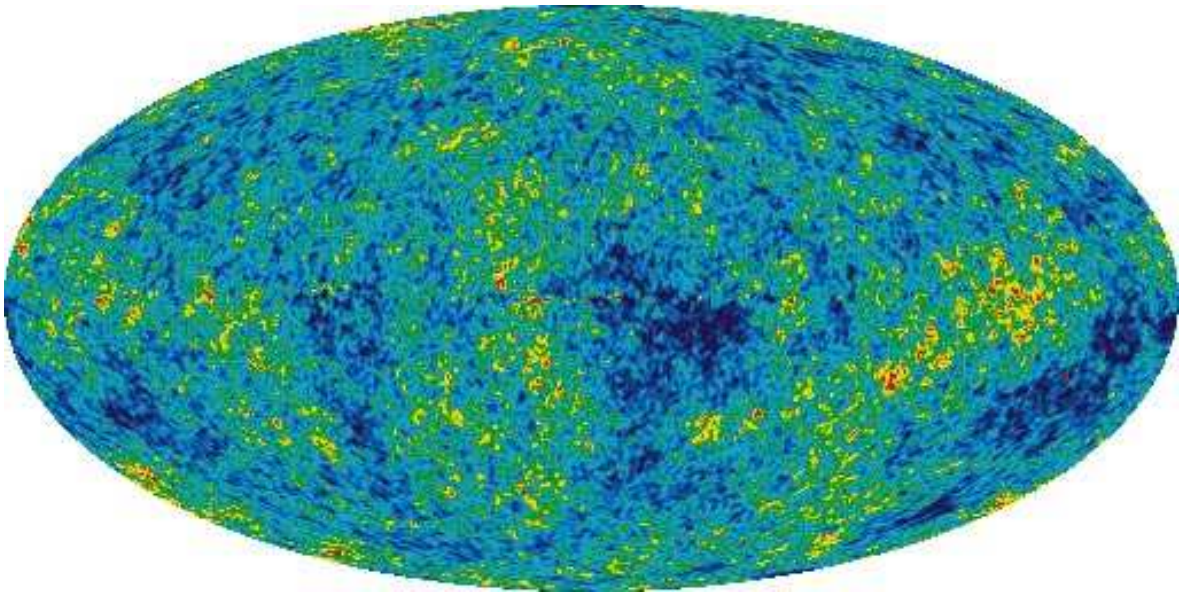}
\end{center}
\caption[CMB temperature anisotropies as seen by the WMAP satellite.]{CMB temperature anisotropies as seen by the WMAP
satellite (five years resuls) \cite{wmap5}. The oval shape is a projection to display the whole sky. The temperature
anisotropies are found to be of the order of 1 part in $10^5$. The background temperature is \hbox{$T_0 = 2.725 \pm 0.002$ K};
regions at that temperature are in very light blue. The hottest regions (in red) correspond to $\Delta T \simeq 200 \mu{\rm 
K}$. The coldest regions (in very dark blue) correspond to $\Delta T \simeq -200 \mu{\rm K}$ (Courtesy of the NASA/WMAP Science 
Team \cite{wmap}).}
\label{wmapsky}
\end{figure*}

\begin{figure*}[!h]
\begin{center}
\includegraphics[scale=0.80]{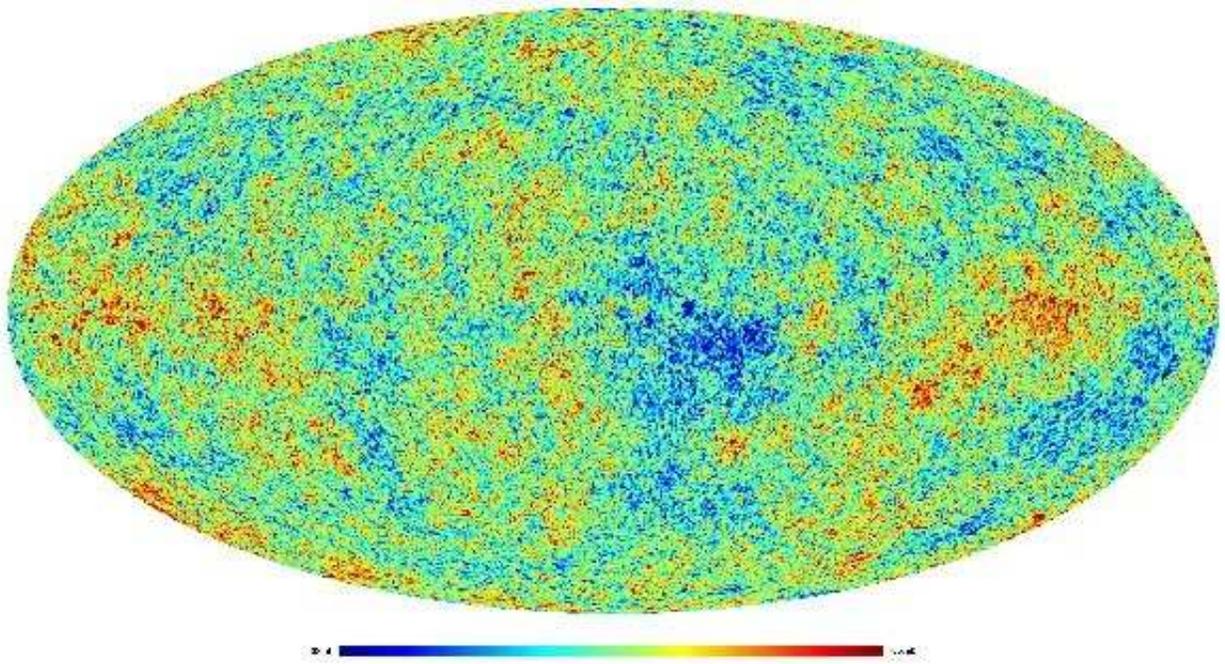}
\end{center}
\caption[Simulation of the CMB temperature anisotropies as seen by the PLANCK satellite.]{Simulation of the CMB temperature
anisotropies as seen by the PLANCK satellite. PLANCK will provide a map of the CMB field at all angular resolutions greater
than 10 arcminutes and with a temperature resolution of the order of 1 part in $10^6$ (ten times better than WMAP) (Courtesy of
ESA's PLANCK mission \cite{planck}).}
\label{plancksky}
\end{figure*}

\begin{figure*}[!h]
\begin{center}
\includegraphics[scale=0.5]{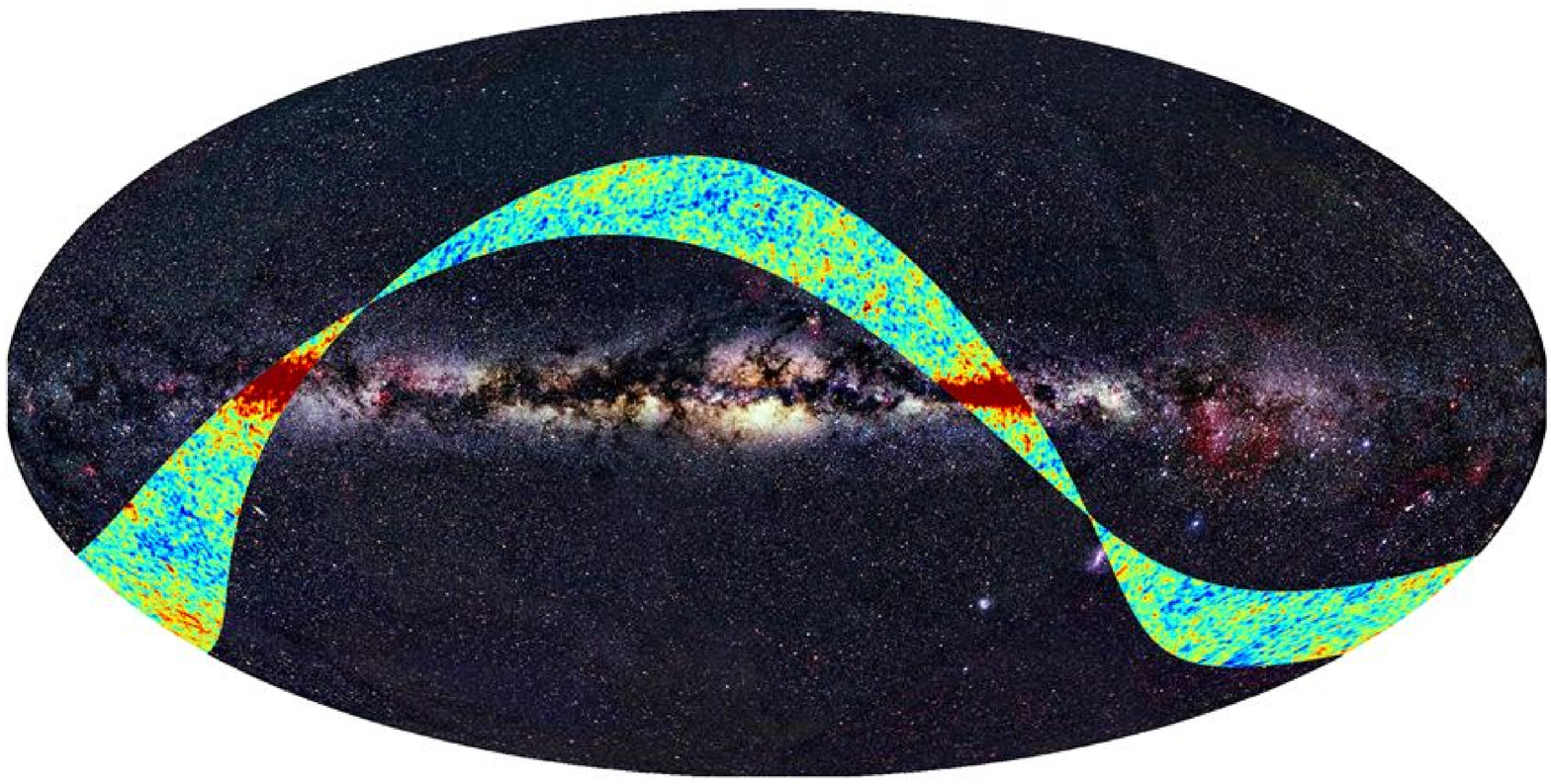}
\end{center}
\caption[The CMB temperature anisotropies as seen by the PLANCK satellite.]{A map of the area of the sky mapped by 
PLANCK during the first light survey. The colours indicate the magnitude of the deviations of the temperature of the Cosmic 
Microwave Background from its average value (red is hotter and blue is colder). (Courtesy of ESA's PLANCK mission
\cite{planck}).}
\label{plancksky2}
\end{figure*}

The introduction of a period of exponential expansion (called inflationary) \cite{albste,guth,linde82a}, prior to
the Big-Bang, brought an elegant solution to the horizon, flatness, and unwanted relics problems that were present
in the original standard cosmological model \cite{albste,guth,kolb,linde82a,riotto1}. In spite of its success at
solving the above mentioned problems, the inflationary period became perhaps more important because of its ability
to stretch the quantum fluctuations of the fields living in the FRW spacetime
\cite{bst,guthpi82,hawking,linde82a,mukhanov1,mukhanovrep,riotto1,starobinsky1}, making them classical
\cite{albrecht1,Burgess:2006jn,grishchuk,guthpi,Kiefer:2008ku,lombardo,lyth84,lythbook,Lyth:2006qz,Martineau:2006ki,Nambu:2008my}
and almost constant soon after horizon exit. They correspond to small inhomogeneities in the energy density and
are responsible, via gravitational attraction, of the large-scale structure seen today in the Universe. If this
scenario turned to be correct, the energy density inhomogeneities should have left their trace in the CMB released
at the time of recombination. Indeed, the Cosmic Background Explorer (COBE) in 1992 \cite{cobe} found and mapped
small anisotropies in the CMB temperature of the order of 1 part in $10^5$ (with average temperature \mbox{$T_0 =
2.725 \pm 0.002$ K} \cite{bennett}), on scales of order thousands of Megaparsecs. With 30 times better angular
resolution and sensitivity than COBE, the Wilkinson Microwave Anisotropy Probe (WMAP) \cite{wmap} confirmed this
picture (see \hbox{Fig. \ref{wmapsky}}), measuring in turn the cosmological parameters with a $1\%$ order
precision \cite{wmap5} on scales of order tens of Megaparsecs. The PLANCK satellite \cite{planck,planck1},
launched in may 2009, will be able to refine these observations (see Fig. \ref{plancksky} and \ref{plancksky2}). With 10 times better
angular resolution and sensitivity than WMAP, PLANCK promises to determine the temperature anisotropies with a
resolution of the order of 1 part in $10^6$, and the cosmological parameters with a $0.1\%$ order precision.

The anisotropies in the CMB temperature\footnote{From now on, and unless otherwise stated, the perturbation $\delta y$ in any
quantity $y$ will be regarded as first-order in cosmological perturbation theory. Unperturbed quantities will be denoted by a
subscript 0 unless otherwise stated.}
$\delta T/T_0$ are directly related to the perturbation in the spatial curvature $\zeta$ (Sachs-Wolfe effect), whose primarily
origin is the stretched quantum fluctuations of one or several scalar fields $\phi_i$ that fill the Universe during inflation
\cite{lythbook,sachs}\footnote{In this and the following expressions the subscripts $k$ stand for the Fourier modes with 
comoving wavenumber $k$.}:
\be
\left(\frac{\delta T}{T_0}\right)_k = -\frac{1}{5} \, \zeta_k \,. \label{connections1}
\ee
The quantity $\zeta$ is related to the perturbation in the intrinsic curvature of space-time slices with uniform energy density
\cite{malikwands}:
\be
^{(3)}R=\frac{4}{a^2}\;\nabla^2\psi\;,
\ee
where $\psi$ is the first order scalar perturbation in the spacial metric.

Astronomers work with the observable quantity $\delta T/T$ and theoretical cosmologists work with $\zeta$. Therefore, we may 
study the statistical porperties of the observed $\delta T/T$ through the spectral functions associated with the primordial
curvature perturbation $\zeta$, whose properties are in general model dependent. Knowing the statistical 
descriptors of $\zeta$ for some particular and well motivated cosmological model proposed for the origin of large scale 
strucutre, we can reject the model or keep it, because some of the statistical descriptors for $\delta T/T$ are known with good 
acuracy or at least have an upper bound \cite{wmap5}.

The statistical properties of the CMB temperature anisotropies can be then described in terms of the spectral functions, like 
the spectrum, bispectrum, trispectrum, etc., of the primordial curvature perturbation $\zeta$. This spectral functions are 
given in terms of other quantities, which have an observational value or an uppper bound. For example, the spectrum $\pz$ is 
parametrized in terms of an amplitude $\calpz\half$, a spectral index $\nz$ and the level of statistical anisotropy $\gz$; the 
bispectrum $\bz$ and trispectrum $\tz$ are parametrized in terms of products of the spectrum $\pz$ and the quantities $\fnl$,  
and $\tnl$ and $\gnl$, respectively. As we will see in the next chapter, the statistical descriptors $\fnl$, $\tnl$ and $\gnl$ 
are usually called levels of non-gaussianity, because non zero values for these quantities imply non-gaussianity in the 
primordial curvature perturbation $\zeta$ as well in the constrast in the temperature of the CMB radiation $\delta T/T$.
The non-gaussian characteristics in the CMB are actually present in the observation \cite{wmap5} as we will see in more
detail in Section \ref{observational}. The status of observation can be summarized as follows\footnote{We are using values 
according of the five year of data from NASA's WMAP satellite \cite{wmap5}}: the spectral amplitude 
$\calpz\half=(4.957\pm 0.094)\times 10^{-5}$ \cite{bunn}, the spectral index $\nz=0.960\pm 0.014$ at $2 \sigma$ \cite{wmap5}, 
the level of non-gaussianity $\fnl$ in the bispectrum is in the range $-9 < f_{NL} < 111$ at $2\sigma$ \cite{wmap5}; and there 
is no observational bound on the levels of non-gaussinity $\tnl$ and $\gnl$ in the trispectrum $\tz$. The amount of statistical 
anisotropy $\gz$ in the spectrum $\pz$ is in the range $\gz \simeq 0.290 \pm 0.093$ \cite{gawe}.

Regarding the statistical descriptors, non-gaussianity in the primordial curvature perturbation $\zeta$ is one of the subjects 
of more interest in modern cosmology, because the non-gaussianity parameters $\fnl$ and $\tnl$ together with the spectrum 
amplitude $A_\zeta$ and spectral index $n_\zeta$ allow us to discriminate between the different models proposed for the origin 
of the
large-scale structure (see for example Refs. \cite{alabidi3,alabidi1,alabidi2}). The most studied and popular 
models are those called the slow-roll models with canonical kinetic terms, because of their 
simplicity and because they easily satisfy the spectral index $\nz$ requierements from observation. However, the usual 
predictions of these models is that the levels of non-gaussianity in the primordial curvature perturbation are expected to be 
unobservable  \cite{battefeld,maldacena,seery7,vernizzi,yokoyama1}.  However, as we will show in chapters \ref{chaptsca} 
and \ref{chaptsca2}, there are some aditional issues that have not been taken into account in the current literature. We study 
these issues to show that it is possible to generate sizeable and observable levels of non-gaussianity in a subclass of small-
field {\it slow-roll} inflationary models with canonical kinetic terms; our main conclussion 
is that if non-gaussianity is detected, the aforementioned models could have strong possibilities to be the ones responsibles 
for the formation of the large-scale structure. 

According to the usual assumption, one or more of these scalar field
perturbations are responsible for the curvature perturbation. In that case, the $n$-point correlators of $\zeta$ are
translationally and rotationally invariants. However, violations of such invariances entail modifications of the usual
definitions for the spectral
functions in terms of the statistical descriptors \cite{acw,armendariz,carroll}.
These violations may be consequences either of the presence of vector field perturbations
\cite{armendariz,bdmr,vc,vc2,RA2,dklr,dkw,dkw2,go,gmv2,gmv,gvnm,himmetoglu3,himmetoglu,himmetoglu2,himmetoglu4,kksy,dkl,koh,ys},
spinor field perturbations \cite{bohmer,shan}, or p-form perturbations \cite{germani,germani2,kobayashi,koivisto,koivisto2}, 
contributing significantly to $\zeta$,
of anisotropic expansion \cite{bamba,bohmer,dechant,gcp,himmetoglu4,kksy,koivisto,ppu1,ppu2,watanabe} or of an
inhomogeneous background \cite{armendariz,carroll,dklr}.
Violation of the statistical isotropy (i.e. violation of the rotational invariance in the
$n$-point correlators of $\zeta$) seems to be present in the data \cite{app,ge,hl,samal} and, although its statistical
significance is still low, the continuous presence of anomalies in every CMB data analysis (see for instance Refs.
\cite{bunn1,dvorkin,dipole2,dipole1,hansen,dipole3,hoftuft,hou,land1,land2,oliveira,schwarz,tegmark}) suggests the evidence
might be decisive in the forthcoming years. The presence of vector fields in the inflationary dynamics is not only important
to be responsible of violations of the statistical isotropy, they also may generate sizeable 
levels of non-gaussianity described by $\fnl$ and $\tnl$; particularly we will show in Chapter \ref{chaptvec}, that including
vector fields allows us to get consistency relations between the statistical descriptors, more precisely between the 
non-gaussianity levels $\fnl$ and $\tnl$ and the amount of statistical anisotropy $\gz$.

Because of the progressive improvement in the accuracy of the satellite measurements described above, it is pertinent to study
the statistical descriptors of the primordial curvature perturbation $\zeta$, for cosmological models of the origin of the 
large-scale structure in the Universe. It is very important because they could be a crucial tool to discriminate between some 
of most usual cosmological models \cite{alabidi1,alabidi2}.

The layout of the thesis is the following: the Chapter \ref{chaptgen} is devoted to study the statistical descriptors for a 
probability distribution function and its relation with the observational parameters, i.e., the spectrum amplitude, the 
spectral index the levels of non-gaussianity $\fnl$ and $\tnl$ in the bispectrum $\bz$ and trispectrum $\tz$ respectively and 
the level of statistical anisotropy in the power spectrum, $\gz$. In this chapter, we also review some generalities of the 
$\dn$ formalism, it has become the standard technique to calculate $\zeta$ and its statistical descriptors. In Chapters 
\ref{chaptsca} and \ref{chaptsca2} we show that it is possible to attain very high, {\it including observable}, values for the 
levels of non-gaussianity $f_{NL}$ and $\tnl$, in a subclass of small-field {\it slow-roll} models of inflation with canonical 
kinetic terms. Comparison with current observationally bounds is made. Chapter \ref{chaptvec} is devoted to study the 
statistical descriptors of the primordial curvature perturbation $\zeta$  when scalar and vector fields perturbations are 
present in the inflationary dynamicc. The levels of non-gaussianity $\fnl$ and $\tnl$ are calculated and related to the level 
of statisitcal anisotropy in the power spectrum, $\gz$. We show that the levels of non-gaussianity may be very high, in some 
cases exceeding the current observationally limit. Finally we conclude in Chapter \ref{chaptconclu}.

%%%%%%%%%%%%%%%%%%%%%%%%%%%%%%%%%%%%%%%%%%%%%%%%%%%%%%%%%%%%%%%%%%%%%%%%%%%%%%%%%%%%%%%%%%%%%%%%%%
%%%%%%%%%%%%%%%%%%%%%%%%%%%%%%%%%%%%%%%%%%%%%%%%%%%%%%%%%%%%%%%%%%%%%%%%%%%%%%%%%%%%%%%%%%%%%%%%%%
\chapter{$\dn$ FORMALISM AND STATISTICAL DESCRIPTORS FOR $\zeta$}\label{chaptgen}       %%%%%%%%%
%%%%%%%%%%%%%%%%%%%%%%%%%%%%%%%%%%%%%%%%%%%%%%%%%%%%%%%%%%%%%%%%%%%%%%%%%%%%%%%%%%%%%%%%%%%%%%%%%%
%%%%%%%%%%%%%%%%%%%%%%%%%%%%%%%%%%%%%%%%%%%%%%%%%%%%%%%%%%%%%%%%%%%%%%%%%%%%%%%%%%%%%%%%%%%%%%%%%%

%%%%%%%%%%%%%%%%%%%%%%%%%%%%%%%%%%%%%%%%%%%%%%%%%%%%%%%
\section{Introduction}                          %%%%%%%
%%%%%%%%%%%%%%%%%%%%%%%%%%%%%%%%%%%%%%%%%%%%%%%%%%%%%%%
The primordial curvature perturbation $\zeta$, as well as the contrast in the temperature of the cosmic microwave background
radiation $\delta T/T$ and the gravitational potential $\Phi_g$, are examples of cosmological functions of space and time being
described by probability distribution functions.  In particular, the probability distribution function $f(\zeta)$ for $\zeta$
has well defined statistical descriptors which depend directly upon the particular inflationary model and that are suitable for
comparison with present observational data. Such a comparison allows us either to reject or to keep particular inflationary
models as those which better represent nature's behaviour.  In this chapter we give a complete
and general description of the $\dn$ formalism including both scalar and vector fields\footnote{ Vector fields
will be responsible of violations of the statistical isotropy, as we will see in Chapter \ref{chaptvec}.}. This formalism is a 
powerful tool and it is commonly used to calculate the primordial curvature perturbation $\zeta$. We also present a 
cosmologically motivated description of the statistical descriptors for probability distribution functions, focusing mainly on 
$f(\zeta)$. Finally, in Section \ref{observational} we give the  observational constraints for $\zeta$.

%%%%%%%%%%%%%%%%%%%%%%%%%%%%%%%%%%%%%%%%%%%%%%%%%%%%%%%%%%%
\section{The $\delta N$ formalism}\label{sdeln} %%%%%%%%%%%
%%%%%%%%%%%%%%%%%%%%%%%%%%%%%%%%%%%%%%%%%%%%%%%%%%%%%%%%%%%
The $\delta N$ formalism \cite{dklr,lms,lr,ss,st,starobinsky} provides a powerful method for calculating $\zeta$ and all its
statistical descriptors at any desired order in cosmological perturbation theory. The $\delta N$ formalism for scalar field
perturbations was given at the linear level in Refs. \cite{ss,starobinsky} and at the non-linear level, which generates  non-
gaussianity, it was described in Refs. \cite{lms,lr}. In a recent paper \cite{dklr} the $\delta N$ formalism was extended to
include vector as well as scalar fields. In this section, we give a brief review of the formalism without assuming 
statistical isotropy and define the primordial curvature perturbation $\zeta$.

In the cosmological standard model \cite{mukhanov} the observable Universe is homogeneous and isotropic, being described by the
unperturbed Friedmann-Robertson-Walker metric whose line element, for a spatially flat universe, looks as follows:
\begin{equation}
ds^2 = -dt^2 + a^2(t) \delta_{ij} dx^i dx^j \,,
\end{equation}
where $a(t)$ is the global expansion parameter, $t$ is the cosmic time, and ${\bf x}$ represents the position in cartesian
spatial coordinates.
The homogeneity and isotropy conditions describe very well the Universe at large scales, but departures from the unperturbed
background are observationally evident at smaller scales.

One way to parametrize the departures from the homogeneous and isotropic background is to include perturbations in the metric,
for which we have to define a slicing and a threading. The slicing will be defined so that the energy density in fixed-$t$
slices of spacetime is uniform. The threading will correspond to comoving fixed-$x$ world lines. With generic coordinates the
perturbed metric of the perturbed universe may be defined as
\be
ds^2 = g_\mn dx^\mu dx^\nu \,.
\ee
To define the cosmological perturbations, one chooses a coordinate
system in the perturbed universe,  and then compares that universe  with an
unperturbed one. The unperturbed universe is taken to be
homogeneous, and  is usually taken to be isotropic as well.
%%%%%%%%%%%%%%%%%%%%%%%%%%%%%%%%%%%%%%%%%%%%%%%%%%%%%%%%%%%%%%%%
\subsection{The curvature perturbation}					%%%%%%%%
%%%%%%%%%%%%%%%%%%%%%%%%%%%%%%%%%%%%%%%%%%%%%%%%%%%%%%%%%%%%%%%%
To define the curvature perturbation, we smooth\footnote{Smoothing a function $g(\bfx)$ means that
$g$ at each location is replaced by its average within a sphere of coordinate
radius $R$ around that position. The averaging may be  done with a smooth
window function such as a gaussian. The smoothed function is supposed to have 
no significant Fourier components with coordinate wavenumbers  $k\gg R^{-1}$, which means that its gradient
at a typical location is at most of order $1/R$. A function $g$ with that property
is said to be `smooth on the scale $R$'.} the metric tensor and the energy-momentum tensor on a comoving scale $R$
and one considers  the super-horizon regime $aR> H\mone$ where $H\equiv \dot a/a$ is the Hubble parameter
and $a(t)$ is the scale factor normalised to 1 at present\footnote{The value of the scale factor at present epoch is usually 
denoted by $a_0$.}. The energy density $\rho$ and 
pressure $P$ are smoothed on the same scale. On the reasonable assumption that the smoothing scale is the biggest relevant
scale,  spatial gradients of the smoothed metric and energy-momentum tensors will be negligible.
As a result,  the evolution  of these quantities at each comoving location will be that of some homogeneous 
`separate universe'. In contrast with earlier works on the separation universe assumption,
we will in this section allow the possibility that the separate universes are anisotropic even though homogoneous.

We consider the  slicing of spacetime with uniform energy density, and the  threading which moves
with the expansion (comoving threading). By virtue of the separate universe assumption, the
threading will be orthogonal to the slicing. The spatial metric  can then be written as
\be
g_{ij}(\bfx,t) \equiv a^2(t)e^{2\zeta(\bfx,t) } \(I e^{2h(\bfx,t)} \)_{ij}
, \label{gij} \ee
where  $I$ is the unit matrix, and the matrix $h$ is traceless so that $Ie^{2h}$ has unit determinant. 
The smoothing scale is chosen to be somewhat shorter than the scales of interest, so that the Fourier components of $\zeta$ on 
those scales is unaffected by the smoothing. The time dependence of the locally defined scale factor
$a(\bfx,t)\equiv a(t) \exp(\zeta)$ defines the rate at which an infinitesimal comoving  volume $\calv$ expands:
$\dot \calv/\calv=3\dot a(\bfx,t)/a(\bfx,t)$.

We split $\ln a$ and $h_{ij}$ into an unperturbed part plus a perturbation:
\bea
\ln a(\bfx,\tau) &\equiv& \ln a(\tau) + \zeta(\bfx,\tau)\,, \\
h_{ij}(\bfx,\tau) &\equiv& h_{ij}(\tau) + \delta h_{ij}(\bfx,\tau)
\,. \eea
The unperturbed parts can be defined as spatial averages within the
observable Universe, but any definition will do as long as it makes the perturbations small
within the observable Universe. If they are small enough, $\zeta$ and $\delta h_{ij}$ can be treated as first-order 
perturbations. That is expected to be the case, with the proviso that a second-order treatment of
$\zeta$ will be necessary to handle its non-gaussianity if that is present at a level corresponding
to $\fnl\lsim 1$ (with the gaussian and non-gaussian components correlated) \cite{lr1}.

Under the reasonable assumption that the Hubble scale $H\mone$ is the biggest relevant
distance scale, the  energy continuity equation $d(\calv \rho)= - P d\calv$ at each 
location is the same as in a homogeneous universe; as far as the evolution of $\rho$
is concerned, we are dealing with a family of separate homogeneous universes. With the additional assumption that the initial 
condition is set by scalar fields during inflation,  the smoothed  $h_{ij}(t)$ is time-independent  after 
smoothing and then the separate universes are  homogeneous as well as isotropic.

Since we are working on slices of uniform $\rho$, the energy continuity equation can
be written 
\be
\dot \rho(t) = - 3 \[ H(t)  + \dot \zeta(\bfx,t) \]
\[ \rho(t) + P(\bfx,t) \]. \label{econt} 
\ee
One write
\be 
P(\bfx,t) = P(t) + \delta P(\bfx,t) 
, \ee
so that $\delta P$ is the pressure perturbation on the uniform density slices, 
and choose $P(t)$ so that the unperturbed quantities satisfy the unperturbed  equation
$\dot\rho = -3H(\rho + P)$. Then
\be
\dot \zeta = - \frac{H\delta P}{\rho + P + \delta P}. \label{zetadot} 
\ee
This gives $\dot\zeta$ if we know $\rho(t)$ and $P(\bfx,t)$. It makes
$\zeta(\bfx)$ time-independent\footnote{Absorving $\dot{\zeta}$ into the unperturbed scale factor.}
during any era when  $P$ is a unique function of  $\rho$  \cite{lms,wmll} (hence uniform on slices of uniform $\rho$).  The 
pressure perturbation is  said to be adiabatic in this case, otherwise it is said to be
non-adiabatic.

The key assumption in the above discussion is that 
in the superhorizon regime certain smoothed quantities
(in this case $\rho$ and $P$)   evolve at each location as they would in an unperturbed
universe. In other words, the evolution of the perturbed universe is that of a family of unperturbed
universes. This is the separate universe assumption, that is useful also in other situations
\cite{lythbook}. 

The primordial curvature perturbation $\zeta$
is directly probed by observation on `cosmological
scales' corresponding to  roughly 
$e^{-15} H_0\mone \lsim k\mone \lsim H_0\mone$\footnote{$H_0$ is the Hubble parameter today.}.
These scale begin to enter the horizon when
 when $T\sim 1\MeV$. The  Universe at that stage
is  radiation dominated to very high accuracy\cite{RDcond3,RDcond4,RDcond1,RDcond2}, 
implying $P=\rho/3$ and a constant curvature perturbation
which we denote simply by $\zeta(\bfx)$, and it is the one constrained by observation as we will see in section 
\ref{observational}. When cosmological scales are the only ones of interest, one should
choose the smoothing scale as $R\sim e^{-15} H_0\mone$. Unless stated otherwise, we make
this choice.

Within a given scenario, $\zeta$ will exist also on smaller inverse wavenumbers, down to some `coherence length'
which might be as low as $k\mone \sim e^{-60}H_0\mone$ (the scale leaving the horizon at the
end of inflation). If one is interested in such scales, the smoothing scale $R$ should be
chosen to be (somewhat less than) the coherence length. 

%%%%%%%%%%%%%%%%%%%%%%%%%%%%%%%%%%%%%%%%%%%%%%%%%%%%%%%%%%%%%%%%
\subsection{The $\delta N$ formula} \label{dnformula}   %%%%%%%%
%%%%%%%%%%%%%%%%%%%%%%%%%%%%%%%%%%%%%%%%%%%%%%%%%%%%%%%%%%%%%%%%
Keeping the comoving threading, we can write the analogue of
 \eq{gij} for a generic slicing:
\be
\tilde g_{ij}(\bfx,\tau) \equiv \tilde a^2(\bfx,\tau) \(Ie^{2\tilde h(\bfx,\tau)} \)_{ij}
, \label{gijtilde} \ee
with again  $Ie^{2\tilde h}$  having unit determinant so that
the rate of volume expansion is $\dot \calv/\calv = 3\tilde a(\bfx,t)$.
Starting with an initial `flat' slicing such that the locally-defined scale
factor is homogeneous, and ending with a slicing of uniform density,
 we then have
\be
\zeta(\bfx,t) = \delta N(\bfx,t)=N(\bfx,t)-N_0(t)
, \ee
where the number of $e$-folds of expansion is defined in terms of the volume
expansion by the usual expression $\dot N = \dot \calv/3\calv$.
The choice of the initial epoch has no effect on $\delta N$, because
the expansion going from one flat slice to another is uniform.
We will choose the  initial epoch to be
a few Hubble times after the smoothing
scale leaves the horizon during inflation.
According to the usual assumption, the
evolution
of the local expansion rate  is determined by the initial values
of one or more of the perturbed  scalar fields $\phi_I$.
Then we can write
\bea
\phi_I(\bfx) &=& \phi_I + \delta\phi_I(\bfx), \\
\zeta(\bfx,t) &=& \delta N(\phi_1(\bfx),\phi_2(\bfx),\ldots,t)
=  N_I(t) \delta\phi_I(\bfx) +\nonumber \\&+& \frac12 N_{IJ}(t)
\delta\phi_I(\bfx) \delta\phi_J (\bfx) + \frac{1}{3\!} N_{IJK}(t)
\delta\phi_I(\bfx) \delta\phi_J (\bfx)\delta\phi_K (\bfx)\ldots , \label{dNsc}
\eea
where $N_I\equiv \pa N/\pa \phi_I$, etc.,\ and the partial derivatives
are evaluated with the fields at their unperturbed values denoted
simply by $\phi_I$. The field perturbations $\delta \phi_I$ in Eq. (\ref{dNsc}) are defined on the `flat' slicing
such that $a(\bfx,t)$ is uniform.

The unperturbed field values
are defined as the spatial averages, over a  comoving
box within which the perturbations are defined. The box size $aL$
should satisfy $LH_0\gg 1$ so that the observable
Universe should fit comfortably inside it \cite{lythbox}. Observations are available  within
the observable universe and, except for the low multipoles of the CMB, all observations
probe scales $k\gg H_0$. To handle them, one should choose the box size as
$L=H_0\mone$ \cite{leblond}. A smaller choice would throw away some of the data while
a bigger choice would make the spatial averages unobservable.
Low multipoles $\ell$ of the CMB anisotropy explore scales of order $H_0\mone/\ell$ not very much smaller than
$H_0\mone$. To handle them  one has to take $L$ bigger than $H_0\mone$.
For most purposes, one should use a  box, such that $\ln(LH_0)$ is just
a few (ie.\ not exponentially large) \cite{klv,lythbox,lythbook}.
When comparing the loop contribution with observation one should normally
set $L=H_0\mone$, except for the low CMB multipoles where one should choose $L\gg H_0\mone$ with
$\ln(kL)\sim 1$. With the choice $L=H_0\mone$, $\ln(kL)\sim 5$ for the scales
explored by the CMB multipoles with $\ell \sim 100$, while $\ln(kL)\sim 10$
for the scales explored by galaxy surveys. Since we are interested in giving orders of
magnitude and simple mathematical expressions, in the current thesis we will set $\ln(kL)\sim 1$\footnote{As we will see in
the next chapters, the $\ln(kL)$ dependence appear when we consider higher order corrections to $\zeta$, explicitly in the 
higher order correlators which describe its statistical properties.}
without loss of generality.

The spatial averages of the scalar fields, that determine $N_I$, etc., and hence
$\zeta$ cannot in general be calculated. Instead they are parameters, that have
to be specified along with the relevant parameters of the action before the
correlators of $\zeta$ can be calculated. The only exception is when $\zeta$
is determined by the perturbation of the inflaton in single-field
inflation. Then, the unperturbed field value when cosmological scales leave
the horizon can be calculated, knowing the number of $e$-folds to the end of
inflation which is determined by the evolution of the scale factor after
inflation.
Although the unperturbed field values cannot be calculated, their mean square
for a random location of the minimal box (ie.\ of the observable Universe)
can sometimes be calculated using
the stochastic formalism \cite{sy}.

On the other hand, if  we suppose that one or more  perturbed vector fields also  affect
the evolution of the local expansion rate, the curvature perturbation, in the simplest case where $\zeta$ is generated by one
scalar field and one vector field and assuming that the anisotropy in the expansion of the Universe is negligible, can be
calculated up to quadratic terms by means of the following truncated expansion \cite{dklr}
\bea
\no\zeta(\bfx)&\equiv&\delta N (\phi(\bfx),A_i(\bfx),t)\\
&=&N_\phi \delta\phi + N_A^i\delta A_i+\frac{1}{2}N_{\phi\phi}(\delta\phi)^2+
N_{\phi A}^i\delta\phi\delta A_i+\frac{1}{2}N_{AA}^{ij}\delta A_i \delta A_j \,,\label{dNvector}
\eea
where
\be
N_\phi\equiv\frac{\partial N}{\partial \phi}\,,\quad
N_A^{i}\equiv\frac{\partial N}{\partial A_i}\,,\quad
N_{\phi\phi} \equiv\frac{\partial^2 N}{\partial \phi^2}\,,\quad
N_{AA}^{ij}\equiv\frac{\partial^2 N}{\partial A_i\partial A_j}\,,
\quad N_{\phi A}^i\equiv\frac{\partial^2 N}{\partial A_i\partial\phi}\,,
\ee
$\phi$ being the scalar field and ${\bf A}$ the vector field, with $i$ denoting the spatial
indices running from 1 to 3. As with the scalar fields, the unperturbed vector field values are defined as averages within the
chosen box.

In these formulas  there is no need to define the basis (triad) for
the components $A_i$. Also, we need not assume that $A_i$ comes from a 4-vector
field, still less from a gauge field.

\subsection{The growth of $\zeta$}

As noted earlier, $\zeta$ is constant during any era when pressure $P$ is a
unique function of energy density $\rho$.
In the simplest scenario, the field
whose perturbation generates  $\zeta$ is the inflaton field $\phi$ in a single-field
model. Then the local value of $\phi$ is supposed to determine the subsequent
evolution of both pressure  and  energy density,
 making $\zeta$ constant from the beginning.

Alternatives to the simplest scenario generate all or part of $\zeta$ at successively
later eras. Such generation is possible during any era,  unless there is sufficiently
complete matter domination ($P=0$) or radiation domination ($\rho=P/3$).
Possibilities in chronological order include generation during

(i) multi-field inflation  \cite{starobinsky},\\  
(ii) at the end of inflation \cite{lythend},\\
(iii) during preheating \cite{klv},\\
(iv) at reheating, and\\
(v) at a second reheating through the curvaton mechanism \cite{curvaton3,luw,curvaton1,curvaton2}.

A vector field cannot replace the scalar field in the simplest scenario,
because unperturbed
inflation with a  single unperturbed vector field will be  very
anisotropic and so will be the resulting curvature perturbation. Even with isotropic
inflation, we will see in Chapter \ref{chaptvec} that a  single
vector field perturbation cannot  be responsible for the entire curvature
perturbation (at least in the scenarios that were discussed in the Ref. \cite{dklr})
because its contribution is highly anisotropic.
It could instead be responsible for part of the curvature perturbation, through any of the mechanisms listed above.

%%%%%%%%%%%%%%%%%%%%%%%%%%%%%%%%%%%%%%%%%%%%%%%%%%%%%%%%%%%%%%%%%%%%%%%%%%%%%%%%%%%%%%%%%%%%%%%%%%%%%%%%%%%%%%
\section{Statistical descriptors for a probability distribution function} \label{descriptors}           %%%%%%
%%%%%%%%%%%%%%%%%%%%%%%%%%%%%%%%%%%%%%%%%%%%%%%%%%%%%%%%%%%%%%%%%%%%%%%%%%%%%%%%%%%%%%%%%%%%%%%%%%%%%%%%%%%%%%
A probability distribution function $f(\zeta)$ for any function of space and time $\zeta({\bf x},t)$ may be understood as the
univocal correspondence between the possible values that $\zeta$ may take throughout the space and the normalised frecuency of
appearences of such values for a given time. Any continous function of $\zeta$ might represent a probability distribution
function as long as $f(\zeta) \geq 0$ and $\int^\infty_{-\infty} f(\zeta) d\zeta = 1$. However, for a particular probability
distribution function, how many independent parameters do we need to completely characterize it in a unique way? And despite
the possible infinite number of parameters required to do this, what is the information encoded in those parameters?

The answers to these questions rely on the moments $m_\zeta(n)$ of the distribution. %($n = 1, ..., \infty$):

For a given probability distribution function $f(\zeta)$, there are an infinite number of moments that work as statistical
descriptors of $\zeta({\bf x},t)$:
\begin{eqnarray}
{\rm the \ mean \ value:} \;\; m_\zeta(1) &\equiv& \langle \zeta \rangle = \int \zeta f(\zeta) d\zeta \,, \\
{\rm the \ variance:} \;\; m_\zeta(2) &\equiv& \int (\zeta - \langle \zeta \rangle)^2 f(\zeta) d\zeta \,, \\
{\rm the \ skewness:} \;\; m_\zeta(3) &\equiv& \frac{\int (\zeta - \langle \zeta \rangle)^3 f(\zeta) d\zeta}
{\big[m_\zeta(2)\big]^{3/2}} \,, \\
{\rm the \ kurtosis:} \;\; m_\zeta(4) &\equiv& \frac{\int (\zeta - \langle \zeta \rangle)^4 f(\zeta) d\zeta}
{\big[m_\zeta(2)\big]^2} \,, \\
&.& \nonumber \\
&.& \nonumber \\
&.& \nonumber \\
&{\rm and \ so \ on.}& \nonumber
\end{eqnarray}
What can we say about $\zeta({\bf x},t)$ from the knowledge of the moments of the distribution? If, for instance, all the odd
moments with $n \geq 3$ (skewness, ... etc) are zero, we can say that the probability distribution function $f(\zeta)$ is even
around the mean value.  If in addition all the even moments with $n \geq 4$ (kurtosis, ... etc) are expressed only as products
of the variance, we can say that the distribution function is gaussian. Indeed, as is well known, the only quantities required
to reproduce a gaussian function are the mean value and the variance:
\begin{eqnarray}
f_{gaussian}(\zeta) \equiv \frac{1}{\sqrt{2\pi m_\zeta(2)}} e^{-(\zeta - m_\zeta(1))^2/2m_\zeta(2)} \,.
\end{eqnarray}
Departures from the exact gaussianity come either from non-vanishing odd moments with $n \geq 3$, in which case
the probability distribution function is non-symmetric around the mean value, or from
higher $n \geq 4$ even moments different to products of the variance, in which case the probability distribution function 
continues to be symmetric around the mean value although its ``peakedness''\footnote{Higher even standarized moments different 
to products of the variance mean more of the variance is due to infrequent extreme deviations, as opposed to frequent modestly-
sized deviations.} is bigger than that for a gaussian function, or from both of them.
A non-gaussian probability distribution function requires then more moments, other than the mean value and the variance, to be
completely reconstructed. Such a reconstruction process is described for instance in Ref. \cite{svw}.

Working in momentum space is especially useful in cosmology because the modes associated with the quantum fluctuations of
scalar fields during inflation become classical once they leave the horizon
\cite{albrecht1,Burgess:2006jn,grishchuk,guthpi,Kiefer:2008ku,lombardo,lyth84,lythbook,Lyth:2006qz,Martineau:2006ki,Nambu:2008my}.
The same applies for the primordial curvature perturbation $\zeta$ which, in addition, is a conserved quantity while staying
outside the horizon if the adiabatic condition is satisfied \cite{lms}. As regards the moments of the probability distribution
function, they have a direct connection with the correlation functions for the Fourier modes $\zeta_{\bf k} = \int d^3k \zeta
({\bf x}) e^{-i {\bf k \cdot x}}$ defined in flat space.

As the $n$-point correlators of $\zeta_{\bf k}$ are generically defined in terms of spectral functions of the wavevectors
involved:
{\small\begin{eqnarray}
{\rm two-point \ correlation} &\rightarrow& {\rm spectrum} \ P_\zeta: \nonumber \\
\langle \zeta_{\bf k_1} \zeta_{\bf k_2} \rangle &\equiv& (2\pi)^3 \delta^3 ({\bf k_1} + {\bf k_2}) P_\zeta(\bfk) \,,
\label{2pc} \\
{\rm three-point \ correlation} &\rightarrow& {\rm bispectrum} \ B_\zeta: \nonumber \\
\langle \zeta_{\bf k_1} \zeta_{\bf k_2} \zeta_{\bf k_3} \rangle &\equiv& (2\pi)^3 \delta^3 ({\bf k_1} + {\bf k_2} + {\bf k_3})
B_\zeta(\bfk_1, \bfk_2, \bfk_3) \,, \label{3pc} \\
{\rm four-point \ correlation} &\rightarrow& {\rm trispectrum} \ T_\zeta: \nonumber \\
\langle \zeta_{\bf k_1} \zeta_{\bf k_2} \zeta_{\bf k_3} \zeta_{\bf k_4} \rangle &\equiv& (2\pi)^3 \delta^3 ({\bf k_1} + {\bf
k_2} + {\bf k_3} + {\bf k_4}) T_\zeta({\bf k_1}, {\bf k_2}, {\bf k_3}, {\bf k_4}) \,, \\
&.& \nonumber \\
&.& \nonumber \\
&.& \nonumber \\
&{\rm and \ so \ on,}& \nonumber
\end{eqnarray}}
the moments of the distribution are then written as momentum integrals of the spectral functions for the modes $\zeta_{\bf k}$:
\begin{eqnarray}
{\rm the \ variance:} \ m_\zeta (2) &=& \int \frac{d^3 k}{(2\pi)^3} P_\zeta (\bfk) \,, \\
{\rm the \ skewness:} \ m_\zeta (3) &=&\frac{ \int \frac{d^3 k_1 \ d^3 k_2}{(2\pi)^6} B_\zeta (\bfk_1,\bfk_2,\bfk_3)}
{\big[\int \frac{d^3 k}{(2\pi)^3} P_\zeta (\bfk) \big]^{3/2}} \,, \\
{\rm the \ kurtosis:} \ m_\zeta (4) &=& \frac{\int \frac{d^3 k_1 \ d^3 k_2 \ d^3 k_3}{(2\pi)^9} T_\zeta ({\bf k_1}, {\bf k_2}, 
{\bf k_3}, {\bf k_4})}{\big[\int \frac{d^3 k}{(2\pi)^3} P_\zeta (\bfk) \big]^2} \,, \\
&.& \nonumber \\
&.& \nonumber \\
&.& \nonumber \\
&{\rm and \ so \ on.}& \nonumber
\end{eqnarray}
Non-gaussianity in $\zeta$ is, therefore, associated with non-vanishing higher order spectral functions, starting from the
bispectrum $B_\zeta$.
%%%%%%%%%%%%%%%%%%%%%%%%%%%%%%%%%%%%%%%%%%%%%%%%%%%%%%%%%%%%%%%%%%%%%%%%%%%%%%%%%%%%%%
\subsection{Statsitical descriptors for primordial curvature perturbation $\zeta$}
Theoretical cosmologists work with $\zeta$. However, astronomers work with observable quantities such as the contrast in the
temperature of the cosmic microwave background radiation $\delta T/T$. The connection between the theoretical cosmologist
quantity $\zeta$ and the astronomer quantity $\delta T/T$ is given by the Sachs-Wolfe effect \cite{sachs} which, at first-order
and for superhorizon scales, looks as follows:
\begin{eqnarray}
\left(\frac{\delta T}{T}\right)_{\bf k} = - \frac{1}{5} \zeta_{\bf k} \,. \label{swe}
\end{eqnarray}
Thus, although it is essential to study the Sachs-Wolfe relation at higher orders, which is far more complicated than Eq.
(\ref{swe}), theoretical cosmologists may study the statistical properties of the observed $\delta T / T$ through the spectral
functions associated with the curvature perturbation $\zeta$\footnote{If we assume that the statistical inhomogeneity is 
present, i.e translational invariance of the n-point correlators of $\zeta$ is broken, it is necesary introduce modifications 
of the usual definitions of the statistical descriptors of the primordial curvature perturbation $\zeta$. For example, if there 
is statistical inhomogeneity then $\langle \zeta_{\bf k_1} \zeta_{\bf k_2} \rangle$ is not proportional to 
$\delta(\bfk_1+\bfk_2)$. In this thesis we will assume statistical homogeneity, so that the statistical descriptors given in 
the last subsection can be correctly applied.}:
\begin{eqnarray}
{\rm mean \ value \ of} \ \delta T/T = 0 &\rightarrow& {\rm mean \ value \ of} \ \zeta = 0 \,, \\
{\rm variance:} \;\; m_{\delta T/T}(2) &\rightarrow& {\rm spectrum:} \;\; P_\zeta (\bfk) \,, \\
{\rm skewness:} \;\; m_{\delta T/T}(3) &\rightarrow& {\rm bispectrum:} \;\; B_\zeta (\bfk_1,\bfk_2,\bfk_3) \,, \\
{\rm kurtosis:} \;\; m_{\delta T/T}(4) &\rightarrow& {\rm trispectrum:} \;\; T_\zeta ({\bf k_1}, {\bf k_2}, {\bf k_3}, {\bf
k_4}) \,, \\
&.& \nonumber \\
&.& \nonumber \\
&.& \nonumber \\
&{\rm and \ so \ on.}& \nonumber
\end{eqnarray}

Now, we will parametrize the spectral functions of $\zeta$ in terms of quantities which are the ones for which
observational bounds are given. Because of the direct connection between these quantities and the moments of the probability
distribution function $f(\zeta)$, we may also call these quantities as the statistical descriptors for $f(\zeta)$.

The bispectrum $B_\zeta$ and trispectrum $T_\zeta$ are parametrized in terms of products of the spectrum $P_\zeta$, and the
quantities $\fnl$ and $\tnl$ and $\gnl$ respectively\footnote{There is actually a sign difference between the $f_{NL}$ defined
here and that defined in Ref. \cite{maldacena}.  The origin of the sign difference lies in the way the observed $f_{NL}$ is
defined \cite{komatsu}, through the Bardeen's curvature perturbation: $\Phi^B = \Phi^B_L + f_{NL} (\Phi^B_L)^2$ with $\Phi^B =
(3/5)\zeta$, and the way $f_{NL}$ is defined in Ref. \cite{maldacena}, through the gauge invariant Newtonian potential: $\Phi^N
= \Phi^N_L + f_{NL} (\Phi^N_L)^2$ with $\Phi^N = -(3/5)\zeta$.} \cite{bl,bsw1,maldacena}:
\begin{eqnarray}
B_\zeta ({\bf k_1},{\bf k_2},{\bf k_3}) &\equiv& \frac{6}{5} f_{NL} ({\bf k_1},{\bf k_2},{\bf k_3}) \left[P_\zeta({\bf k_1})
P_\zeta({\bf k_2}) + {\rm c. \ p.} \right] \,, \label{bfp} \\
T_\zeta ({\bf k_1}, {\bf k_2}, {\bf k_3}, {\bf k_4}) &\equiv& \frac{1}{2} \tau_{NL} ({\bf k_1},{\bf k_2},{\bf k_3}, {\bf k_4})
\left[P_\zeta({\bf k_1}) P_\zeta({\bf k_2}) P_\zeta({\bf k_1} + {\bf k_4}) + {\rm c. \ p.} \right] + \nonumber \\
&& + \frac{54}{25} g_{NL} ({\bf k_1},{\bf k_2},{\bf k_3}, {\bf k_4}) \left[P_\zeta({\bf k_1}) P_\zeta({\bf k_2}) P_\zeta({\bf
k_3}) + {\rm c. \ p.} \right] \,, \nonumber \\
&& \label{tfp}
\end{eqnarray}
where {\rm c. p.} means {\rm cyclic  permutations}. Higher order spectral functions would be parametrized in an analogous way.

By virtue of the reality condition $\zeta(-\bfk)=\zeta^*(\bfk)$,
an equivalent definition of the spectrum is
\be
\vev{\zeta({\bf k_1})\zeta^*({\bf k_2})}
= (2\pi)^3 \delta^3({\bf k_1}-{\bf k_2}) \pz(\bfk) \,.
\label{spectrum2} \ee
Setting ${\bf k_1}={\bf k_2}$ the left hand side is $\vev{|\zeta(\bfk)|^2}$.
It follows that the  the spectrum is positive and nonzero.
Even if $\pz(\bfk)$  is anisotropic, the reality condition requires
$\pz(\bfk)=\pz(-\bfk)$. The spectrum will therefore be of the form \cite{acw}
\be
\pz(\bfk) = \pz\su{iso}(k) \[ 1 + g_\zeta
 (\hat{\bfd}\cdot \hat \bfk)^2 + \cdots \]
\,, \label{astadef}
\ee
where $\pz\su{iso}(k)$ is the average over all directions, $\hat{\bfd}$ is some  unit vector, $\hat \bfk$ is a unit vector
along $\bfk$ and $\gz$ is the level of statistical anisotropy. The homogeneity and isotropy requirements at large scales imply
that the spectrum $P_\zeta$ and bispectrum $B_\zeta$ are functions of the wavenumbers only. For the trispectrum $T_\zeta$ and
the other higher order spectral functions, the momentum dependence also involves the direction of the wavevectors.

On the other hand if the $n$-point correlators are also invariant under rotations (statistical isotropy) the spectral functions
$P_\zeta ({\bf k}) \equiv (2\pi^2 / k^3) \calp_\zeta ({\bf k})$ and $B_\zeta ({\bf k_1},{\bf k_2},{\bf k_3})$ depend only on
the magnitude of the wavevectors \cite{acw}. In this case the spectrum $P_\zeta$ is parametrized in terms of an amplitude
$\mathcal{P}^{1/2}_\zeta$ and a spectral index $n_\zeta$ which measures the deviation from an exactly scale-invariant spectrum
\cite{lythbook}:
\begin{equation}
P_\zeta (k) \equiv \frac{2\pi^2}{k^3} \mathcal{P}_\zeta \left(\frac{k}{aH}\right)^{n_\zeta - 1} \,. \label{asidef}
\end{equation}
Given the present observational status, $n_\zeta$, $f_{NL}$,  $\tau_{NL}$, $\gnl$ and $\gz$ are the statistical 
descriptors that discriminate among models for the origin of the large-scale structure once $\mathcal{P}^{1/2}_\zeta$ has been 
fixed to the observed value.  Since non-vanishing higher order spectral functions such as $B_\zeta$ and $T_\zeta$ imply 
non-gaussianity in the primordial curvature perturbation $\zeta$, the statistical descriptors $f_{NL}$, $\tau_{NL}$ and $\gnl$ 
are usually called the levels of non-gaussianity.

%%%%%%%%%%%%%%%%%%%%%%%%%%%%%%%%%%%%%%%%%%%%%%%%%%%%%%%%%%%%%%%%%%%%%%%%%%%%%%
\section{Observational constraints} \label{observational}              %%%%%%%
%%%%%%%%%%%%%%%%%%%%%%%%%%%%%%%%%%%%%%%%%%%%%%%%%%%%%%%%%%%%%%%%%%%%%%%%%%%%%%

Direct observation, coming from the anisotropy of the CMB and the
inhomogeneity of the galaxy distribution, gives information on what are
called cosmological scales \cite{wmap5}. These correspond to a range
$\Delta \ln k\sim 10$ or so downwards from the scale $k\mone\sim H_0\mone$
that corresponds to the size of the observable Universe.

\subsection{Spectrum and non-gaussianity}
Observational results concerning the spectrum $\calpz$ are generally
obtained with the assumption of statistical isotropy ($\gz=0$), but they would
not be greatly affected by the inclusion of anisotropy at the $10\%$ level.

NASA's COBE-satellite provided us with a reliable value for the spectral amplitude $\mathcal{P}^{1/2}_\zeta$ \cite{bunn}:
$\mathcal{P}^{1/2}_\zeta = (4.957 \pm 0.094) \times 10^{-5}$ which is usually called the COBE normalisation.  As regards the
spectral index, the latest data release and analysis from the WMAP satellite shows that $n_\zeta = 0.960 \pm 0.014$
\cite{wmap5} which rejects exact scale invariance at more than $2\sigma$. Such a result has been extensively used to constrain
inflation model building \cite{alabidi1}, and although several classes of inflationary models have been ruled out through the
spectral index, lots of models are still allowed; that is why it is so important an appropiate knowledge of the statistical
descriptors $f_{NL}$ and $\tau_{NL}$.
Present observations show that the primordial curvature perturbation $\zeta$ is almost, but not completely, gaussian.  The
level of non-gaussianity $f_{NL}$ in the bispectrum $B_\zeta$, after five years of data from NASA's WMAP satellite, is in the 
range $-9 < f_{NL} < 111$ at $2\sigma$ \cite{wmap5}. There is at
present no observational bound on the level of non-gaussianity $\tau_{NL}$ in the trispectrum $T_\zeta$ although it was
predicted that COBE should either detect it or impose the lower bound $|\tau_{NL}| \lsim 10^8$ \cite{bl,okamoto}. It is
expected that future WMAP data releases will either detect non-gaussianity or reduce the bounds on $f_{NL}$ and $\tau_{NL}$ at
the $2\sigma$ level to $|f_{NL}|\lsim 40$ \cite{komatsu} and $|\tau_{NL}|\lsim
2 \times 10^{4}$ \cite{kogo} respectively. The ESA's PLANCK satellite \cite{planck,planck1}, launched in 2009, promises to
reduce the bounds to $|f_{NL}|\lsim 10$ \cite{komatsu} and $|\tau_{NL}|\lsim 560$  \cite{kogo} at the $2\sigma$ level if non-
gaussianity is not detected. In addition, by studying the 21-cm emission spectral line in the cosmic neutral Hydrogen prior to
the era of reionization, it is also possible to know about the levels of non-gaussianity $f_{NL}$ and $\tau_{NL}$;
the 21-cm background anisotropies capture information about the primordial non-gaussianity better than any high resolution map
of cosmic microwave background radiation: an experiment like this could reduce the bounds on the non-gaussianity levels to $|
f_{NL}|\lsim 0.2$ \cite{cooray1,cooray2} and $|\tau_{NL}|\lsim 20$ \cite{cooray2} at the $2\sigma$ confidence. Finally, it is
worth stating that there have been recent claims about the detection of non-gaussianity in the bispectrum $B_\zeta$ of $\zeta$
from the WMAP 3-year data \cite{jeong,yadav}. Such claims, which report a rejection of $f_{NL} = 0$ at more that $2\sigma$
($26.9 < f_{NL} < 146.7$), are based on the estimation of the bispectrum while using some specific foreground masks.  The WMAP
5-year analysis \cite{wmap5} shows a similar behaviour when using those masks, but reduces the significance of the results when
other more conservative masks are included allowing again the possibility of exact gaussianity.

\subsection{Statistical anisotropy and statistical inhomogeneity}\label{obsvec}

Taking all the uncertainties into account, observation is consistent with statistical anisotropy and statistical
inhomogeneity allowing either of these things at around the $10\%$ level. Indeed, some recent papers 
\cite{app,gawe,ge,hl,samal} claim for the presence of statistical anisotropy in the five-year data from the NASA's WMAP 
satellite \cite{wmap}. In this section we briefly review what is known.

\begin{figure}
\begin{center}
\begin{tabular}{ccc}
\includegraphics[width=5.1cm,height=6.3cm]{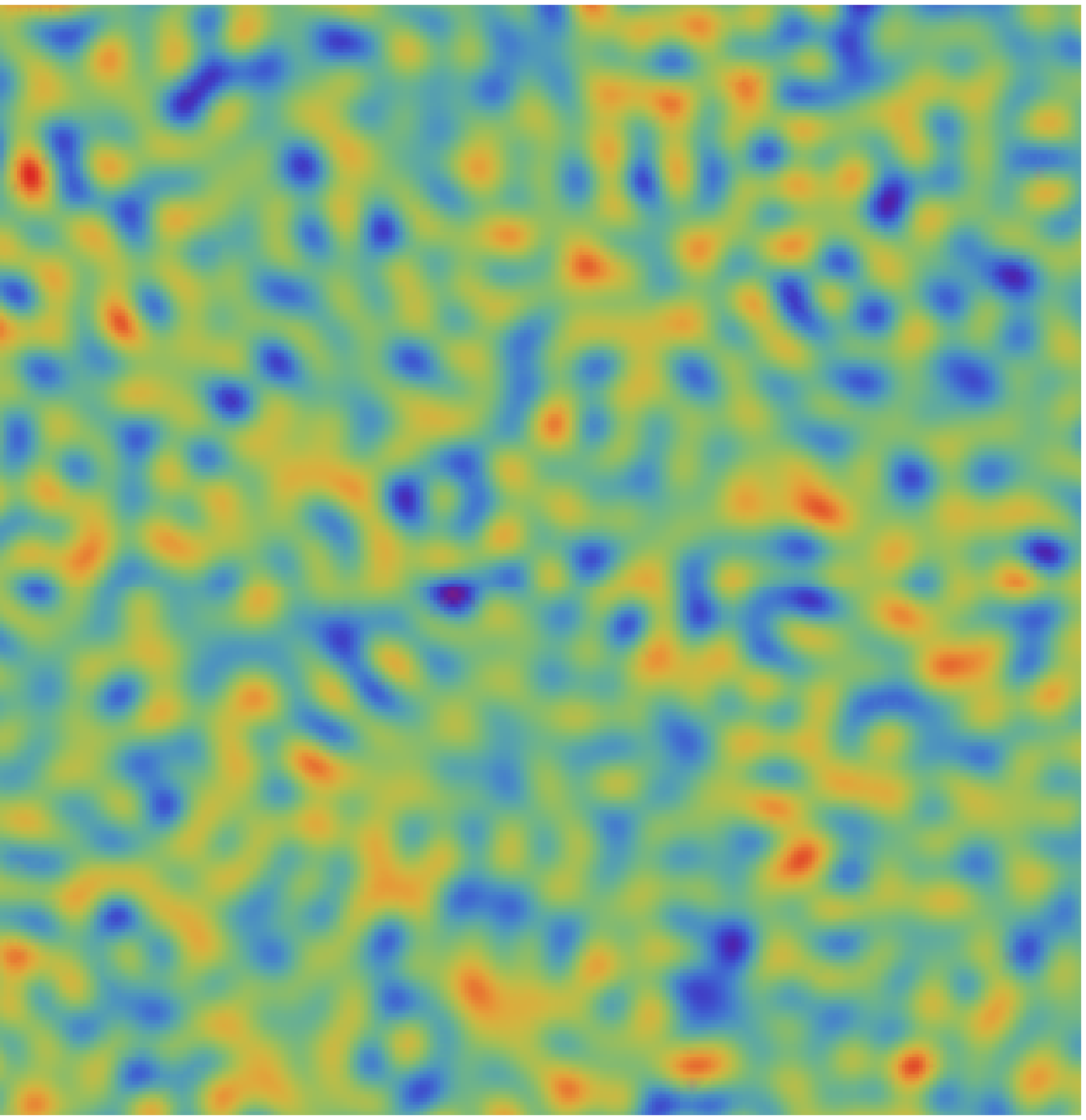} & \includegraphics[width=5.1cm,height=6.3cm]{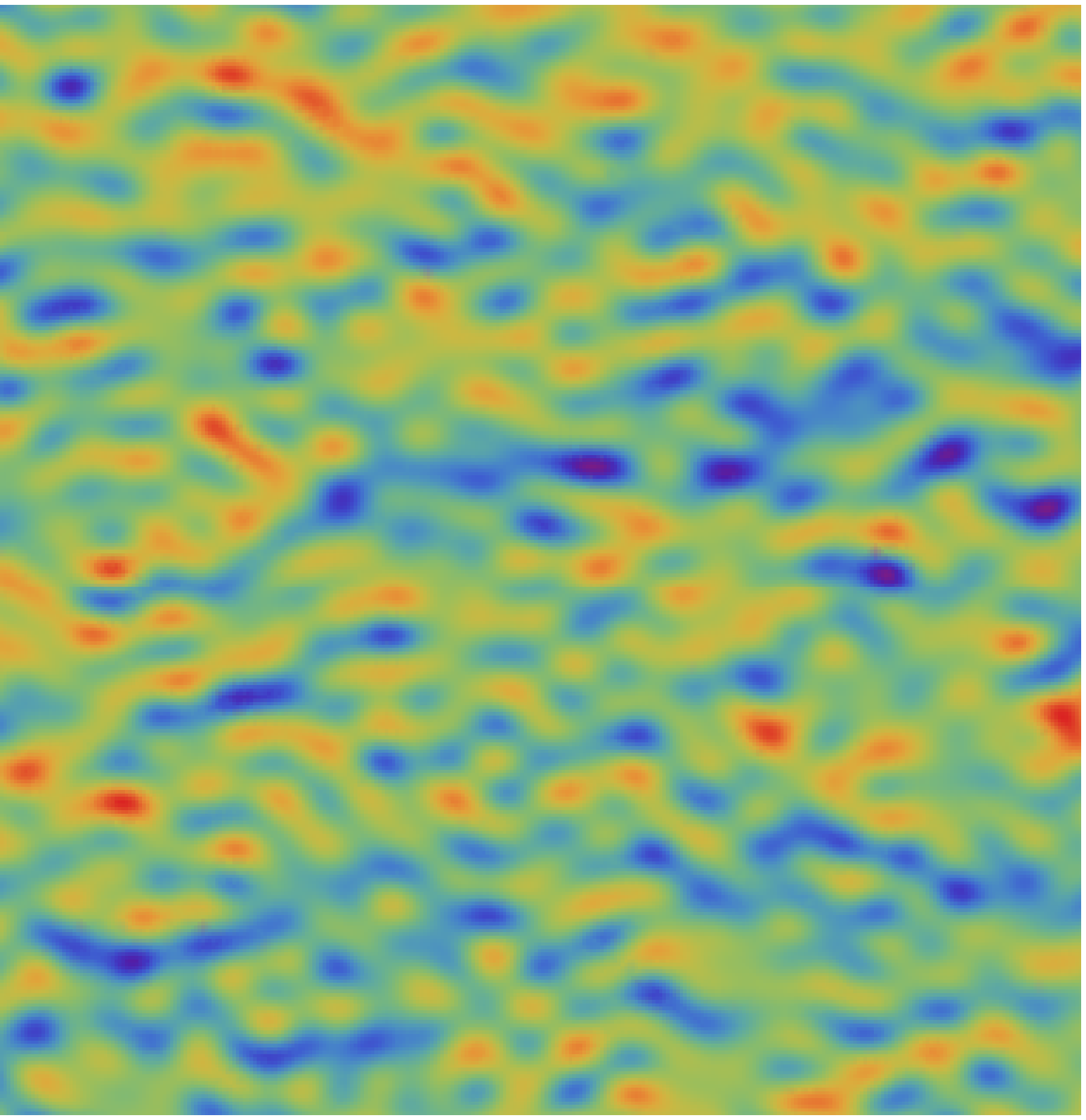} & 
\includegraphics[width=5.1cm,height=6.3cm]{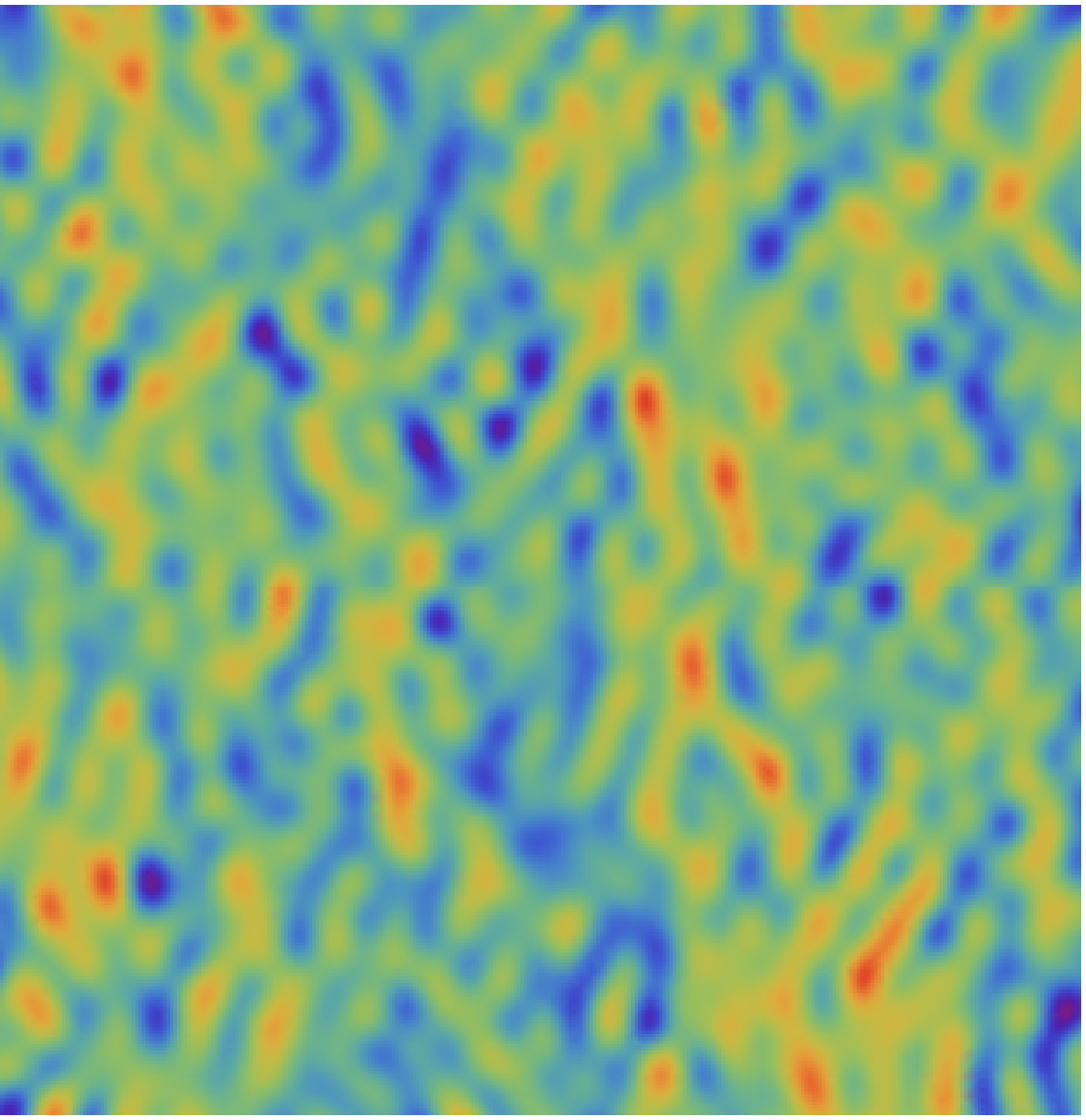} \\
(a) & (b) & (c)
\end{tabular}
\end{center}
\caption[Simulation corresponding to statistical isotropy and statistical anisotropy]{Statistical anisotropy. (a). This simulation
corresponds to the statistical isotropic case $g_\zeta = 0$.  (b). In contrast, this simulation corresponds to the statistical
anisotropic case with $g_\zeta =1$, $\hat{\bfd}$ pointing along the horizontal direction and setting to zero the 
isotropic part. (c). Same as in (b) but with $\hat{\bfd}$ pointing along the vertical direction. (Courtesy of Mindaugas 
Kar\v{c}iauskas).} \label{aniso}
\end{figure}

Assuming statistical homogeneity of the curvature perturbation,
a recent study \cite{gawe} of the CMB temperature perturbation
finds weak evidence for statistical anisotropy (see Fig. \ref{aniso} for an example of who is statitiscal anisotropy). They 
keep only the leading (quadrupolar) term of \eq{astadef}:
\be
\calpz(\bfk) = \calp_\zeta\su{iso}(k) \( 1 + g_\zeta
(\hat{\bfd}\cdot \hat \bfk)^2 \) \,,
\label{curvquad} \ee
and find $\gz \simeq 0.290 \pm 0.031$ which rules out statistical isotropy at more than $9 \sigma$.
Nevertheless, the preferred direction lies near the plane of the solar system, which makes
the authors of Ref. \cite{gawe} believe that this effect could be due to an unresolved
systematic error (among other possible systematic errors which have not been demonstrated
either to be the source of this statistical anisotropy nor to be completely uncorrelated
\cite{gawe}).

Even if the result found in Ref. \cite{gawe} turns out to be due to a systematic error, some
forecasted constraints on $\gz$ show that the statistical anisotropy subject is worth
studying \cite{pullen}: $|g_\zeta| \lsim 0.1$ for the NASA's WMAP satellite \cite{wmap} if there is no detection, and $|
g_\zeta| \lsim 0.02$ for the ESA's PLANCK satellite \cite{planck} if there is no detection.
There is at present no bound on statistical anisotropy of the 3-point or higher correlators.

In  some different studies, the mean-square CMB perturbation in opposite
hemispheres has been measured, to see if there is any difference between
hemispheres.  Quite recent works \cite{dipole2,dipole1,dipole3} find a
difference of order ten percent, for a certain choice of the hemispheres, with statistical
significance at the $99\%$ level. Given the difficulty of handling systematic
uncertainties it would be premature to regard the evidence for this
hemispherical anisotropy as completely overwhelming. Nevertheless, what would hemispherical anisotropy  imply for the curvature
perturbation? Focussing on a small patch of sky, the  statistical anisotropy of the curvature perturbation
implies that the mean-square temperature perturbation within a
given small patch will {\em in general} depend on the direction of that patch.
This is because the mean square within such a patch depends
upon the mean square of the curvature perturbation in a small planar region of space
perpendicular to the line of sight located at last scattering. But the mean-square temperature
will be {\em the same} in patches at opposite directions in the sky, because they explore the curvature perturbation
$\zeta(\bfk)$ in the same $\bfk$-plane and the spectrum $\calpz(\bfk)$
is  invariant under the change $\bfk\to-\bfk$. It follows that statistical
anisotropy of the curvature perturbation cannot by itself generate a
hemispherical anisotropy.  We may then conclude that hemispherical
anisotropy of the CMB temperature requires statistical {\em inhomogeneity} of the
curvature perturbation. Then $\vev{\zeta({\bf k_1})\zeta({\bf k_2})}$ is not proportional to $\delta^3({\bf k_1}+{\bf k_2})$
\cite{carroll}.

%%%%%%%%%%%%%%%%%%%%%%%%%%%%%%%%%%%%%%%%%%%%%%
\section{Conclusions}			%%%%%%%%%%%%%%
%%%%%%%%%%%%%%%%%%%%%%%%%%%%%%%%%%%%%%%%%%%%%%
In this chapter we have presented the statistical description of primordial curvature perturbation $\zeta$. This framework 
allows us to define statistical descriptors, which provides a bridge between theory and observation. The relevant parameters 
that parametrize these statistical descritors are: the spectrum amplitude $\calpz\half$, the spectral index $\nz$, the levels of 
non-gaussianity $\fnl$, $\tnl$ and $\gnl$ and the level of statistical anisotropy $\gz$. Some of these parameters have an upper 
bound from observation, so it is very important to study theoretical models that successfully reproduce these observations. To 
study these theoretical aspects, we have presented a poweful tool to calculate the primordial cuvature pertubation $\zeta$ 
and  all its statistical descriptors; such a tool is usully called the $\dn$ formalism. In the next two chapters we will see 
how to work out this formalism.

%%%%%%%%%%%%%%%%%%%%%%%%%%%%%%%%%%%%%%%%%%%%%%%%%%%%%%%%%%%%%%%%%%%%%%%%%%%%%%%%%%%%%%%%%%%%%%%%%%%%%%%%%%%%
%%%%%%%%%%%%%%%%%%%%%%%%%%%%%%%%%%%%%%%%%%%%%%%%%%%%%%%%%%%%%%%%%%%%%%%%%%%%%%%%%%%%%%%%%%%%%%%%%%%%%%%%%%%%
\chapter[PRIMORDIAL NON- GAUSSIANITY IN SLOW-ROLL INFLATION: THE BISPECTRUM]                        %%%%%%%%
{ON THE ISSUE OF THE $\zeta$ SERIES CONVERGENCE AND LOOP CORRECTIONS IN THE GENERATION 		        %%%%%%%%
OF OBSERVABLE PRIMORDIAL NON- GAUSSIANITY IN SLOW-ROLL INFLATION: THE BISPECTRUM}\label{chaptsca}   %%%%%%%%
%%%%%%%%%%%%%%%%%%%%%%%%%%%%%%%%%%%%%%%%%%%%%%%%%%%%%%%%%%%%%%%%%%%%%%%%%%%%%%%%%%%%%%%%%%%%%%%%%%%%%%%%%%%%
%%%%%%%%%%%%%%%%%%%%%%%%%%%%%%%%%%%%%%%%%%%%%%%%%%%%%%%%%%%%%%%%%%%%%%%%%%%%%%%%%%%%%%%%%%%%%%%%%%%%%%%%%%%%

\section{Introduction}

Since COBE \cite{cobe} discovered and mapped the anisotropies in the temperature of the cosmic microwave background radiation 
\cite{smooth}, many balloon and satellite experiments have refined
the measurements of such anisotropies, reaching up to now an amazing combined precision. The COBE sequel has continued with the 
WMAP satellite \cite{wmap} which has been able to measure the temperature angular power spectrum up to the third peak with 
unprecedent precision \cite{hinshaw}, and increase the level of sensitivity to primordial non-gaussianity in the bispectrum by 
two orders of magnitude compared to COBE \cite{wmap1,wmap5}.  The next-to-WMAP satellite, PLANCK \cite{planck}, which
was launched in may of 2009, is expected to precisely measure the temperature angular power spectrum up to the eighth 
peak \cite{planck1}, and improve the level of sensitivity to primordial non-gaussianity in the bispectrum by one order of 
magnitude compared to WMAP \cite{komatsu}.

Because of the progressive improvement in the accuracy of the satellite measurements %observational state-of-the-art
described above, it is pertinent to study cosmological inflationary models that generate significant (and observable) levels of 
non-gaussianity. An interesting way to address the problem involves the $\delta\textit{N}$ formalism 
\cite{dklr,lms,lr,ss,st,starobinsky}, which can be employed to give the levels of non-gaussianity $f_{NL}$ \cite{lr} and 
$\tau_{NL}$ \cite{alabidi2,bl} in the bispectrum $B_\zeta$ and trispectrum $T_\zeta$ of the primordial curvature 
perturbation $\zeta$ respectively. Such non-gaussianity levels are given, for slow-roll inflationary models, in terms of the 
local evolution of the universe under consideration, as well as of the $n$-point correlators, evaluated a few Hubble times 
after horizon exit, of the perturbations $\delta\phi_{i}$ in the scalar fields that determine the dynamics of such a universe 
during inflation.  

In the $\delta\textit{N}$ formalism for slow-roll inflationary models, the primordial curvature perturbation 
$\zeta(\textbf{x},t)$ is written as a Taylor series in the scalar field perturbations $\delta\phi_{i}(\textbf{x},t_\star)$ evaluated a few Hublle times after horizon exit\footnote{This equation is similar to one in \eq{dNsc}, the difference 
is that here we are redefined it so that $\langle \zeta(t,\bfx)=0\rangle$.}, 
\begin{eqnarray}
\zeta(t,\textbf{x})&=&\sum_{I}N_{I}(t)\delta\phi_{I}(\textbf{x},t_\star) - \sum_{I}N_{I}(t) \langle\delta\phi_{I}
(\textbf{x},t_\star)\rangle + \nonumber \\
&&+\frac{1}{2}\sum_{IJ}N_{IJ}(t)\delta\phi_{I}(\textbf{x},t_\star)\delta\phi_{J}(\textbf{x},t_\star)-\frac{1}{2}\sum_{IJ}N_{IJ}(t) \langle\delta\phi_{I}(\textbf{x},t_\star)\delta\phi_{J}(\textbf{x},t_\star)\rangle + \nonumber \\
&&+\frac{1}{3!}\sum_{IJK}N_{IJK}(t)\delta\phi_{I}(\textbf{x},t_\star)\delta\phi_{J}(\textbf{x},t_\star)\delta\phi_{k}(\textbf{x},t_\star)\no\\& -& \frac{1}{3!}\sum_{IJK}N_{IJK}(t)\langle\delta\phi_{I}(\textbf{x},t_\star)\delta\phi_{J}(\textbf{x},t_\star)\delta\phi_{K}(\textbf{x},t_\star)\rangle + \nonumber \\
&&+...\;,
\end{eqnarray}
It is in this way that the correlation 
functions of $\zeta$ (for instance, $\langle\zeta_{\bf k_{1}}\zeta_{\bf k_{2}}\zeta_{\bf k_{3}}\rangle$) can be obtained in 
terms of series, as often happens in Quantum Field Theory where the probability amplitude is a series whose possible truncation 
at any desired order is determined by the coupling constants of the theory. A highly relevant question is that of whether the 
series for $\delta N$ converges in cosmological perturbation theory and whether it is possible in addition to find some 
quantities that determine the possible truncation of the series, which in this sense would be analogous to the coupling 
constants in Quantum Field Theory.  In general such  quantities will depend on the specific inflationary model; the series then 
cannot be simply truncated at some order until one is sure that 
it does indeed converge, and besides, one has to be careful not to forget any term that may be leading in the series even if it 
is of higher order in the coupling constant. This issue has not been investigated in the present literature, and generally the 
series has been truncated to second- or third-order neglecting in addition terms that could be the leading ones 
\cite{alabidi1,alabidi2,battefeld,bl,bsw1,lr,ss,seery3,vernizzi,yokoyama2,yokoyama1,zaballa}. 

The most studied and popular inflationary models nowadays are those of the slow-roll variety with canonical kinetic terms 
\cite{lyth6,lythbook,lyth5}, because of their simplicity and because they easily satisfy the spectral index requirements for 
the generation of large-scale structures. One of the usual predictions from inflation and the theory of cosmological 
perturbations 
is that the levels of non-gaussianity in the primordial perturbations are expected to be unobservably small when considering 
this class of models \cite{battefeld,li,maldacena,seery3,seery5,seery4,seery7,vernizzi,yokoyama1,zaballa}\footnote{One possible 
exception is the two-field slow-roll model analyzed in Ref. \cite{alabidi1} (see also Refs. \cite{bernardeu2,bernardeu1}) where 
{\it observable, of order one, values for} $f_{NL}$ are generated for
a reduced window parameter associated with
the initial field values when taking into account only the tree-level terms in both $P_\zeta$ and $B_\zeta$. However, such a 
result seems to be incompatible with the general expectation, proved in Ref. \cite{vernizzi}, of $f_{NL}$ being of order the 
slow-roll parameters, and {\it in consequence unobservable}, for two-field slow-roll models with separable potential when 
considering only the tree-level terms both in $P_\zeta$ and $B_\zeta$.  The origin of the discrepancy could be understood by 
conjecturing that the trajectory in field space, for the models in Refs. \cite{alabidi1,bernardeu2,bernardeu1}, seems to be 
sharply curved, being quite near a saddle point; such a condition is required, according to Ref. \cite{vernizzi}, to generate 
$f_{NL} \sim \mathcal{O}(1)$. \label{laila}}. This fact leads us to analyze the cosmological perturbations in the framework of 
first-order cosmological perturbation theory. Non-gaussian characteristics are then suppressed since the non-linearities in the 
inflaton potential and in the metric perturbations are not taken into account.  The non-gaussian characteristics are actually 
present and they are made explicit if second-order \cite{lr1} or higher-order corrections are considered.

The whole literature that encompasses the slow-roll inflationary models with canonical kinetic terms reports that the non-
gaussianity level $f_{NL}$ is expected to be very small, being of the order of the slow-roll parameters $\epsilon_i$ and 
$\eta_i$, ($\epsilon_i, |\eta_i| \ll 1$) \cite{battefeld,maldacena,seery7,vernizzi,yokoyama1}. These works have not taken into 
account either the convergence of the series for $\zeta$ nor the possibility that loop corrections dominate over the tree level 
ones in the $n$-point correlators. Our main result in this chapter is the recognition of the possible
convergence of the $\zeta$ series, and the
existence of some ``coupling constants'' that determine the possible truncation of the $\zeta$ series at any desired order.
When this situation is encountered in a subclass of small-field
{\it slow-roll} inflationary models with canonical kinetic terms,
the one-loop corrections may dominate the series when calculating either the spectrum $P_\zeta$, or the bispectrum $B_\zeta$. 
This in turn {\it may generate sizeable and observable levels of non-gaussianity} in total contrast with the general claims 
found in the present literature.

The layout of the chapter is the following:  Section \ref{conver} is devoted to the issue of the $\zeta$ series convergence
and loop corrections in the framework of the $\delta N$ formalism. The presentation of the current knowledge about primordial 
non-gaussianity in slow-roll inflationary models is given in Section \ref{ngsr}.  A particular subclass of small-field slow-
roll inflationary
models is the subject of Section \ref{model} as it is this subclass of models that generate significant levels of non-
gaussianity. The available parameter space for this subclass of models is constrained in Section \ref{rest} by taking into
account some  observational requirements such as the COBE normalisation, the scalar spectral tilt, and the minimal amount of
inflation. Another requirement of methodological nature,  the possible tree-level or one-loop dominance in $P_\zeta$ and/or
$B_\zeta$, is considered in this section. The level of non-gaussianity $f_{NL}$ in the bispectrum $B_\zeta$ is calculated in 
Section \ref{fnl} for models where $\zeta$ is generated during inflation; a comparison with the current literature is made. 
Section \ref{seccou} is devoted to central issues in the consistency of the approach followed such as satisfying necessary 
conditions for the convergence of the $\zeta$ series and working in a perturbative regime. Finally in Section \ref{conclusca}
we conclude. As regards the level of non-gaussianity $\tau_{NL}$ in the trispectrum $T_\zeta$, it will be studied in the 
following Chapter.

%%%%%%%%%%%%%%%%%%%%%%%%%%%%%%%%%%%%%%%%%%%%%%%%%%%%%%%%%%%%%%%%%%%%%%%%%%%%%%%%%%
\section{$\zeta$ series convergence and loop corrections}\label{conver}		%%%%%%
%%%%%%%%%%%%%%%%%%%%%%%%%%%%%%%%%%%%%%%%%%%%%%%%%%%%%%%%%%%%%%%%%%%%%%%%%%%%%%%%%%
In order to calculate $\zeta (t,{\bf x})$ from Eq. (\ref{dNsc}), we need information about the physical
content of the Universe at times $t$ and $t_{\rm in}$. By choosing the initial time $t_{\rm in}$ a few Hubble
times after the cosmologically relevant scales leave the horizon during inflation $t_{\rm in} = t_\star$, and the
final time $t$ corresponding to a slice of uniform energy density, we recognize that $N$, for slow-roll inflationary
models, is completely parametrized by the values a few Hubble times after horizon exit of the scalar fields $\phi_i$
present during inflation and the energy density at the time at which one wishes to calculate $\zeta$:
\begin{equation}
\zeta (t,{\bf x}) \equiv N(\rho(t),\phi_1(t_\star,{\bf x}),\phi_2(t_\star,{\bf x}), ...) - N(\rho(t),\phi_1(t_\star),
\phi_2(t_\star), ...) \,.
\end{equation}
The previous expression can be Taylor-expanded around the unperturbed background values for the scalar fields $\phi_i$
and suitably redefined so that $\langle \zeta(t,{\bf x}) \rangle = 0$. Thus,
\begin{eqnarray}
\zeta(t,\textbf{x})&=&\sum_{I}N_{I}(t)\delta\phi_{I}(\textbf{x},t_\star) - \sum_{I}N_{I}(t) \langle\delta\phi_{I}
(\textbf{x},t_\star)\rangle + \nonumber \\
&&+\frac{1}{2}\sum_{IJ}N_{IJ}(t)\delta\phi_{I}(\textbf{x},t_\star)\delta\phi_{J}(\textbf{x},t_\star)-\frac{1}{2}\sum_{IJ}N_{IJ}
(t) \langle\delta\phi_{I}(\textbf{x},t_\star)\delta\phi_{J}(\textbf{x},t_\star)\rangle + \nonumber \\
&&+\frac{1}{3!}\sum_{IJK}N_{IJK}(t)\delta\phi_{I}(\textbf{x},t_\star)\delta\phi_{J}(\textbf{x},t_\star)\delta\phi_{k}
(\textbf{x},t_\star)\no\\& -& \frac{1}{3!}\sum_{IJK}N_{IJK}(t)\langle\delta\phi_{I}(\textbf{x},t_\star)\delta\phi_{J}
(\textbf{x},t_\star)\delta\phi_{K}(\textbf{x},t_\star)\rangle + \nonumber \\
&&+...\;, \label{Nexp}
\end{eqnarray}
where the $\delta \phi_i (t_\star,{\bf x})$ are the scalar field perturbations in the flat slice a few Hubble
times after horizon exit, whose spectrum amplitude is given by \cite{bunch}
\begin{equation}
\mathcal{P}^{1/2}_{\delta \phi_{i}} =
\frac{H_\star}{2\pi}
\,,
\end{equation}
and the notation for the $N$ derivatives is $N_i\equiv\frac{\partial N}{\partial\phi_{i}}$,
$N_{ij}\equiv\frac{\partial^{2}N}{\partial\phi_{i}\partial\phi_{j}}$, and so on.

The expression in Eq. (\ref{Nexp}) has been used to calculate the statistical descriptors of $\zeta$ at
any desired order in cosmological perturbation theory by consistently truncating the series \cite{lr}.
For instance, by truncating the series at first order, the amplitude of the spectrum $P_\zeta$ of $\zeta$
defined in Eqs. (\ref{2pc}) and (\ref{asidef}) is given by \cite{ss}
\begin{equation}
\mathcal{P}_\zeta = \left(\frac{H_\star}{2\pi}\right)^2 \sum_i N_i^2 \,, \label{adnf}
\end{equation}
which in turn gives the well known formula for the spectral index \cite{ss}:
\begin{equation}
n_\zeta - 1 = -2\epsilon - 2m_P^2 \frac{\sum_{ij} V_i N_j N_{ij}}{V \sum_i N_i^2} \,, \label{ndnf}
\end{equation}
where a subindex $i$ in $V$ means a derivative with respect to the $\phi_i$ field, and
being $\epsilon$ one of the slow-roll parameters defined by $\epsilon = -\dot{H}/H^2$, $m_P = (8\pi G)^{-2}$
the reduced Planck mass, and $V$ the scalar inflationary potential.
Analogously, the level of non-gaussianity $f_{NL}$ in the bispectrum $B_\zeta$ of $\zeta$ defined in Eqs. (\ref{3pc})
and (\ref{bfp}) is obtained by truncating the series at second order and assuming that the scalar field perturbations
$\delta \phi_i$ are perfectly gaussian \cite{lr}:
\begin{equation}
\frac{6}{5} f_{NL} = \frac{\sum_{ij} N_i N_j N_{ij}}{\left[\sum_i N_i^2\right]^2} +  \mathcal{P}_\zeta
\frac{\sum_{ijk} N_{ij} N_{jk} N_{ki}}{\left[\sum_i N_i^2\right]^3} \ln(kL) \,. \label{fdnf}
\end{equation}
In the last expression the $\ln(kL)$ factor is of order one, $L$ being the infrared cutoff when calculating the
stochastic properties in a minimal box \cite{bernardeu4,lythbox}.

The truncated series methodology has proved to be powerful and reliable at reproducing successfully the level of
non-gaussianity $f_{NL}$ in single-field slow-roll models \cite{seery5} and in the curvaton scenario \cite{bl}.
Nevertheless, for more general models, how reliable is it to truncate the series at some order? In the first place,
from Eq. (\ref{Nexp}) it is impossible to know whether the series converges until the $N$ derivatives are explicitly
calculated and the convergence radius is obtained;  obviously if the series is not convergent at all, the expansion in Eq.
(\ref{Nexp}) is meaningless.  Without any proof of the contrary, the current assumption in the literature
\cite{alabidi3,alabidi2,battefeld,bl,bsw1,lr,ss,seery3,vernizzi,yokoyama2,yokoyama1,zaballa} has been that the
$\zeta$ series is convergent. In addition,
supposing that the convergence radius is finally known, the truncation at any desired order would again be meaningless if some
leading terms in the series get excluded.
Such a situation might easily happen if each ${\bf x}$-dependent term in the $\zeta$ series is considered smaller
than the previous one, which indeed is the standard assumption
\cite{alabidi3,alabidi2,battefeld,bl,bsw1,lr,ss,seery3,vernizzi,yokoyama2,yokoyama1,zaballa}, but which is not a universal
fact.

When studying the series through a diagrammatic approach \cite{byrnes1}, in an analogous way to that for Quantum Field Theory
via Feynman diagrams, the first-order terms in the spectral functions are called the tree-level terms. Examples of these
tree-level terms are those in Eqs. (\ref{adnf}) and (\ref{ndnf}), and the first one in Eq. (\ref{fdnf}). Higher-order
corrections, such as that which contributes with the second term in Eq. (\ref{fdnf}), are called the loop terms because they
involve internal momentum integrations. The statistical descriptors of $\zeta$ has been so far studied by naively neglecting
the loop corrections against the tree-level terms
\cite{alabidi3,alabidi2,battefeld,bl,bsw1,lr,ss,seery3,vernizzi,yokoyama2,yokoyama1,zaballa};
nevertheless, as might happen in Quantum Field Theory, eventually some loop corrections could be bigger than the tree-level
terms, so it is essential to properly study the possible $n$-loop dominance in the spectral functions.

%%%%%%%%%%%%%%%%%%%%%%%%%%%%%%%%%%%%%%%%%%%%%%%%%%%%%%%%%%%%%%%%%%%%%%
\section{Non-gaussianity in slow-roll inflation}  \label{ngsr} %%%%%%%
%%%%%%%%%%%%%%%%%%%%%%%%%%%%%%%%%%%%%%%%%%%%%%%%%%%%%%%%%%%%%%%%%%%%%%
The most frequent class of inflationary models found in the literature are those which satisfy the so called slow-roll
conditions, as these very simple models easily meet the spectral index observational requirements discussed in Subsection
\ref{observational} for the generation of large-scale structures.

The slow-roll conditions for single-field inflationary models with canonical kinetic terms read
\begin{eqnarray}
\dot{\phi}^2 &\ll& V(\phi) \,, \label{1stsrc} \\
|\ddot{\phi}| &\ll& |3H\dot{\phi}| \,, \label{2ndsrc}
\end{eqnarray}
where  $\phi$ is the inflaton field and $V(\phi)$ is the scalar field potential.
On defining the slow-roll parameters $\epsilon$ and $\eta_\phi$ as \cite{lythbook}
\begin{eqnarray}
\epsilon &\equiv& -\frac{\dot{H}}{H^2} \,, \label{1stsrp} \\
\eta_\phi &\equiv& \epsilon - \frac{\ddot{\phi}}{H\dot{\phi}} \,, \label{2ndsrp}
\end{eqnarray}
the slow-roll conditions in Eqs. (\ref{1stsrc}) and (\ref{2ndsrc}) translate into strong constraints for the slow-roll
parameters: $\epsilon, |\eta_\phi| \ll 1$, which actually become $\epsilon, |\eta_\phi| \lsim 10^{-2}$ in view of Eq.
(\ref{ndnf}) for single-field inflation:
\begin{equation}
n_\zeta - 1 = 2\eta_\phi - 6\epsilon \,,
\end{equation}
and the observational requirements presented in Subsection \ref{observational}.

Multifield slow-roll models may also be characterized by a set of slow-roll parameters which generalize those in Eqs.
(\ref{1stsrp}) and (\ref{2ndsrp}) \cite{lyth5}:
\begin{eqnarray}
\epsilon_i &\equiv& \frac{m_P^2}{2} \left(\frac{V_i}{V}\right)^2 \,, \\
\eta_i &\equiv& m_P^2 \frac{V_{ii}}{V} \,.
\end{eqnarray}
By writing the slow-roll parameters in terms of derivatives of the scalar potential, as in the latter two expressions,
we realize that the slow-roll conditions require very flat potentials to be met.

The level of non-gaussianity $f_{NL}$ in slow-roll inflationary models with canonical kinetic terms has been
studied both for single-field \cite{maldacena} and for multiple-field inflation \cite{battefeld,vernizzi,yokoyama1},
assuming $\zeta$ series convergence and considering only the tree-level terms both in $P_\zeta$ and in $B_\zeta$.
What these works find is a strong dependence on the slow-roll parameters $\epsilon_i$ and $\eta_i$; for instance,
Ref. \cite{maldacena} gives us for single-field models:
\begin{equation}
\frac{6}{5} f_{NL} = \epsilon(1 + f) + 2\epsilon - \eta_\phi \,,
\end{equation}
where $f$ is a function of the shape of the wavevectors triangle within the range $0 \leq f \leq 5/6$. Refs.
\cite{bartolo,lythbox} show that in such a case the inclusion of loop corrections is unnecessary because the latter are so
small compared to the tree-level terms. Thus, $f_{NL}$ in single-field models with canonical kinetic terms is slow-roll
suppressed and, therefore, unobservably small.  As regards the multifield models, $f_{NL}$ was shown, first in the case of two-
field inflation with separable potential \cite{vernizzi} and later for multiple-field inflation with separable \cite{battefeld}
and non-separable \cite{yokoyama1} potentials, to be a rather complex function of the slow-roll parameters and the scalar
potential that in most of the cases ends up being slow-roll suppressed. Only for models with a sharply curved trajectory in
field space might the $f_{NL}$ be at most of order one, the only possible examples to date being the models of Refs.
\cite{alabidi3,bernardeu2,bernardeu1,byrnes3}.
Again, such predictions are based on the assumptions that the $\zeta$ series is convergent and that the tree-level terms are
the leading ones, so they might be badly violated if loop corrections are considered.

Following a parallel treatment to that in Ref. \cite{vernizzi}, the level of non-gaussianity $\tau_{NL}$ is calculated in Ref.
\cite{seery3} for multifield slow-roll inflationary models with canonical kinetic terms, separable potential, and assuming
convergence of the $\zeta$ series and tree-level dominance.  From reaching similar conclusions to those found for the $f_{NL}$
case, the $\tau_{NL}$ is slow-roll suppressed in most of the cases although it might be of order one if the trajectory in field
space is sharply curved. Nevertheless, as we will shown in the next chapter, there may be a big enhancement in $\tnl$ if loop
corrections are taken into account.

Finally, it is worth mentioning that there are other classes of models where the levels of non-gaussianity $f_{NL}$ and
$\tau_{NL}$ are big enough to be observable.  Some of these models correspond to general langrangians with non-canonical
kinetic terms ($k$-inflation \cite{armendariz}, DBI inflation \cite{silverstein}, ghost inflation \cite{arkani}, etc.), where
the sizeable levels of non-gaussianity have mostly a quantum origin, i.e. their origin relies on the quantum correlators of the
field perturbations a few Hubble times after horizon exit. Non-gaussianity in $B_\zeta$ has been studied in these models for
the single-field case \cite{chen,seery6} and also for the multifield case \cite{arroja2,gao,langlois1,langlois2}. A recent
paper discusses the non-gaussianity in $T_\zeta$ for these general models for single-field inflation \cite{arroja}. In
contrast, there are some other models where the large non-gaussianities have their origin in the field dynamics at the end of
inflation \cite{bernardeu5,lythend}; nice examples of this proposal are studied for instance in Refs.
\cite{matsuda2,matsuda1,sasakin2,sasakin1}. However, since the inflationary models of the slow-roll variety with canonical
kinetic terms are the simplest, the most popular, and the best studied so far although, in principle, the non-gaussianity
statistical descriptors are too small to ever be observable, it is very interesting to consider the possibility of having an
example of such models which does generate {\it sizeable and observable values for $f_{NL}$ and $\tnl$}. This appealing
possibility will be the subject of the following sections and also the next chapter.

%%%%%%%%%%%%%%%%%%%%%%%%%%%%%%%%%%%%%%%%%%%%%%%%%%%%%%%%%%%%%%%%%%%%%%%%%%%%%%%%%%%%%%%%%
\section{A subclass of small-field slow-roll inflationary models} \label{model}		%%%%%
%%%%%%%%%%%%%%%%%%%%%%%%%%%%%%%%%%%%%%%%%%%%%%%%%%%%%%%%%%%%%%%%%%%%%%%%%%%%%%%%%%%%%%%%%

According to the classification of inflationary models proposed in Ref. \cite{dodelson2}, the small-field models are those of
the form that would be expected as a result of spontaneous symmetry breaking, with a field initially near an unstable
equilibrium point (usually taken to be at the origin) and rolling toward a stable minimum $\langle \phi \rangle \neq 0$.  Thus,
inflation occurs when the field is small relative to its expectation value $\phi \ll \langle \phi \rangle$.  Some interesting
examples are the original models of new inflation \cite{albste,linde}, modular inflation from string theory
\cite{dimopoulos}, natural inflation \cite{freese}, and hilltop inflation \cite{boubekeur2}. As a result, the inflationary
potential for small-field models may be taken as
\begin{equation}
V = \sum_i \Lambda_i \left[1 - \left(\frac{\phi_i}{\mu_i}\right)^p \right] \,,
\end{equation}
where the subscript $i$ here denotes the relevant quantities of the $i$th field, $p$ is the same for all fields, and
$\Lambda_i$ and $\mu_i$ are the parameters describing the height and tilt of the potential of the $i$th field.

\begin{figure*} [t]
\begin{center}
\includegraphics[width=10cm,height=15cm,angle=-90]{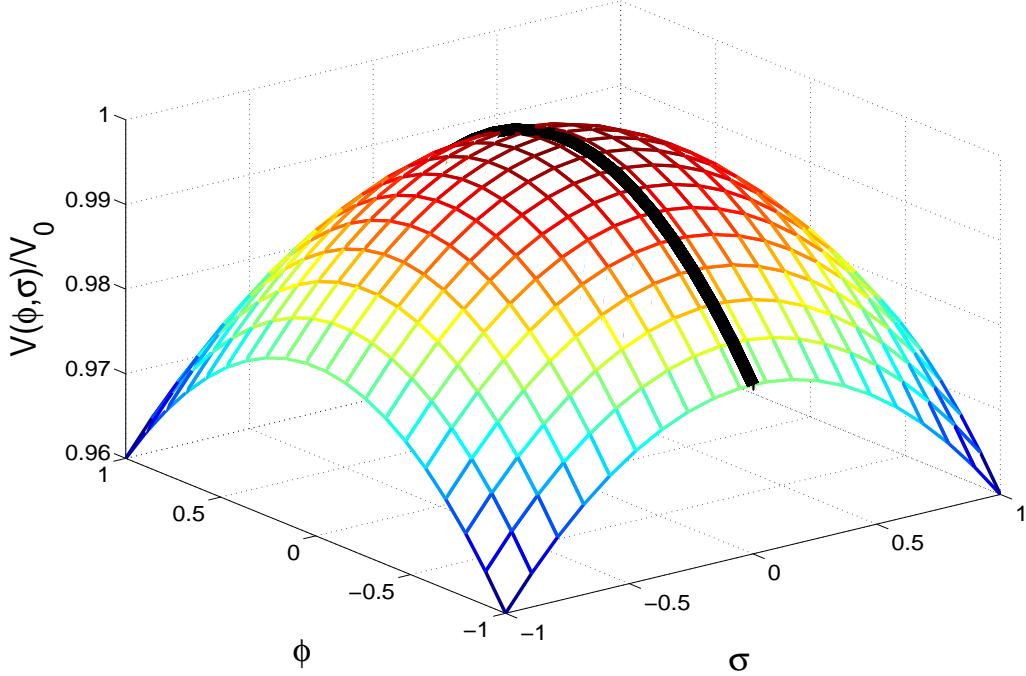}
\end{center}
\caption[Our small-field slow-roll potential of Eq. (\ref{pot}) with $\eta_\phi,\eta_\sigma < 0$]{Our small-field slow-roll
potential of Eq. (\ref{pot}) with $\eta_\phi,\eta_\sigma < 0$. The inflaton starts near the maximum and moves away from the
origin following the $\sigma = 0$ trajectory depicted with the solid black line. (This figure has been taken from Ref.
\cite{alabidi1}).}
\label{potential}
\end{figure*}

While Ref. \cite{ahmad} studies the spectrum of $\zeta$ for general values of the parameter $p$ and an arbitrary number of
fields, assuming $\zeta$ series convergence and tree-level dominance, we will specialize to the $p=2$ case for two fields
$\phi$ and $\sigma$:
\begin{equation}
V = V_0\left(1+\frac{1}{2}\eta_\phi \frac{\phi^2}{m_P^2} +\frac{1}{2}\eta_\sigma \frac{\sigma^2}{m_P^2}\right) \,, \label{pot}
\end{equation}
where we have traded the expressions
\begin{eqnarray}
\Lambda_1 + \Lambda_2 &{\rm for}& V_0 \,, \\
%and the parameters
\frac{\Lambda_1}{\mu_1^2} &{\rm for}& -V_0 \frac{\eta_\phi}{2m_P^2} \,,
\end{eqnarray}
and
\begin{equation}
\frac{\Lambda_2}{\mu_2^2} \; \; \; {\rm for} \; \; \; -V_0 \frac{\eta_\sigma}{2m_P^2} \,.
\end{equation}
By doing this, and assuming that the first term in Eq. (\ref{pot}) dominates, $\eta_\phi < 0$ and $\eta_\sigma < 0$ become the
usual $\eta$ slow-roll parameters associated with the fields $\phi$ and $\sigma$.

We have chosen for simplicity the $\sigma=0$ trajectory (see Fig. \ref{potential}) since in that case the potential in Eq.
(\ref{pot}) reproduces for some number of e-folds the hybrid inflation scenario \cite{linde2} where $\phi$ is the inflaton and
$\sigma$ is the waterfall field. Non-gaussianity in such a model has been studied in Refs.
\cite{alabidi3,enqvist1,lr,lr1,vaihkonen,zaballa}; in particular, Ref. \cite{lr} used a one-loop correction to conjecture that
$f_{NL}$ in this model would be sizeable {\it only if} $\zeta$ {\it was not} generated during inflation, which turns out not to
be a necessary requirement as we will show later \cite{cogollo,valenzuela}. Ref. \cite{alabidi3}, in contrast, works only at tree-level with the same
potential as Eq. (\ref{pot}) but relaxing the $\sigma = 0$ condition, finding that values for $f_{NL} \sim \mathcal{O}(1)$ are
possible for a small set of initial conditions and assuming a saddle-point-like form for the potential ($\eta_\phi < 0$ and
$\eta_\sigma > 0$).

%%%%%%%%%%%%%%%%%%%%%%%%%%%%%%%%%%%%%%%%%%%%%%%%%%%%%%%%%%%%%%%%%%%%%%%%%%%%%%%%%%%%%
\section{Constraints for having a reliable parameter space} \label{rest}	%%%%%%%%%
%%%%%%%%%%%%%%%%%%%%%%%%%%%%%%%%%%%%%%%%%%%%%%%%%%%%%%%%%%%%%%%%%%%%%%%%%%%%%%%%%%%%%

We will explore now the constraints that the model must satisfy before we calculate the level of non-gaussianity $f_{NL}$. Our guiding idea will be the consideration of the role that the tree-level terms and one-loop corrections in $P_\zeta$ and 
$B_\zeta$ have in the determination of the available parameter space.  Only after calculating $f_{NL}$ in
Section \ref{fnl} will we come back to the discussion of the consistency of the approach followed in the present section by
studying the $\zeta$ series convergence and the validity of the truncation at one loop level.
%%%%%%%%%%%%%%%%%%%%%%%%%%%%%%%%%%%%%%%%%%%%%%%%%%%%%%%%%%%%%%%%%%%%%%%%%%
\subsection{Tree-level or one-loop dominance: $\fnl$}\label{domsecfnl}
%%%%%%%%%%%%%%%%%%%%%%%%%%%%%%%%%%%%%%%%%%%%%%%%%%%%%%%%%%%%%%%%%%%%%%%%%%
Since we are considering a slow-roll regime, the evolution of the background $\phi$ and $\sigma$ fields in such a case is given
by the Klein-Gordon equation
\begin{equation}
\ddot{\phi} + 3H\dot{\phi} + V_\phi = 0 \,,
\end{equation}
supplemented with the slow-roll condition in Eq. (\ref{2ndsrc}).  This leads to
\begin{eqnarray}
\phi(N) &=& \phi_\star \exp(-N\eta_\phi) \,, \label{srp} \\
\sigma(N) &=& \sigma_\star \exp(-N\eta_\sigma) \label{srs}\,,
\end{eqnarray}
so the potential above leads to the following derivatives of $N$ with respect to $\phi_\star$ and $\sigma_\star$ for the
$\sigma = 0$ trajectory:\footnote{When calculating the $N$-derivatives, we have considered that the final time corresponds to a
slice of uniform energy density. This means, in the slow-roll approximation, that $V$ is homogeneous.}
\begin{eqnarray}
&&N_\phi = \frac{1}{\eta_\phi \phi_\star} \,, \hspace{4mm}  N_\sigma = 0 \,, \label{1d} \\
&&N_{\phi \phi} = -\frac{1}{\eta_\phi \phi_\star^2} \,, \hspace{4mm} N_{\phi \sigma} = 0 \,, \hspace{4mm} N_{\sigma \sigma} =
\frac{\eta_\sigma}{\eta_\phi^2 \phi_\star^2} \exp[2N (\eta_\phi - \eta_\sigma)] \,, \label{2d} \\
&&N_{\phi \phi \phi} = \frac{2}{\eta_\phi \phi_\star^3}\,, \hspace{4mm} N_{\phi \phi \sigma} = 0\,, \hspace{4mm} N_{\sigma
\sigma \phi} = -\frac{2\eta_\sigma^2}{\eta_\phi^3 \phi_\star^3} \exp[2N (\eta_\phi - \eta_\sigma)]\,, \hspace{3mm} N_{\sigma
\sigma \sigma} = 0\,, \label{3d} \\
&&... \nonumber \\
&&{\rm and \ so \ on}. \nonumber
\end{eqnarray}
By means of the $\delta N$ formalism, we can make use of the above formulae to calculate the spectrum and the bispectrum of the
curvature perturbation including the tree-level and the one-loop contributions when $|\eta_\sigma| > |\eta_\phi|$ (see Appendix
\ref{app}).  This is the interesting case since, as will be shown in Section \ref{fnl}, it generates sizeable values for
$f_{NL}$.  Following the results in Appendix \ref{app},  we will write down just the leading terms to the tree-level and one-
loop contributions given in Eqs. (\ref{pt1}), (\ref{1lfpd}), (\ref{tlbd1}) and (\ref{1lfbd}).
\bea
{\mathcal P}_\zeta^{tree}&=&\frac{1}{\eta_\phi^2\phi_\star^2}\left(\frac{H_\star}{2 \pi}\right)^2 \,,\label{pt}\\
{\mathcal P}_\zeta^{ 1-loop}&=&\frac{\eta_\sigma^2}{\eta_\phi^4\phi_\star^4}\exp[4N(\eta_\phi-\eta_\sigma)]
\left(\frac{H_\star}{2 \pi}\right)^4\ln(kL) \,,\label{pl}\\
B_\zeta^{tree} &=& -\frac{1}{\eta_\phi^3 \phi_\star^4} \left(\frac{H_\star}{2\pi}\right)^4 4\pi^4 \left(\frac{\sum_i k_i^3}
{\prod_i k_i^3}\right) \,, \label{bt} \\
B_\zeta^{1-loop} &=& \frac{\eta_\sigma^3}{\eta_\phi^6 \phi_\star^6} \exp[6N(\eta_\phi - \eta_\sigma)] \left(\frac{H_\star}
{2\pi}\right)^6 \ln(kL) 4\pi^4 \left(\frac{\sum_i k_i^3}{\prod_i k_i^3}\right) \,. \label{bl}
\eea
Because of the exponential factors in Eqs. (\ref{pl}) and (\ref{bl}), it might be possible that the one-loop corrections
dominate over $P_\zeta$ and/or $B_\zeta$.  There are three posibilities in complete connection with the position of the $\phi$
field when the cosmologically relevant scales are exiting the horizon:
%%%%%%%%%%%%%%%%%%%%%%%%%%%%%%%%%%%%%%%%%%%%%%%%%%%%%%%%%%%%%%%%%%%%%%%%%%%%%%%%%%%%%%%%%%%%%%%%%%%%%%
\subsubsection{Both $B_\zeta$ and $P_\zeta$ are dominated by the one-loop corrections}
Comparing Eqs. (\ref{pt}) with (\ref{pl}) and Eqs. (\ref{bt}) with (\ref{bl}) we require in this case that
\begin{eqnarray}
\frac{\eta_\sigma^2}{\eta_\phi^2} \exp[4N(|\eta_\sigma|-|\eta_\phi|)] &\gg& \frac{1}{\frac{1}{\phi_\star^2}
\left(\frac{H_\star}{2\pi}\right)^2} \,, \\
\frac{\eta_\sigma^3}{\eta_\phi^3} \exp[6N(|\eta_\sigma|-|\eta_\phi|)] &\gg& \frac{1}{\frac{1}{\phi_\star^2}
\left(\frac{H_\star}{2\pi}\right)^2} \,,
\end{eqnarray}
in which case only the first inequality is required. Employing the definition for the tensor to scalar ratio $r$ \cite{lyth6}:
\begin{equation}
r \equiv \frac{\mathcal{P}_T}{\mathcal{P}_\zeta} = \frac{\frac{8}{m_P^2} \left(\frac{H_\star}{2\pi}\right)^2}
{\mathcal{P}_\zeta} \,, \label{defr}
\end{equation}
$\mathcal{P}_T^{1/2}$ being the amplitude of the spectrum for primordial gravitational waves,
we can write such an inequality as
\begin{equation}
%&&
\left(\frac{\phi_\star}{m_P}\right)^2 \ll \frac{r \mathcal{P}_\zeta}{8} \frac{\eta_\sigma^2}{\eta_\phi^2} \exp[4N(|
\eta_\sigma|-|\eta_\phi|)] \,.
\label{lowphi}
\end{equation}
From now on we will name the parameter window described by Eq. (\ref{lowphi}) as the low $\phi_\star$ region since the latter
represents a region of allowed values for $\phi_\star$ limited by an upper bound.
%%%%%%%%%%%%%%%%%%%%%%%%%%%%%%%%%%%%%%%%%%%%%%%%%%%%%%%%%%%%%%%%%%%%%%%%%%%%%%%%%%%%%%%%%%%%%%%%%%%%%%
\subsubsection{$B_\zeta$ dominated by the one-loop correction and $P_\zeta$ dominated by the tree-level term} \label{bz1l}
Comparing Eqs. (\ref{pt}) with (\ref{pl}) and Eqs. (\ref{bt}) with (\ref{bl}) we require in this case that
\begin{eqnarray}
\frac{\eta_\sigma^2}{\eta_\phi^2} \exp[4N(|\eta_\sigma|-|\eta_\phi|)] &\ll& \frac{1}{\frac{1}{\phi_\star^2}
\left(\frac{H_\star}{2\pi}\right)^2} \,, \label{wn} \\
\frac{\eta_\sigma^3}{\eta_\phi^3} \exp[6N(|\eta_\sigma|-|\eta_\phi|)] &\gg& \frac{1}{\frac{1}{\phi_\star^2}
\left(\frac{H_\star}{2\pi}\right)^2} \,,
\end{eqnarray}
which combines to give, employing the definition for the tensor to scalar ratio $r$ introduced in Eq. (\ref{defr}),
\begin{equation}
%&&
\frac{r \mathcal{P}_\zeta}{8} \frac{\eta_\sigma^2}{\eta_\phi^2} \exp[4N(|\eta_\sigma|-|\eta_\phi|)] \ll \left(\frac{\phi_\star}
{m_P}\right)^2 \ll \frac{r \mathcal{P}_\zeta}{8} \frac{\eta_\sigma^3}{\eta_\phi^3} \exp[6N(|\eta_\sigma|-|\eta_\phi|)] \,. %\\
\label{intc}
\end{equation}
From now on we will name the parameter window described by Eq. (\ref{intc}) as the intermediate $\phi_\star$ region since the
latter represents a region of allowed values for $\phi_\star$ limited by both an upper and a lower bound.
%%%%%%%%%%%%%%%%%%%%%%%%%%%%%%%%%%%%%%%%%%%%%%%%%%%%%%%%%%%%%%%%%%%%%%%%%%%%%%%%%%%%%%%%%%%%%%%%%%%%%%
\subsubsection{Both $B_\zeta$ and $P_\zeta$ are dominated by the tree-level terms}
Comparing Eqs. (\ref{pt}) with (\ref{pl}) and Eqs. (\ref{bt}) with (\ref{bl}) we require in this case that
\begin{eqnarray}
\frac{\eta_\sigma^2}{\eta_\phi^2} \exp[4N(|\eta_\sigma|-|\eta_\phi|)] &\ll& \frac{1}{\frac{1}{\phi_\star^2}
\left(\frac{H_\star}{2\pi}\right)^2} \,, \\
\frac{\eta_\sigma^3}{\eta_\phi^3} \exp[6N(|\eta_\sigma|-|\eta_\phi|)] &\ll& \frac{1}{\frac{1}{\phi_\star^2}
\left(\frac{H_\star}{2\pi}\right)^2} \,,
\end{eqnarray}
in which case only the second inequality is required.  Employing the definition for the tensor to scalar ratio $r$ introduced
in Eq. (\ref{defr}), we can write such an inequality as
\begin{eqnarray}
\left(\frac{\phi_\star}{m_P}\right)^2 \gg \frac{r \mathcal{P}_\zeta}{8} \frac{\eta_\sigma^3}{\eta_\phi^3} \exp[6N(|
\eta_\sigma|-|\eta_\phi|)] \,. \label{highphi}
\end{eqnarray}
From now on we will name the parameter window described by Eq. (\ref{highphi}) as the high $\phi_\star$ region, since the
latter represents a region of allowed values for $\phi_\star$ limited by a lower bound.
%%%%%%%%%%%%%%%%%%%%%%%%%%%%%%%%%%%%%%%%%%%%%%%%%%%%%%%%%%%%%%%%%%%
\subsection{Spectrum normalisation condition} \label{secnorm}
%%%%%%%%%%%%%%%%%%%%%%%%%%%%%%%%%%%%%%%%%%%%%%%%%%%%%%%%%%%%%%%%%%%
The model must satisfy the COBE normalisation on the spectrum amplitude $\mathcal{P}^{1/2}_\zeta$ \cite{bunn} considering that
$\zeta$ is assumed to be generated during inflation\footnote{The scenario where $\zeta$ is assumed not to be 
generated during inflation will be presented in the next chapter.}. There exist two possibilities discussed right below.
%%%%%%%%%%%%%%%%%%%%%%%%%%%%%%%%%%%%%%%%%%%%%%%%%%%%%%%%%%%%%%%%%%%%%%%%%%%%%%%%%%%%%%%%%%%%%%%%%%%%%%
\subsubsection{$\zeta$ generated during inflation and $P_\zeta$ dominated by the one-loop correction}
According to Eq. (\ref{pl}), and the tensor to scalar ratio $r$ definition in Eq. (\ref{defr}), we have in this case
\begin{eqnarray}
\mathcal{P}_\zeta^{1-loop} &=& \frac{\eta_\sigma^2}{\eta_\phi^4 \phi_\star^4} \exp[4N(|\eta_\sigma| - |\eta_\phi|)]
\left(\frac{H_\star}{2\pi}\right)^4 \ln(kL) \nonumber \\
&=& \frac{\eta_\sigma^2}{\eta_\phi^4} \exp[4N(|\eta_\sigma| - |\eta_\phi|)] \left(\frac{m_P}{\phi_\star}\right)^4 \left(\frac{r
\mathcal{P}_\zeta}{8}\right)^2 \ln(kL) \,,
\end{eqnarray}
which reduces to
\begin{equation}
\left(\frac{\phi_\star}{m_P}\right)^4 = \left(\frac{r}{8}\right)^2 \mathcal{P}_\zeta \frac{\eta_\sigma^2}{\eta_\phi^4}
\exp[4N(|\eta_\sigma| - |\eta_\phi|)] \ln(kL) \,,
\end{equation}
where $\mathcal{P}_\zeta$ must be replaced by the observed value presented in Subsection \ref{observational}.
%%%%%%%%%%%%%%%%%%%%%%%%%%%%%%%%%%%%%%%%%%%%%%%%%%%%%%%%%%%%%%%%%%%%%%%%%%%%%%%%%%%%%%%%%%%%%%%%%%%%%%
\subsubsection{$\zeta$ generated during inflation and $P_\zeta$ dominated by the tree-level term} \label{zipt}
According to Eq. (\ref{pt}), and the tensor to scalar ratio $r$ definition in Eq. (\ref{defr}), we have in this case
\begin{eqnarray}
\mathcal{P}_\zeta^{tree} &=& \frac{1}{\eta_\phi^2 \phi_\star^2} \left(\frac{H_\star}{2\pi}\right)^2 \nonumber \\
&=& \frac{1}{\eta_\phi^2} \left(\frac{m_P}{\phi_\star}\right)^2 \frac{r \mathcal{P}_\zeta}{8} \,, \label{cx1}
\end{eqnarray}
which reduces to
\begin{equation}
\left(\frac{\phi_\star}{m_P}\right)^2 = \frac{1}{\eta_\phi^2} \frac{r}{8} \,. \label{normt}
\end{equation}
Notice that in such a situation, the value of the $\phi$ field when the cosmologically relevant scales are exiting the horizon
depends exclusively on the tensor to scalar ratio $r$, once $\eta_\phi$ has been fixed by the spectral tilt constraint as we
will see later.
%%%%%%%%%%%%%%%%%%%%%%%%%%%%%%%%%%%%%%%%%%%%%%%%%%%%%%%%%%%%%
\subsection{Spectral tilt constraint} \label{sectilt}
%%%%%%%%%%%%%%%%%%%%%%%%%%%%%%%%%%%%%%%%%%%%%%%%%%%%%%%%%%%%
The combined 5-year WMAP + Type I Supernovae + Baryon Acoustic Oscillations data \cite{wmap5} constrain the value for the
spectral tilt as
\begin{equation}
n_\zeta - 1 = -0.040 \pm 0.014 \,.
\end{equation}
Here again we have two possibilities:  $\mathcal{P}_\zeta$ is dominated either by the one-loop correction or by the tree-level
term:
%%%%%%%%%%%%%%%%%%%%%%%%%%%%%%%%%%%%%%%%%%%%%%%%%%%%%%%%%%%%%%%%%%%%%%%%%%%%%%%%%%%%%%%%%%%%%%%%%%%%%%
\subsubsection{$P_\zeta$ dominated by the one-loop correction} \label{indexkom}
In this case the usual spectral index formula at tree-level \cite{ss} gets modified to account for the leading one-loop
correction:
\begin{equation}
n_\zeta - 1 = -4\epsilon - 2m_P^2 \frac{\sum_{ijk} V_k N_{ijk} N_{ij}}{V \sum_{ij} N_{ij} N_{ij}} + \left[\ln
(kL)\right]^{-1}\,.
\end{equation}
By making use of the derivatives in Eqs. (\ref{1d}), (\ref{2d}), and (\ref{3d}), we have
\begin{equation}
n_\zeta - 1 = -4\epsilon + 4\eta_\sigma + \left[\ln (kL)\right]^{-1}\,, \label{nskomat}
\end{equation}
which implies that the observed value for $n_\zeta$ is never reproduced
in view of $\ln (kL) \sim \mathcal{O} (1)$.  Moreover, when calculating the running spectral index $dn_\zeta/d \ln k$ from Eq.
(\ref{nskomat}),
we obtain
\begin{equation}
\frac{dn_\zeta}{d \ln k} = -\left[\ln(kL)\right]^{-2} \,,
\end{equation}
which rules out the possibility that $P_\zeta$ is dominated by the one-loop correction since the calculated $dn_\zeta/d \ln k$
is far from the observationally allowed $2\sigma$ range of values:  $-0.0728 < dn_\zeta/d \ln k < 0.0087$ 
\cite{wmap5}\footnote{We thank Eiichiro Komatsu for pointing out to us the dependence of $\nz$ and $d\nz/d\ln k$ on 
$\ln(kL)$.}.
%%%%%%%%%%%%%%%%%%%%%%%%%%%%%%%%%%%%%%%%%%%%%%%%%%%%%%%%%%%%%%%%%%%%%%%%%%%%%%%%%%%%%%%%%%%%%%%%%%%%%%
\subsubsection{$P_\zeta$ dominated by the tree-level term}
Now the usual spectral index formula \cite{ss} applies:
\begin{equation}
n_\zeta - 1 = -2\epsilon - 2m_P^2 \frac{\sum_{ij} V_i N_j N_{ij}}{V \sum_i N_i^2} \,,
\end{equation}
giving the following result once the derivatives in Eqs. (\ref{1d}), (\ref{2d}), and (\ref{3d}) have been used:
\begin{equation}
n_\zeta - 1 = -2\epsilon + 2\eta_\phi \,.
\end{equation}
The efect of the $\epsilon$ parameter may be discarded in the previous expression since, as often happens in small-field models
\cite{alabidi2,boubekeur2}, $\epsilon$ is negligible being much less than $|\eta_\sigma|$:
\begin{equation}
\epsilon = \frac{m_P^2}{2} \frac{V_\phi^2 + V_\sigma^2}{V^2} = |\eta_\phi| \left[\frac{1}{2} |\eta_\phi| \left(\frac{\phi}
{m_P}\right)^2\right] \ll |\eta_\phi| < |\eta_\sigma| \,,
\end{equation}
according to the prescription that the potential in Eq. (\ref{pot}) is dominated by the constant term.
Thus, by using the central value for $n_\zeta - 1$, we get
\begin{equation}
\eta_\phi = -0.020 \,. \label{tiltt}
\end{equation}
%%%%%%%%%%%%%%%%%%%%%%%%%%%%%%%%%%%%%%%%%%%%%%%%%%%%%%%
\subsection{Amount of inflation} \label{secamount}
%%%%%%%%%%%%%%%%%%%%%%%%%%%%%%%%%%%%%%%%%%%%%%%%%%%%%%
Because of the characteristics of the inflationary potential in Eq. (\ref{pot}), inflation is eternal in this model.  However,
Ref. \cite{naruko} introduced the multi-brid inflation idea of Refs. \cite{sasakin2,sasakin1} so that the potential in Eq.
(\ref{pot}) is achieved during inflation
while a third field $\rho$ acting as a waterfall field is stabilized in $\rho = 0$. During inflation $\rho$ is heavy and it is
trapped with a vacuum expectation value equal to zero, so neglecting it during inflation is a good approximation. The end of
inflation comes when the effective mass of $\rho$ becomes negative, which is possible to obtain if $V_0$ in the potential of
Eq. (\ref{pot}) is replaced by
\begin{equation}
V_0 = \frac{1}{2} G(\phi,\sigma) \rho^2 + \frac{\lambda}{4} \left(\rho^2 - \frac{\Sigma^2}{\lambda} \right)^2 \,,
\label{endcoup}
\end{equation}
where
\begin{equation}
G(\phi,\sigma) = g_1\phi^2 + g_2\sigma^2 \,, \label{hypend}
\end{equation}
$\Sigma$ has some definite value and $g_1$, $g_2$, and $\lambda$ some coupling constants. The end of inflation actually happens
on a hypersurface defined in general by  \cite{naruko,sasakin1}
\begin{equation}
G(\phi,\sigma) = \Sigma^2 \,.
\end{equation}
In general the hypersurface in Eq. (\ref{hypend}), defined by the end of inflation condition,
is not a surface of uniform energy density.  Because the $\delta N$ formalism requires the
final slice to be of uniform energy density\footnote{See for instance Ref. \cite{cogollo}.},
we need to add a small correction term to the amount of expansion up to the surface where
$\rho$ is destabilised.  In addition, the end of inflation is inhomogeneous, which
generically leads to different predictions from those obtained during inflation for the
spectral functions \cite{alabidiend,lythend,salem}. In particular, large levels of
non-gaussianity may be obtained by tunning the free parameters of the model. Specifically,
by making $g_1/g_2 \ll 1$, large values for $f_{NL}$ and $\tau_{NL}$ are obtained due to the
end of inflation mechanism rather than due to the dynamics during slow-roll inflation
\cite{alabidiend,huang,naruko}.

Ref. \cite{bch2} chose instead the case $g_1^2/g_2^2 = \eta_\phi/\eta_\sigma$ such that the surface where $\rho$ is
destabilised corresponds to a surface of uniform energy density\footnote{Ref. \cite{bch2} studies as well the case where $g_1^2
= g_2^2$, but we are not going to consider it here.}. In this case all of the spectral functions are the same as those
calculated in this thesis (see Refs. \cite{bch2,cogollo,valenzuela}), which in turn are valid at the final hypersurface of uniform energy
density during slow-roll inflation. Thus, we have a definite mechanism to end inflation which, nevertheless, leaves intact the
non-gaussianity generated during inflation.

It is well known that the number of e-folds of expansion from the time the cosmological scales exit the horizon to the end of
inflation is presumably around but less than 62 \cite{dodelson,lythbook,mukhanov,weinberg3}.  The slow-roll evolution of the
$\phi$ field in Eq. (\ref{srp}) tells us that such an amount of inflation is given by
\begin{equation}
N = -\frac{1}{\eta_\phi} \ln\left(\frac{\phi_{end}}{\phi_\star}\right) \lsim 62 \,, \label{end}
\end{equation}
where $\phi_{end}$ is the value of the $\phi$ field at the end of inflation. Such a value depends noticeably on the coupling
constants in Eq. (\ref{endcoup}).  We will in this chapter not concentrate on the allowed parameter window for $g_1$, $g_2$, and
$\lambda$. Instead, we will give an upper bound on $\phi$ during inflation, for the $\eta_\phi < 0$ case, consistent with the
potential in Eq. (\ref{pot}) and the end of inflation mechanism described above. %For the cases where $\eta_\phi >0$, we will
%not give any further details about the end of inflation since they are not necessary as we will see in the following sections.

Keeping in mind the results of Ref. \cite{armendariz2} which say that the ultraviolet cutoff in cosmological perturbation
theory could be a few orders of magnitude bigger than $m_P$, we will tune the coupling constants in Eq. (\ref{endcoup}) so that
inflation for $\eta_\phi < 0$ comes to an end when $|\eta_\phi|\phi^2/2m_P^2 \sim 10^{-2}$. This allows us to be on the safe 
side (avoiding large modifications to the potential coming from ultraviolet cutoff-suppressed non-renormalisable terms, and 
keeping the potential dominated by the constant $V_0$ term). Coming back to Eq. (\ref{end}), we get then
\begin{equation}
N = \frac{1}{|\eta_\phi|} \ln\left[\left(\frac{2\times10^{-2}}{|\eta_\phi|}\right)^{1/2}\frac{m_P}{\phi_\star}\right] \lsim 62
\,, \label{amount}
\end{equation}
which leads to
\begin{equation}
\frac{\phi_\star}{m_P} \gsim \left(\frac{2\times10^{-2}}{|\eta_\phi|}\right)^{1/2}\exp(-62|\eta_\phi|) \,. \label{amountc}
\end{equation}
%%%%%%%%%%%%%%%%%%%%%%%%%%%%%%%%%%%%%%%%%%%%%%%%%%%%%%%%%%%%%%%%%%
\section{Non-Gaussianity: $f_{NL}$} \label{fnl}			%%%%%%%%%%
%%%%%%%%%%%%%%%%%%%%%%%%%%%%%%%%%%%%%%%%%%%%%%%%%%%%%%%%%%%%%%%%%%
In this section we will calculate the level of non-gaussianity represented in the parameter $f_{NL}$ by taking into account the
constraints presented in Subsections \ref{secnorm}, \ref{sectilt}, and \ref{secamount}, and the different $\phi_\star$ regions
discussed in Subsection \ref{domsecfnl}.

\subsection{The low $\phi_\star$ region}
This case is of no observational interest because $P_\zeta$ dominated by the one-loop correction is already ruled out by the
observed spectral index and its running as shown in Subsection \ref{indexkom}.  In addition, the generated non-gaussianity is
so big that it causes violation of the observational constraint $f_{NL} > -9$:
\begin{equation}
\frac{6}{5} f_{NL} = \frac{B_\zeta^{1-loop}}{4\pi^4 \frac{\sum_i k_i^3}{\prod_i k_i^3} (\mathcal{P}_\zeta^{1-loop})^2} =
-[\mathcal{P}_\zeta \ln(kL)]^{-1/2} \sim -2 \times 10^4 \,,
\end{equation}
according to the expressions in Eqs. (\ref{asidef}), (\ref{bfp}), (\ref{pl}), and (\ref{bl}).

We want to remark that, although it is of no observational interest, this case represents the first example of large non-
gaussianity in the bispectrum $B_\zeta$ of $\zeta$ for a slow-roll model of inflation with canonical kinetic terms. It is funny
to realize that the model in this case is additionally ruled out because the observational constraint on $f_{NL}$ is violated
{\it by an excess} and not by a shortfall as is currently thought \cite{battefeld,maldacena,seery7,vernizzi,yokoyama1}.

\subsection{The intermediate $\phi_\star$ region}

The level of non-gaussianity, according to the expressions in Eqs. (\ref{asidef}), (\ref{bfp}), (\ref{pt}), and (\ref{bl}), is
in this case given by
{\small\begin{eqnarray}
\frac{6}{5} f_{NL} &=& \frac{B_\zeta^{1-loop}}{4\pi^4 \frac{\sum_i k_i^3}{\prod_i k_i^3} (\mathcal{P}_\zeta^{tree})^2} =
\frac{\eta_\sigma^3}{\eta_\phi^2 \phi_\star^2} \exp[6N(|\eta_\sigma|-|\eta_\phi|)] \left(\frac{H_\star}{2\pi}\right)^2 \ln(kL)
\nonumber \\
&=& \frac{\eta_\sigma^3}{\eta_\phi^2} \exp[6N(|\eta_\sigma|-|\eta_\phi|)] \left(\frac{m_P}{\phi_\star}\right)^2 \frac{r
\mathcal{P}_\zeta}{8} \ln(kL) \nonumber \\
&=& \eta_\sigma^3 \exp[6N(|\eta_\sigma|-|\eta_\phi|)] \mathcal{P}_\zeta \ln(kL) \,, \label{poseta} \\
\Rightarrow \frac{6}{5} f_{NL} &\approx& -2.457 \times 10^{-9} |\eta_\sigma|^3 \exp[300 \ \ln(5.657 \times 10^{-2} r^{-1/2})
\left(|\eta_\sigma| - 0.020\right)] \,,
\end{eqnarray}}
where in the last line we have used expressions from Eqs. (\ref{normt}), (\ref{tiltt}), and (\ref{amount}).

Now, by implementing the spectral tilt constraint in Eq. (\ref{tiltt}) in the spectrum normalisation constraint in Eq.
(\ref{normt}) and the amount of inflation constraint in Eq. (\ref{amountc}), we conclude that the tensor to scalar ratio $r$ is
bounded from below:  $r \gsim 2.680 \times 10^{-4}$.

\begin{figure*} [t]
\begin{center}
\includegraphics[width=15cm,height=10cm]{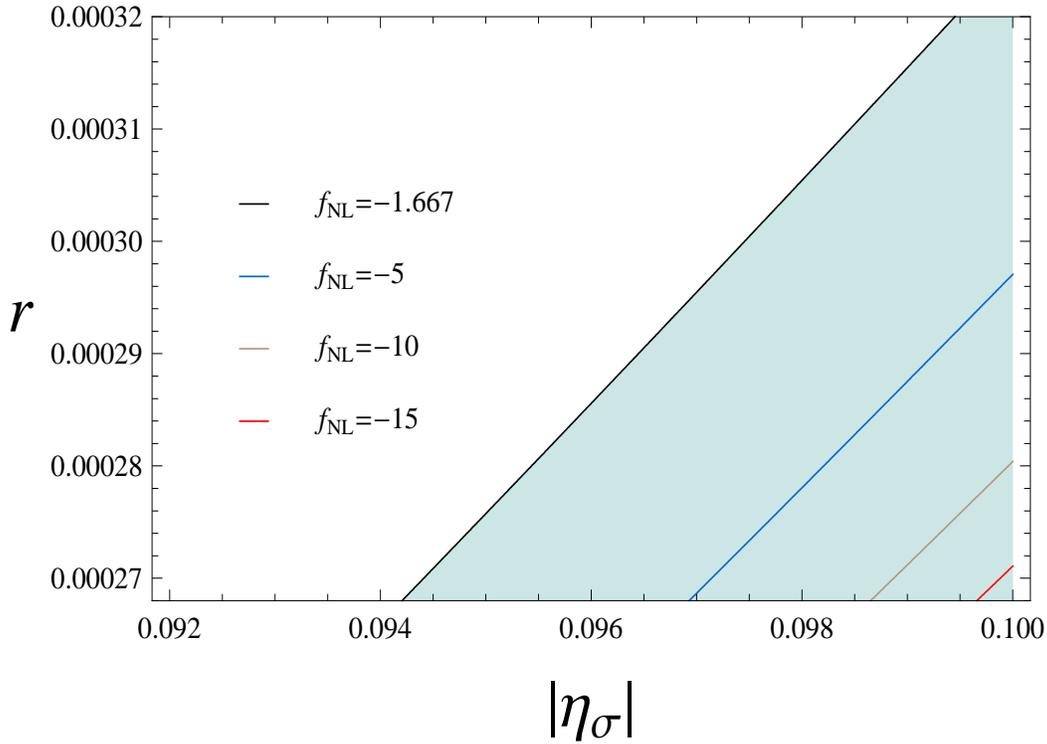}
\end{center}
\caption[Contours of $f_{NL}$ in the $r$ vs $|\eta_\sigma|$ plot]{Contours of $f_{NL}$ in the $r$ vs $|\eta_\sigma|$ plot.
The intermediate (high) $\phi_\star$ region corresponds to the shaded (white) region. The WMAP (and also PLANCK)
observationally allowed $2\sigma$ range of values for negative $f_{NL}$, $-9 < f_{NL}$, is completely inside the intermediate
$\phi_\star$ region. Notice that the boundary line between the high and the intermediate $\phi_\star$ regions matches almost
exactly the $f_{NL} = -1.667$ line. }
\label{fig32}
\end{figure*}

In the plot $r$ vs $|\eta_\sigma|$ in figure \ref{fig32}, we show lines of constant $f_{NL}$ corresponding to the values $f_{NL}
= -5, -10, -15$.  We also show the high and intermediate $\phi_\star$ regions in agreement with the constraint in Eq.
(\ref{intc}):
\be
\frac{r \mathcal{P}_\zeta}{8} \frac{\eta_\sigma^2}{\eta_\phi^2} \exp[4N(|\eta_\sigma|-|\eta_\phi|)] \ll \left(\frac{\phi_\star}
{m_P}\right)^2 \ll \frac{r \mathcal{P}_\zeta}{8} \frac{\eta_\sigma^3}{\eta_\phi^3} \exp[6N(|\eta_\sigma|-|\eta_\phi|)] \,,
\nonumber
\ee
\be
\Rightarrow\;\; 8.139 \times 10^6 \ll |\eta_\sigma|^3 \exp[300 \ln(5.657 \times 10^{-2} r^{-1/2})\left(|\eta_\sigma| -
0.020\right)] \ll 8.210 \times 10^{12} \,.
\ee
As is evident from the plot, the WMAP (and also PLANCK) observationally allowed $2\sigma$ range of values for negative
$f_{NL}$, $-9 < f_{NL}$, is completely inside the intermediate $\phi_\star$ region as required.  More negative values for
$f_{NL}$, up to $f_{NL} = -20.647$ are consistent within our framework for the intermediate $\phi_\star$ region, but they are
ruled out from observation. Nevertheless, like for the low $\phi_\star$ region studied above, it is interesting to see a slow-
roll inflationary model with canonical kinetic terms where the observational restriction on $f_{NL}$ may be violated {\it by an
excess} and not by a shortfall. So we conclude that {\it if} $B_\zeta$ {\it is dominated by the one-loop correction but}
$P_\zeta$ {\it is dominated by the tree-level term, sizeable non-gaussianity is generated even if} $\zeta$ {\it is generated
during inflation}.  We also conclude, from looking at the small values that the tensor to scalar ratio $r$ takes in figure
\ref{fig32} compared with the present technological bound $r \gsim 10^{-3}$ \cite{friedman}, that {\it for non-gaussianity to 
be observable in this model, primordial gravitational waves must be undetectable}.

Notice that in oder to get positive values for $f_{NL}$, which is observationally more interesting in view of the results
presented in Subsection \ref{observational}, $\eta_\sigma$ should be positive according to Eq. (\ref{poseta}).  However, being
$\eta_\phi$ negative in order to reproduce the observed spectral tilt, the argument of the exponential in Eq. (\ref{poseta})
would be negative, making the $f_{NL}$ obtained too small to be observationally interesting\footnote{We thank Eiichiro 
Komatsu for questioning us about this issue.}.  As regards the general case, in
view of the previous reason being model dependent, we may only say that in order to get $f_{NL}$ positive when $B_\zeta$ is
dominated by the one-loop corrections, $B_\zeta$ should be positive (based on the definition of $f_{NL}$ in Eq. (\ref{bfp}))
which means that the maximum between $N_{\sigma \sigma}$ and $N_{\phi \phi}$ should be positive in view of Eq. (\ref{poskomb}).

Finally we want to point out that, by reducing our model to the single-field case, the consistency relation between $f_{NL}$
and $n_\zeta$ presented in Ref. \cite{creminelli}: $f_{NL} \sim \mathcal{O} (n_\zeta -1)$ is not violated since in that case
$B_\zeta$ is never dominated by the one-loop corrections for slow-roll inflation as demonstrated in Ref. 
\cite{lythbox}\footnote{We thank Filippo Vernizzi for questioning us about this issue.}. Thus,
the level of non-gaussianity $f_{NL}$ for our model reduced to the single-field case is described by the high $\phi_\star$
region as shown below.
%%%%%%%%%%%%%%%%%%%%%%%%%%%%%%%%%%%%%%%%%%%%%%%%%%%%
\subsection{The high $\phi_\star$ region}		%%%%
%%%%%%%%%%%%%%%%%%%%%%%%%%%%%%%%%%%%%%%%%%%%%%%%%%%%
This case is of no observational interest because, according to the expressions in Eqs. (\ref{asidef}), (\ref{bfp}),
(\ref{pt}), (\ref{bt}), and (\ref{tiltt}), the non-gaussianity generated is too small to be observable:
\begin{equation}
\frac{6}{5} f_{NL} = \frac{B_\zeta^{tree}}{4\pi^4 \frac{\sum_i k_i^3}{\prod_i k_i^3} (\mathcal{P}_\zeta^{tree})^2} = -\eta_\phi
= 0.020 \,\label{fnlslowroll},
\end{equation}
in agreement with the consistency relation of Ref. \cite{creminelli} for our model reduced to the single-field case, and with
the general expectations of Refs. \cite{battefeld,maldacena,seery7,vernizzi,yokoyama1} for slow-roll inflationary models with
canonical kinetic terms where only the tree-level contributions are considered.
%%%%%%%%%%%%%%%%%%%%%%%%%%%%%%%%%%%%%%%%%%%%%%%%%%%%%%%%%%%%%%%%%%%%%%%%%%%%%%%%%%%%%%%%%%%%%%
\section{Convergence of the $\zeta$ series and perturbative regime} \label{seccou}       %%%%%
%%%%%%%%%%%%%%%%%%%%%%%%%%%%%%%%%%%%%%%%%%%%%%%%%%%%%%%%%%%%%%%%%%%%%%%%%%%%%%%%%%%%%%%%%%%%%%
In Sections \ref{rest} and \ref{fnl} we have worked up to the one-loop diagrams in order to constrain the
parameter space and find the level of non-gaussianity $f_{NL}$.
It is time then to address the issue of the $\zeta$ series convergence and justify the existence of a perturbative regime so
that the truncation of the series up to the one-loop order, for the model we have considered, is valid. A way to do that is by
rederiving the $\zeta$ series in terms of $\delta \phi_\star$ and $\delta \sigma_\star$ by equating the unperturbed scalar
potential to the perturbed one at the final time $t$;  this of course is valid in view of the first slow-roll condition in Eq.
(\ref{1stsrc}) and the final slice being one of uniform energy density:
\begin{eqnarray}
&&V_0 \left \{1 + \frac{1}{2} \eta_\phi \frac{\phi_\star^2}{m_P^2} \exp[-2N\eta_\phi] + \frac{1}{2} \eta_\sigma
\frac{\sigma_\star^2}{m_P^2} \exp[-2N\eta_\sigma] \right \} \nonumber \\
&&= V_0 \left \{1 + \frac{1}{2} \eta_\phi \frac{(\phi_\star + \delta \phi_\star)^2}{m_P^2} \exp[-2(N+\delta N)\eta_\phi] +
\frac{1}{2} \eta_\sigma \frac{(\sigma_\star + \delta \sigma_\star)^2}{m_P^2} \exp[-2(N + \delta N)\eta_\sigma] \right \} \,.
\nonumber \\
&&
\end{eqnarray}
From the previous expression it follows that
\begin{eqnarray}
&&\eta_\phi \phi_\star^2 \exp[-2N\eta_\phi] + \eta_\sigma \sigma_\star^2 \exp[-2N\eta_\sigma] \nonumber \\
&&= \eta_\phi (\phi_\star + \delta \phi_\star)^2 \exp[-2(N+\delta N)\eta_\phi] + \eta_\sigma (\sigma_\star + \delta
\sigma_\star)^2 \exp[-2(N + \delta N)\eta_\sigma] \,,  \label{1con}
\end{eqnarray}
which is easier to handle in terms of variables $x$ and $y$ defined as
\begin{eqnarray}
x &\equiv& \frac{\delta \phi_\star}{\phi_\star} \,, \label{xvdef} \\
y &\equiv& \left[\frac{\eta_\sigma^3}{\eta_\phi^3} \frac{\sigma_\star^2}{\phi_\star^2} \left(1 + \frac{\delta \sigma_\star}
{\sigma_\star}\right)^2 \exp[2N(\eta_\phi-\eta_\sigma)]\right]^{1/2} \,. \label{yvdef}
\end{eqnarray}
Thus, the exponentials factors contaning $N$ (but not $\delta N$) are completely absorbed in $y$, and the expression in Eq.
(\ref{1con}) looks as follows:
\begin{equation}
1 + \frac{\eta_\phi^2}{\eta_\sigma^2} \frac{1}{\left(1 + \frac{\delta \sigma_\star}{\sigma_\star}\right)^2} y^2 = (1 + x)^2
\exp [-2 \delta N \eta_\phi] + \frac{\eta_\phi^2}{\eta_\sigma^2} y^2 \exp[-2 \delta N \eta_\sigma] \,. \label{xyme}
\end{equation}
If we were able to solve for $\delta N$ in Eq. (\ref{xyme}) in terms of $\eta_\phi$, $\eta_\sigma$, $x$, and $y$ (after making
$\sigma_\star = 0$), we could Taylor-expand around $x=0$ and $y=0$ reproducing the ${\bf x}$-dependent part of Eq.
(\ref{Nexp}).  This would be really good because the Taylor expansion would look so clean, in the sense that all the concerning
exponential factors contaning $N$ which appear explicitely in Eq. (\ref{Nexp}) would already be absorbed in $y$, that the issue
of truncating at some specific order in $\delta \phi_\star$ and $\delta \sigma_\star$ would be simply justified by requiring $|
x| \ll 1$ and $|y| \ll 1$.
Nevertheless, as is seen in Eq. (\ref{xyme}), it is impossible to solve for $\delta N$ in terms of $\eta_\phi$, $\eta_\sigma$,
$x$, and $y$ unless we make a Taylor expansion of the exponential functions aroud $\delta N = 0$: %By making such an expansion
%we obtain %from Eq. (\ref{xyme})
\begin{eqnarray}
0 &=&  \left \{\left[(1 + x)^2 - 1\right] + \frac{\eta_\phi^2}{\eta_\sigma^2} y^2 \left[1 - \frac{1}{\left(1 + \frac{\delta
\sigma_\star}{\sigma_\star} \right)^2} \right] \right \}+ \nonumber \\
&&+\delta N  \left[-2\eta_\phi (1 + x)^2 - 2\frac{\eta_\phi^2}{\eta_\sigma} y^2 \right] + \delta N^2  \left[2\eta_\phi^2 (1 +
x)^2 + 2\eta_\phi^2 y^2 \right] + ... \,. \label{2con}
\end{eqnarray}
Notice that the Taylor expansion of the exponential functions is always convergent whatever the arguments of the exponentials
are.  Moreover, if the Taylor expansion derived from a function $f(x)$ converges, it converges precisely to $f(x)$
\cite{spivak}.  Thus, the expression in Eq. (\ref{2con}) is actually the same as the expression in Eq. (\ref{xyme}).

Now, solving for $\delta N$ in terms of $\eta_\phi$, $\eta_\sigma$, $x$, and $y$, although possible in view of the expression
in Eq. (\ref{2con}), is not an easy business. That is why we will truncate the series in Eq. (\ref{2con}) up to second order in
$\delta N$ and solve the resultant quadratic equation\footnote{The truncation up to second order in $\delta N$ has been chosen
in order to have complete consistency with the order of the variables $x$ and $y$ in Eq. (\ref{xyme}).}. Notice that, since
$\zeta \equiv \delta N - \langle \delta N \rangle$ and $\zeta \sim 10^{-5}$, we may truncate the series in Eq. (\ref{2con}) up
to whatever order we wish and still reproduce $\zeta$ to high accuracy. Thus, the solution for the quadratic equation coming 
from the series in Eq. (\ref{2con}) after truncation at second order is:
\begin{eqnarray}
\delta N &\approx& \Big \{ \frac{1}{2} \left(1 + x \right)^2 + \frac{1}{2} \frac{\eta_\phi}{\eta_\sigma} y^2 \pm \Big \{
\left[\frac{1}{2} \left(1 + x \right)^2 + \frac{1}{2} \frac{\eta_\phi}{\eta_\sigma} y^2 \right]^2 - \nonumber \\
&& - \frac{1}{2} \left \{ \left[\left(1 + x \right)^2 - 1\right] + \frac{\eta_\phi^2}{\eta_\sigma^2} y^2 \left[1 - \frac{1}
{\left(1 + \frac{\delta \sigma_\star}{\sigma_\star}\right)^2} \right] \right \} \left[\left(1 + x \right)^2 + y^2  \right] \Big
\}^{1/2} \Big \} \times \nonumber \\
&& \times \left \{\eta_\phi \left[ \left(1 + x \right)^2 + y^2 \right] \right \}^{-1} \,. \label{dnxy}
\end{eqnarray}
If in addition we make Taylor expansions of the square root and the factor in the third line of the previous expression around
$x=0$ and $y=0$:
\begin{eqnarray}
&&\left \{ \left[\frac{1}{2} \left(1 + x \right)^2 + \frac{1}{2} \frac{\eta_\phi}{\eta_\sigma} y^2 \right]^2 - \frac{1}{2}
\left \{ \left[\left(1 + x \right)^2 - 1\right] + \frac{\eta_\phi^2}{\eta_\sigma^2} y^2 \left[1 - \frac{1}{\left(1 +
\frac{\delta \sigma_\star}{\sigma_\star}\right)^2} \right] \right \} \left[\left(1 + x \right)^2 + y^2  \right] \right \}^{1/2}
\nonumber \\
&&= \frac{1}{2} - x^2 + \frac{\eta_\phi}{2\eta_\sigma} \left \{ 1 - \frac{\eta_\phi}{\eta_\sigma} \left[1 - \frac{1}{\left(1 +
\frac{\delta \sigma_\star}{\sigma_\star}\right)^2} \right] \right \} y^2 + ... \,, \label{sqrtexp} \\
&&\left \{\eta_\phi \left[ \left(1 + x \right)^2 + y^2 \right] \right \}^{-1} = \frac{1}{\eta_\phi} \left[1 - 2x + 3x^2 -y^2 +
... \right] \,, \label{-1exp}
\end{eqnarray}
introducing them into Eq. (\ref{dnxy}), we end up with the following power series for $\delta N$:
\begin{equation}
\delta N \approx \frac{1}{\eta_\phi} \left( x - \frac{x^2}{2} + \frac{\eta_\phi^2}{2\eta_\sigma^2} y^2 + ... \right) \,,
\label{dnexxy}
\end{equation}
where the $\pm$ symbol is changed to the $-$ sign so that $\delta N$ remains a perturbation, and the trajectory $\sigma = 0$ is
chosen. Coming back to the variables $\delta \phi_\star$ and $\delta \sigma_\star$ we see that Eq. (\ref{dnexxy}) reproduces
the ${\bf x}$-dependent part of Eq. (\ref{Nexp}) in view of Eqs. (\ref{1d}) and (\ref{2d}) up to second order in $\delta
\phi_\star$ and $\delta \sigma_\star$:
\begin{equation}
\delta N \approx \frac{\delta \phi_\star}{\eta_\phi \phi_\star} - \frac{1}{2\eta_\phi} \left(\frac{\delta \phi_\star}
{\phi_\star}\right)^2 + \frac{\eta_\sigma}{2\eta_\phi^2} \left(\frac{\delta \sigma_\star}{\phi_\star}\right)^2 \exp \left[2N
\left(\eta_\phi - \eta_\sigma \right) \right] + ... \,.
\end{equation}
Eq. (\ref{dnexxy}), although reliable only up to second order, tells us that the expected behaviour of $\delta N$ in terms of
$\eta_\phi$, $\eta_\sigma$, $x$, and $y$ is indeed obtained.  Moreover, from our previous discussion we know that $\delta N$
can be exactly written in terms of a series of $x$ and $y$ withouth the explicit appearance of the concerning exponential
factors containing $N$. This is indeed partially confirmed up to third order when introducing Eqs. (\ref{1d}), (\ref{2d}), and
(\ref{3d}) into the ${\bf x}$-dependent part of Eq. (\ref{Nexp}):
\begin{equation}
\delta N =\frac{1}{\eta_\phi} \left( x - \frac{x^2}{2} + \frac{\eta_\phi^2}{2\eta_\sigma^2} y^2 + \frac{x^3}{3} -
\frac{\eta_\phi}{3\eta_\sigma} xy^2 + ... \right) \,.  \label{dntrun}
\end{equation}
The bottom line of this discussion is that we have been able to identify two quantities that determine the truncation of the
series up to some specific order.  These two quantities are $x$ and $y$ which we could identify as the ``coupling constants''
of the theory in the context of Quantum Field Theory.  By making $|x| \ll 1$ and $|y| \ll 1$ we can see from Eq. (\ref{dntrun})
that all the terms higher than second order in $x$ and $y$ are subleading compared to the second-order ones.  As regards the
first-order terms compared to the second-order ones, we see that the latter are not necessarily subleading compared to the
former because of the non-existence of the first-order $y$ term and in view of $|y/x| \lsim 1600$ from Eqs. (\ref{xvdef}) and
(\ref{yvdef}) and the values for $\eta_\phi$, $\eta_\sigma$ and $N$ considered in Sections \ref{rest},
\ref{fnl} and \ref{taonl}. In the language of the Feynman-like diagrams \cite{byrnes1}, truncating $\delta N$ in Eq.
(\ref{dntrun}) up to second order in $x$ and $y$ means considering only the leading diagrams at tree level and one loop which
is what we have done in Sections \ref{rest}, \ref{fnl}. In fact, $|x| \ll 1$ and $|y| \ll 1$ mean that
\be
|x| \equiv \left|\frac{\delta \phi_\star}{\phi_\star}\right| \approx \left(\frac{H_\star}{2\pi} \right) \frac{1}{\phi_\star}
\ll 1 \,,
\label{converg1}
\ee
\be
|y|\equiv \left \{ \frac{\eta_\sigma^3}{\eta_\phi^3} \frac{\delta \sigma_\star^2}{\phi_\star^2} \exp [2N (\eta_\phi -
\eta_\sigma)] \right \}^{1/2} \approx \left \{ \frac{\eta_\sigma^3}{\eta_\phi^3} \left(\frac{H_\star}{2\pi}\right)^2 \frac{1}
{\phi_\star^2} \exp [2N (\eta_\phi - \eta_\sigma)] \right \}^{1/2} \ll 1 \,,\label{converg2}
\ee
which are well satisfied for the cases when $P_\zeta$ is dominated by the tree-level term (see Subsection \ref{zipt} - Eq.
(\ref{cx1}) and Subsection \ref{bz1l} - Eq. (\ref{wn})):
\be
\left(\frac{H_\star}{2\pi} \right) \frac{1}{\phi_\star} = |\eta_\phi| \mathcal{P}_\zeta^{1/2} \approx 10^{-6} \,, \\
\ee
\be
\left \{ \frac{\eta_\sigma^3}{\eta_\phi^3} \left(\frac{H_\star}{2\pi}\right)^2 \frac{1}{\phi_\star^2} \exp [2N (\eta_\phi -
\eta_\sigma)] \right \}^{1/2} \ll \left \{ \frac{\eta_\sigma}{\eta_\phi} \exp [-2N (|\eta_\sigma| - |\eta_\phi|)] \right
\}^{1/2}  \lsim 2 \,.
\ee
By explicitly calculating the two-loop and three-loop diagrams for $P_\zeta$ and $B_\zeta$, and employing the results of Ref.
\cite{jarnhus}, we have checked that the conditions $|x| \ll 1$ and $|y| \ll 1$ efectively make these diagrams subleading
compared to the leading ones at one-loop level.

Finally, we will discuss the convergence of the $\zeta$ series in view of Eqs. (\ref{dnxy}), (\ref{sqrtexp}), and
(\ref{-1exp}).  We first note that the series in Eq. (\ref{sqrtexp}) is always convergent.  As regards the series in Eq.
(\ref{-1exp}), it will not be convergent at all while the function
\begin{equation}
\frac{1}{(1+x)^2 + y^2} \approx \frac{1}{(1+x)^2 + B^2 x^2} \,,  \label{fB}
\end{equation}
with
\begin{equation}
B = \left(\frac{\eta_\sigma}{\eta_\phi}\right)^{3/2} \exp [N(\eta_\phi - \eta_\sigma)] \,,
\end{equation}
does not satisfy the following necessary condition \cite{spivak}:  for the Taylor series around $x=0$ of a function $f(x)$ to
be convergent, it is necessary that the extension $f(z)$ to the complex plane of $f(x)$ is continous in a neighbourhood of
$z=0$.  If this is the case, and the Taylor series of $f(z)$ is indeed convergent, the convergence circle must either match or
be inside the aforementioned neighbourhood. Of course, this is not a sufficient condition, but at least gives us a constraint
on the possible values that $x$ may take.

Applying this condition to the expression in Eq. (\ref{fB}), we see that the extension of this function to the complex plane
has poles for $(1+z)^2 = -B^2 z^2$ which leads to
\begin{equation}
z = \frac{\pm iB - 1}{B^2 + 1} \,.
\end{equation}
Therefore, the extension to the complex plane of Eq. (\ref{fB}) is continous for
\begin{equation}
|z| < \frac{B^2 + 1}{B^2 + 1} = 1\,,
\end{equation}
so the necessary condition for the convergence of the series in Eq. (\ref{-1exp}), and therefore for the convergence of the
series in Eq. (\ref{dnxy}) which is what we are interested in, is given by $|x| < 1$.  Thus, such a necessary condition for the
convergence of the $\zeta$ series is automatically satisfied once we choose $|x| \ll 1$, as we have seen above it is required
for working in a perturbative regime.
%%%%%%%%%%%%%%%%%%%%%%%%%%%%%%%%%%%%%%%%%%%%%%%%%%%%%%%%%%%%%%%
\section{Conclusions} \label{conclusca}	     %%%%%%%%%%%%%%%%%%
%%%%%%%%%%%%%%%%%%%%%%%%%%%%%%%%%%%%%%%%%%%%%%%%%%%%%%%%%%%%%%%
Is it reasonable to study the primordial curvature perturbation $\zeta$ by identifying it
with a truncated $\delta N$ series expansion? Is it actually possible to cut out with
confidence such a series at some specific order? Is it true that all the slow-roll
inflationary models with canonical kinetic terms produce primordial non-gaussianity
supressed by the slow-roll parameters? Are the loop corrections in cosmological perturbation
theory always smaller than the tree-level terms?  We have addressed these questions in this
chapter, answering all of them by paying particular attention to a special slow-roll inflationary model.
The $\zeta$ series expansion is indeed a powerful tool to study the statistical
descriptors of $\zeta$; nevertheless, we should seek for the convergence radius in order
not to obtain results that actually have nothing to do with $\zeta$. We may cut out the
series but, to be completely sure about the precision of our approximations, we have to
study the conditions for the existence of a perturbative regime. Non-gaussianity in
slow-roll inflationary models with canonical kinetic terms is not always suppressed by the
slow-roll parameters; we have seen this at tree-level for $f_{NL}$ in
Refs. \cite{alabidi3,byrnes3}, and considering loop corrections for $f_{NL}$ in
the prersent chapter and in Refs. \cite{cogollo,bch2}. Particulary, we shown in this chapter that it is possible to attain very 
high, including observable, values for the level of non-gaussianity $f_{NL}$ associated with the bispectrum $B_\zeta$ of the 
primordial curvature perturbation $\zeta$, in a subclass of small-field slow-roll models of inflation with canonical 
kinetic terms. Such a result was obtained by taking care of loop corrections both in the 
spectrum $P_\zeta$ and the bispectrum $B_\zeta$ and assuming that the latter can dominate over the former; of course, this 
possibility is model dependent. More precisely, we can say that if $B_\zeta$ is dominated by 
the one-loop correction but $P_\zeta$ is dominated by the tree-level term, sizeable non-gaussianity is generated even if 
$\zeta$ is generated during inflation. What is interesting is that this kind of particular models are populary known to predict 
too small values for the level of non-gaussianity $\fnl$, as small as the  slow-roll parameters.  Finally, as far as we have 
investigated, the loop corrections in cosmological perturbation theory are not always smaller than the tree-level terms; in 
fact, when they become the leading contributions, a surprising phenomenology appears in front of our eyes.
%%%%%%%%%%%%%%%%%%%%%%%%%%%%%%%%%%%%%%%%%%%%%%%%%%%%%%%%%%%%%%%%%%%%%%%%%%%%%%%%%%%%%%%%%%%%%%%%%%%%%%%%%%%%
%%%%%%%%%%%%%%%%%%%%%%%%%%%%%%%%%%%%%%%%%%%%%%%%%%%%%%%%%%%%%%%%%%%%%%%%%%%%%%%%%%%%%%%%%%%%%%%%%%%%%%%%%%%%
\chapter[PRIMORDIAL NON- GAUSSIANITY IN SLOW-ROLL INFLATION: THE TRISPECTRUM]					    %%%%%%%%
{ON THE ISSUE OF THE $\zeta$ SERIES CONVERGENCE AND LOOP CORRECTIONS IN THE GENERATION 		        %%%%%%%%
OF OBSERVABLE PRIMORDIAL NON- GAUSSIANITY IN SLOW-ROLL INFLATION: THE TRISPECTRUM}\label{chaptsca2}  %%%%%%%%
%%%%%%%%%%%%%%%%%%%%%%%%%%%%%%%%%%%%%%%%%%%%%%%%%%%%%%%%%%%%%%%%%%%%%%%%%%%%%%%%%%%%%%%%%%%%%%%%%%%%%%%%%%%%
%%%%%%%%%%%%%%%%%%%%%%%%%%%%%%%%%%%%%%%%%%%%%%%%%%%%%%%%%%%%%%%%%%%%%%%%%%%%%%%%%%%%%%%%%%%%%%%%%%%%%%%%%%%%
%%%%%%%%%%%%%%%%%%%%%%%%%%%%%%%%%%
\section{Introduction}     %%%%%%%
%%%%%%%%%%%%%%%%%%%%%%%%%%%%%%%%%%
The primordial curvature perturbation $\zeta$ \cite{dodelson,lythbook,mukhanov,weinberg3}, and its $\delta N$ 
expansion\footnote{By ``$\delta N$ expansion'' we mean approximating $\delta N$ by a power series expansion in the initial 
conditions. By ``$\delta N$ formula'' we mean the statement that to lowest order in spatial gradients $\zeta \equiv \delta N$. 
These conventions have been and will be used throughout the text.} \cite{dklr,lms,lr,ss,st,starobinsky}, was the subject of 
study in a previous chapter (see also \cite{cogollo}). We were interested in how well the convergence of the $\zeta$ series was 
understood, and if the traditional arguments to cut out the $\zeta$ series at second order \cite{lr,zaballa}, keeping only 
the tree-level terms to study the statistical descriptors of $\zeta$ 
\cite{alabidi1,battefeld,byrnes3,bsw1,seery3,vernizzi,yokoyama1,yokoyama2,yokoyama3}, were reliable\footnote{We follow the 
terminology of Ref. \cite{byrnes1} to identify the tree-level terms and the loop contributions in a diagrammatic 
approach. The associated diagrams are called {\it Feynman-like diagrams}.}. We argued that a previous study of the $\zeta$ 
series convergence, the viability of a perturbative regime, and the relative weight of the loop contributions against the tree-
level terms, were completely necessary and in some cases surprising. For instance, the levels of non-gaussianity $f_{NL}$ and 
$\tau_{NL}$ in the bispectrum $B_\zeta$ and trispectrum $T_\zeta$ 
of $\zeta$ respectively, for slow-roll inflationary models with canonical kinetic terms \cite{lyth6,lythbook,lyth5}, are 
usually 
thought to be of order $\mathcal{O} (\epsilon_i,\eta_i)$ \cite{battefeld,vernizzi,yokoyama1}\footnote{See however Refs. 
\cite{alabidi1,byrnes3}.} and $\mathcal{O} (r)$ \cite{seery3,ssv}\footnote{See however Refs. \cite{slri,bch2}.} respectively, 
were $\epsilon_i$ and $\eta_i$ are the slow-roll parameters with $\epsilon_i,|\eta_i| \ll 1$ \cite{lyth5} 
and $r$ is the tensor to scalar ratio \cite{lyth6} with $r < 0.22$ at the $95 \%$ confidence level \cite{wmap5}. However, in 
order to reach such a conclusion, 
generic models were used where the loop contributions are comparatively suppressed and, therefore, the truncated $\delta N$ 
expansion may be used. Of course exceptions may occur, and in those cases it is crucial to check up to what order the truncated 
$\delta N$ expansion may be used, and which loop contributions are larger than the tree-level terms.  In any of these cases, 
general models or exceptions, the question regarding the representation of $\zeta$ by the $\delta N$ expansion is a matter to 
discuss.
 
Refs. \cite{alabidi1,byrnes3} show that large, {\it and observable}, non-gaussianity in $B_\zeta$ is indeed possible for 
certain classes of {\it slow-roll} models with {\it canonical} kinetic terms and special trajectories in field space, relying 
only on the tree-level terms.  Ref. \cite{bch2} does the same for $B_\zeta$ and $T_\zeta$ but this time arguing that the loop 
corrections are always suppressed against the tree-level terms if the quantum fluctuations of the scalar fields do not 
overwhelm the classical evolution. Nonetheless, although the resultant phenomenology from papers in Refs. 
\cite{alabidi1,byrnes3,bch2} is very interesting, the classicality argument used in Ref. \cite{bch2} 
is very conservatively stated leading to too strong conclusions as we will argue later in this chapter.
More research remains to be done to understand the role of the quantum diffusion and, being this beyond the scope of the 
present chapter, we will leave the discussion for a future research project. 
We addressed the $\zeta$ series convergence and the existence of a perturbative regime in the previous chapter, showing how important the requirements to guarantee those conditions are. Moreover, we showed that for a 
subclass of small-field {\it slow-roll} inflationary models with {\it canonical} kinetic terms, the one-loop correction to 
$B_\zeta$ might be much larger than the tree-level terms, giving as a result large, {\it and observable}, non-gaussianity 
parameterised by $f_{NL}$.  The present chapter extends the analysis presented in the previous one to $T_\zeta$ showing, for 
the first time, that {\it large and observable} non-gaussianity parameterised by $\tau_{NL}$ is possible in {\it slow-roll} 
inflationary models with {\it canonical} kinetic terms due to loop effects, in total contrast with the usual belief based on 
the results of Refs. \cite{seery3,ssv}. In order to properly identify the non-gaussianity levels found in previous chapter 
and in the present one with those that are constrained by observation, we comment on the probability that an observer in an 
ensemble of realizations of the density field in our scenario sees a non-gaussian distribution. As we will show such a 
probability is non-negligible for the concave downward potential, making indeed the observation of the non-gaussianity studied 
in this chapter quite possible. 

The layout of the chapter is the following: in Section \ref{model2} we make some aditional comments about ot the slow-roll 
inflationary model that exhibits large levels of non-gaussianity when loop corrections are considered. This model was 
described in more detail in Section \ref{model}. 
In Section \ref{class} we study the impact of the quantum fluctions of the 
scalar fields on their classical evolution.  As a result we argue how the loop suppression proof given in Ref. \cite{bch2} does
not apply to our model. Section \ref{prob} studies the probability of realizing the scenario proposed in this thesis for a 
typical observer.  Section \ref{constraints} is devoted to the reduction of the available parameter window by taking into 
account some restrictions of methodological and physical nature. The level of non-gaussianity $\tau_{NL}$ in the trispectrum 
$T_\zeta$ is calculated in Section \ref{taonl} for models where $\zeta$ is, or is not, generated during inflation; a 
comparison with the current literature and the results found in the previous chapter for $f_{NL}$ is done.  In 
Section \ref{fnlafter} the level of non-gaussianity $f_{NL}$ in the bispectrum $B_\zeta$ is calculated for models where $\zeta$ 
is not generated during inflation.  Finally, Section \ref{conclusca2} presents the conclusions.
%%%%%%%%%%%%%%%%%%%%%%%%%%%%%%%%%%%%%%%%%%%%%%%%%%%%%%%%%%%%%%%%%%%%%%%%%%%%%%%%%%%%%
\section{A quadratic two-field slow-roll model of inflation} \label{model2}	%%%%%%%%%
%%%%%%%%%%%%%%%%%%%%%%%%%%%%%%%%%%%%%%%%%%%%%%%%%%%%%%%%%%%%%%%%%%%%%%%%%%%%%%%%%%%%
We will give in this Section some relevants remarks about the inflationary potential studied in Section \ref{model} and given by
\begin{equation}
V = V_0\left(1+\frac{1}{2}\eta_\phi \frac{\phi^2}{m_P^2} +\frac{1}{2}\eta_\sigma \frac{\sigma^2}{m_P^2}\right) \,, \label{pot1}
\end{equation}
where $\phi$ and $\sigma$ are the inflaton fields and $m_P$ is the reduced Planck mass.
By assuming that the first term in Eq. (\ref{pot1}) dominates, $\eta_\phi$ and $\eta_\sigma$ become the usual $\eta$ slow-roll 
parameters associated with the fields $\phi$ and $\sigma$.

We have chosen the $\sigma=0$ trajectory since this case is the easiest to work from the point of view of the analytical 
calculations and because it gives the most 
interesting results. In addition, such a trajectory for the potential in Eq. (\ref{pot1}) reproduces for some number of e-folds 
(for $\eta_\phi,\eta_\sigma > 0$) the hybrid inflation scenario \cite{linde2} where $\phi$ is the inflaton and $\sigma$ is the 
waterfall field.  We will analyze in Section \ref{prob} the probability for an observer to live in a region where $\sigma=0$ 
for the concave downward potential. Non-gaussianity in the bispectrum $B_\zeta$ of $\zeta$ for this kind of model has been 
studied in Refs. \cite{alabidi1,byrnes3,bch2,cogollo,enqvist1,lr,lr1,vaihkonen,zaballa}; in particular, Ref. 
\cite{cogollo} shows that the one-loop correction dominates over the tree-level terms if $\eta_\phi,\eta_\sigma < 0$ and $|
\eta_\sigma| > |\eta_\phi|$, generating in this way large values for $f_{NL}$ even if $\zeta$ {\it is} generated during 
inflation. Refs. \cite{alabidi1,byrnes3}, in contrast, work only at tree-level with the same potential as Eq. (\ref{pot1}) but 
relaxing the $\sigma = 0$ condition, finding that large values for $f_{NL}$ are possible for a small set of initial conditions. 
Ref. \cite{bch2} improves the analysis in Ref. \cite{byrnes3}, this time taking into account also the trispectrum $T_\zeta$ of 
$\zeta$ and the role of the loop corrections.  According to that reference, large values for $\tau_{NL}$ are also possible for 
a small set of initial conditions if the tree-level terms dominate over the loop corrections.  Moreover, it is claimed that 
loop corrections for this model are always suppressed against the tree-level terms if the quantum fluctuations of the fields 
are subdominant against their classical evolution.  The opposite case seems to happen for some narrow range of initial 
conditions including $\sigma_\star = 0$, which is the case studied in this chapter. 
As we will argue in Section \ref{class}, the classicality condition in Ref. \cite{bch2} is expressed in a very conservative 
way, leading to too strong and non-general conclusions.  Dominance of loop corrections is then safe from the classical vs 
quantum condition, allowing the interesting large levels of non-gaussianity discussed in Chapter \ref{chaptsca} and in the 
present one.

Following the results in Appendix \ref{app},  we will write down the leading terms to the spectrum, bispectrum, 
and trispectrum of the primordial curvature perturbation $\zeta$ including the tree-level and one-loop contributions given in 
Eqs. (\ref{pt}), (\ref{pl}), (\ref{bt}), (\ref{bl}), (\ref{ttap}) and (\ref{tlap}):
\bea
{\mathcal P}_\zeta^{tree}&=&\frac{1}{\eta_\phi^2\phi_\star^2}\left(\frac{H_\star}{2 \pi}\right)^2 \,,\label{pt2}\\
{\mathcal P}_\zeta^{ 1-loop}&=&\frac{\eta_\sigma^2}{\eta_\phi^4\phi_\star^4}\exp[4N(\eta_\phi-\eta_\sigma)]
\left(\frac{H_\star}{2 \pi}\right)^4\ln(kL) \,,\label{pl2}\\
B_\zeta^{tree} &=& -\frac{1}{\eta_\phi^3 \phi_\star^4} \left(\frac{H_\star}{2\pi}\right)^4 4\pi^4 \left(\frac{\sum_i k_i^3}
{\prod_i k_i^3}\right) \,, \label{bt2} \\
B_\zeta^{1-loop} &=& \frac{\eta_\sigma^3}{\eta_\phi^6 \phi_\star^6} \exp[6N(\eta_\phi - \eta_\sigma)] \left(\frac{H_\star}
{2\pi}\right)^6 \ln(kL) 4\pi^4 \left(\frac{\sum_i k_i^3}{\prod_i k_i^3}\right) \,, \label{bl2}\\
T_\zeta^{tree}&=& \frac{1}{\eta_\phi^4\phi_\star^6}\left(\frac{H_\star}{2 \pi}\right)^6\left[\frac{2\pi^2}{k_2^3}\frac{2\pi^2}
{k_4^3}\frac{2\pi^2}{|{\bf k}_3+{\bf k}_4|^3} + 11 \ {\rm permutations}\right] \,,\label{tt2}\\
T_\zeta^{ 1-loop}&=&\frac{\eta_\sigma^4}{\eta_\phi^8\phi_\star^8}\exp[8N(\eta_\phi-\eta_\sigma)] \left(\frac{H_\star}{2
\pi}\right)^8\ln(kL) \ 4 \Big[ \frac{2\pi^2}{k_2^3}\frac{2\pi^2}{k_4^3}\frac{2\pi^2}{|{\bf k}_3+{\bf k}_4|^3} + \nonumber \\
&&+11 \ {\rm permutations} \Big] \,.\label{tl2}
\eea
where $L$ is the infrared cutoff chosen so that the quantities are calculated in a minimal
box \cite{bernardeu4,lythbox}. Except when considering low CMB multipoles, the box size should
be set at $L \sim H_0$ \cite{klv,leblond},
giving $\ln(kL) \sim \mathcal{O} (1)$ for relevant cosmological scales.

The important factor in the loop corrections is the exponential.  This exponential function is directly related to the 
quadratic form of the potential with a leading constant term.  It will give a large contribution if $\eta_\phi > \eta_\sigma$.  
In Chapter \ref{chaptsca}, we chose the concave downward potential in order to satisfy the spectral tilt constraint, which 
makes $\eta_\phi < 0$, while keeping $|\eta_\sigma| > |\eta_\phi|$.  In this chapter we will consider the same case.
%%%%%%%%%%%%%%%%%%%%%%%%%%%%%%%%%%%%%%%%%%%%%
\section{Classicality} \label{class}	%%%%%
%%%%%%%%%%%%%%%%%%%%%%%%%%%%%%%%%%%%%%%%%%%%%
Ref. \cite{bch2} argues in Appendices A and B how, by imposing the requirement that the quantum fluctuations of the fields 
around their background values do not overwhelm the respective classical evolutions, the loop corrections to $P_\zeta$, 
$B_\zeta$, and $T_\zeta$ are suppressed against the tree-level terms. The proof relies on the fact that, if the classicality 
condition is satisfied, the second-order terms in the $\delta N$ expansion in \eq{Nexp},
are subleading against the first-order terms. This in turn implies
\begin{eqnarray}
\frac{P_\zeta^{1-loop}}{P_\zeta^{tree}} \ll 1 \,, \\
\frac{B_\zeta^{1-loop}}{B_\zeta^{tree}} \ll 1 \,, \\ 
\frac{T_\zeta^{1-loop}}{T_\zeta^{tree}} \ll 1 \,,
\end{eqnarray}
as explicitly stated in Eqs. (A.16-A.19) of Ref. \cite{bch2}.  In addition, under the same assumptions, higher order 
corrections in the spectral functions $P_\zeta$, $B_\zeta$, and $T_\zeta$ are always subleading against the one-loop 
corrections and, therefore, subleading against the tree-level terms. This conclusion is obtained if the $\delta N$ expansion 
may be truncated at fourth order. However, what is the classicality condition employed in Ref. \cite{bch2}?

Assuming slow-roll evolution for each field $\phi_i$, which is valid only if the quantum fluctuation $\delta \phi_i$ is by far 
smaller than the classical evolution $\Delta \phi_i$, the classical change in the $\phi_i$ field during a Hubble time around 
horizon exit is
\begin{equation}
\Delta \phi_i (t_\star) \approx -\frac{V_i (\phi)}{3H_\ast^2 \sqrt{6}} \,, 
\end{equation}
where $V_i$ denotes the derivative of the potential with respect to the $i$-th field. Comparing the latter expression with the 
quantum fluctuation
\begin{equation}
\delta \phi_i (t_\star) \approx \frac{H_\ast}{2\pi} \,, 
\end{equation}
and requiring that $\Delta \phi_i$ is much larger than $\delta \phi_i$, we get
\begin{equation}
|\dot{\phi_i}|_\star \gg \sqrt{\frac{3}{2\pi^2}} H_\ast^2 \,. \label{classcondhy}
\end{equation}
For our quadratic two-field slow-roll model of inflation, where the slow-roll evolution is given by Eqs. 
(\ref{srp})-(\ref{srs}), the previous expression translates into
\begin{equation}
|\phi_i|_\star \gg \sqrt{\frac{3}{2\pi^2}} \left|\frac{H_\ast}{\eta_i}\right| \,, 
\end{equation}
which is the one given in Eq. (A.1) of Ref. \cite{bch2}. Such a condition is equivalent to
\begin{equation}
\left|\frac{\delta \phi_i}{\phi_i}\right|_\star \ll \left|\frac{\eta_i}{\sqrt{6}}\right| \,, \label{classcond} 
\end{equation}
which is the one given in Eq. (A.2) of Ref. \cite{bch2}. Thus, under this condition, the trajectory $\sigma = 0$ studied in 
this thesis seems not to be well described by the slow-roll approximation and, therefore, the 
obtained results based in the $\delta N$ formalism would not be reliable.

This classicality argument given in Ref. \cite{bch2} is too conservatively stated. To see why it is like that,
we may reason in the following way for general inflationary models: 
for any point along the background classical trajectory in field space it is possible to rotate the
field axes so that, instantaneously, there is an inflaton (or `adiabatic') field that points
along the trajectory and some light `entropy' fields which point in orthogonal directions
\cite{gordon}. The quantum fluctuations for the entropy fields are non-vanishing but the
classical evolution for each of these fields is zero.  Since the condition in
Eq. (\ref{classcondhy}) is not formulated in any particular field parameterisation, we may
argue that for any multi-field inflationary model the application of this condition would
lead to a background inflationary trajectory dominated by the quantum evolution. Thus,
slow-roll conditions would always be impossible to apply. 
A more general classicality condition should still be the one in Eq. (\ref{classcondhy}) but
only applied to the adiabatic field and not to the entropy fields. In that respect the
$\sigma = 0$ trajectory studied in this thesis is safe from
large quantum fluctuations since the classicality condition in Eq. (\ref{classcondhy})
applied only to the $\phi$ field is extremely well satisfied as long as $P_\zeta \ll 1$,
which is actually the case for single field slow-roll inflation \cite{thesis}. Indeed, a
much better way of stating the classicality condition is the
following:\footnote{We acknowledge Misao Sasaki for pointing out to us this idea.} if the
inflationary trajectory must be dominated by the classical motion of the fields, then the
perturbation in the amount of inflation, due to the quantum fluctuations of the fields,
must be negligible:
\begin{equation}
\delta N \ll 1 \,. 
\end{equation}
By virtue of the $\delta N$ formalism, this expression is simply satisfied if the free
parameters of the inflationary model under consideration are chosen so that the COBE
normalisation ($\mathcal{P}_\zeta^{1/2} \approx 5 \times 10^{-5}$ \cite{bunn}) is satisfied,
which is always the case. Nevertheless, we understand that the role of quantum diffusion is
of great importance (see for instance Refs. \cite{garcia-bellido,randall}), and a dedicated study of this issue is
left for a future research project.

%%%%%%%%%%%%%%%%%%%%%%%%%%%%%%%%%%%%%%%%%%%%%
\section{Probability} \label{prob}     %%%%%%
%%%%%%%%%%%%%%%%%%%%%%%%%%%%%%%%%%%%%%%%%%%%%

The main purpose of this chapter is to identify regions in the parameter space with high levels of primordial non-gaussianity. 
Then, we proceed to compare the obtained non-gaussianity with observation. In order to do the latter, we first need to realize 
what the probability is for a typical observer to live in a universe where the inflationary trajectory is the one studied in 
this thesis: $\sigma = 0$.  This is particularly relevant for the concave downward potential where the background trajectory 
$\sigma = 0$ is unstable.

In the context of quantum cosmology, the probability of quantum creation of a closed universe is proportional to 
\cite{lindeprob,vilenkin2,vilenkin1,vilenkin3}
\begin{equation}
P \sim \exp \left(-\frac{24\pi^2 m_P^4}{V}\right) \,, 
\end{equation}
which means that the universe can be created if $V$ is not too much smaller than the Planck density.  Thus, for our concave 
downward potential, having chosen the field contributions to $V$ in Eq. (\ref{pot}) to be negligible is good because it 
increases the probability. In addition, within a set of initial conditions for $\phi$, the most probable initial condition for 
$\sigma$ is $\sigma = 0$.  The $\phi = 0$ trajectory is also highly probable but, since we are assuming $|\eta_\sigma| > |
\eta_\phi|$, the $\sigma = 0$ trajectory is more probable. This of course implies that the levels of non-gaussianity obtained 
in this thesis may be observable.
%%%%%%%%%%%%%%%%%%%%%%%%%%%%%%%%%%%%%%%%%%%%%%%%%%%%%%%%%%%%%%%%%%%%%%%%%%%%
\section{Reducing the available parameter window} \label{constraints}	%%%%
%%%%%%%%%%%%%%%%%%%%%%%%%%%%%%%%%%%%%%%%%%%%%%%%%%%%%%%%%%%%%%%%%%%%%%%%%%%%
The analysis of the observed spectral index and the $\zeta$ series convergence was given in Subsection \ref{secnorm} and 
in Section \ref{seccou}, respectively.  Regarding the existence of a perturbative regime, we have to add to the discussion in 
Section \ref{seccou} that, by cutting out the series in \eq{dntrun} at second-order, just one 
Feynman-like diagram per spectral function of $\zeta$ is necessary to study the loop corrections to these spectral 
functions\footnote{This is assuming that the Feynman-like diagrams containing $n$-point correlators of the field perturbations 
with $n \geq 3$ are subdominat against the diagrams containing only two-point correlators.  For the trispectrum this does not 
happen when the tree-level terms dominate over the loop corrections \cite{seery3,ssv}. However, for the cases considered in 
this chapter, when the loop corrections in the trispectrum dominate over the tree-level terms generating in turn large values for 
$\tau_{NL}$, the diagrams containing $n$-point correlators of the field perturbations with $n \geq 3$ are expected to be 
subdominat because of their dependence on the slow-roll parameters \cite{jarnhus}.}. That is why in the previous chapter 
there was just one leading diagram for the one-loop correction to $P_\zeta$ (Fig. \ref{Ldis}a), as well as one leading 
diagram for the one-loop correction to $B_\zeta$ (Fig. \ref{olbd}a). When applied 
to $T_\zeta$, this analysis shows that the only diagrams to consider are the one in Fig. \ref{tf}a for the tree-level terms, 
and the one in Fig. \ref{tf}b for the loop corrections. Such diagrams lead to the expressions in Eqs. (\ref{ttap}) and 
(\ref{tlap}) for $T_\zeta^{tree}$ and $T_\zeta^{ 1-loop}$.

In the following, we will give the relevant information for $T_\zeta$ when $\zeta$ {\it is} generated during inflation, and for 
both $B_\zeta$ and $T_\zeta$ when $\zeta$ {\it is not} generated during inflation.

\subsection{Tree-level or one-loop dominance: $\tnl$}\label{domsectnl}

The exponencial factors in Eqs. (\ref{pl2}) and (\ref{tl2}) open up the possibility that the loop corrections dominate over
$\mathcal{P}_\zeta$ and/or $T_\zeta$. There are three posibilities:

\subsubsection{Both $T_\zeta$ and $\mathcal{P}_\zeta$ are dominated by the one-loop corrections}
Comparing Eqs. (\ref{pt2}) with (\ref{pl2}) and Eqs. (\ref{tt2}) with (\ref{tl2}) we requiere in this case that
\bea
\frac{\eta_\sigma^2}{\eta_\phi^2} \exp[4N(\eta_\phi-\eta_\sigma)] &\gg& \frac{1}{\frac{1}{\phi_\star^2} \left(\frac{H_\star}
{2\pi}\right)^2} \,, \\
4\frac{\eta_\sigma^4}{\eta_\phi^4} \exp[8N(\eta_\phi-\eta_\sigma)] &\gg& \frac{1}{\frac{1}{\phi_\star^2} \left(\frac{H_\star}
{2\pi}\right)^2} \,,
\eea
in which case only the first inequality is required. Employing the definition for the tensor to scalar ratio $r$ introduced in
\eq{defr},
we can write such inequality as
\be
\left(\frac{\phi_\star}{m_P}\right)^2 \ll \frac{r \mathcal{P}_\zeta}{8} \frac{\eta_\sigma^2}{\eta_\phi^2}
\exp[4N(\eta_\phi-\eta_\sigma)] \,.
\label{ptloop}
\ee
From now on we will name the parameter window described by Eq. (\ref{ptloop}) as the low $\phi_\star$ $T$-region\footnote{The
$T$ in $T$-region is introduced in order to differentiate explicitly between these regions and those found in the subsection 
\ref{domsecfnl} for $B_\zeta$.}, since the latter represents a region of allowed values for $\phi_\star$ limited by
an upper bound.

\subsubsection{$T_\zeta$ dominated by the one-loop correction and $\mathcal{P}_\zeta$ dominated by the tree-level term}
\label{t1lptresub}
Comparing Eqs. (\ref{pt2}) with (\ref{pl2}) and Eqs. (\ref{tt2}) with (\ref{tl2}) we requiere in this case that
\bea
\frac{\eta_\sigma^2}{\eta_\phi^2} \exp[4N(\eta_\phi-\eta_\sigma)] &\ll& \frac{1}{\frac{1}{\phi_\star^2} \left(\frac{H_\star}
{2\pi}\right)^2} \,, \label{t1lptre} \\
4\frac{\eta_\sigma^4}{\eta_\phi^4} \exp[8N(\eta_\phi-\eta_\sigma)] &\gg& \frac{1}{\frac{1}{\phi_\star^2} \left(\frac{H_\star}
{2\pi}\right)^2} \,,
\eea
which combine to give, employing the definition for the tensor to scalar ratio $r$ introduced in Eq. (\ref{defr}),
\be
\frac{r \mathcal{P}_\zeta}{8} \frac{\eta_\sigma^2}{\eta_\phi^2} \exp[4N(\eta_\phi-\eta_\sigma)] \ll \left(\frac{\phi_\star}
{m_P}\right)^2 \ll \frac{r \mathcal{P}_\zeta}{8} \frac{4\eta_\sigma^4}{\eta_\phi^4} \exp[8N(\eta_\phi-\eta_\sigma)] \,.
\label{intct}
\ee
From now on we will name the parameter window described by Eq. (\ref{intct}) as the intermediate $\phi_\star$ $T$-region, since
the latter represents a region of allowed values for $\phi_\star$ limited by both an upper and a lower bound.

\subsubsection{Both $T_\zeta$ and $\mathcal{P}_\zeta$ are dominated by the tree-level terms}

Comparing Eqs. (\ref{pt2}) with (\ref{pl2}) and Eqs. (\ref{tt2}) with (\ref{tl2}) we requiere in this case that
\bea
\frac{\eta_\sigma^2}{\eta_\phi^2} \exp[4N(\eta_\phi-\eta_\sigma)] &\ll& \frac{1}{\frac{1}{\phi_\star^2} \left(\frac{H_\star}
{2\pi}\right)^2} \,, \\
4\frac{\eta_\sigma^4}{\eta_\phi^4} \exp[8N(\eta_\phi-\eta_\sigma)] &\ll& \frac{1}{\frac{1}{\phi_\star^2} \left(\frac{H_\star}
{2\pi}\right)^2} \,,
\eea
in which case only the second inequality is required.  Employing the definition for the tensor to scalar ratio $r$ introduced
in Eq. (\ref{defr}), we can write such an inequality as
\bea
\left(\frac{\phi_\star}{m_P}\right)^2 \gg \frac{r \mathcal{P}_\zeta}{8} \frac{4\eta_\sigma^4}{\eta_\phi^4}
\exp[8N(\eta_\phi-\eta_\sigma)] \,.
\label{highphit}
\eea
From now on we will name the parameter window described by Eq. (\ref{highphit}) as the high $\phi_\star$ $T$-region, since the
latter represents a region of allowed values for $\phi_\star$ limited by a lower bound.

\subsection{The normalisation of the spectrum}

Either $\zeta$ is or is not generated during inflation, we must satisfy the appropriate spectrum normalisation condition. There 
exists four possibilities; however, it was shown in Subsection \ref{secnorm} that the case where $\zeta$ is generated during 
inflation and $\mathcal{P}_\zeta$ is dominated by the one-loop correction, is no of observational interest since it is 
impossible to reproduce the observed spectral index and its running. We also showed in this Subsection, that when $\zeta$ is 
generated during inflation and $\mathcal{P}_\zeta$ is dominated by the tree-level correction, the available 
parameter region is given by \eq{normt}. The other two possibilities are discussed right below.

\subsubsection{$\zeta$ not generated during inflation and $\mathcal{P}_\zeta$ dominated by the one-loop correction}
According to Eqs. (\ref{pl2}) and (\ref{defr}) we have in this case
\bea
\mathcal{P}_\zeta^{1-loop} &=& \frac{\eta_\sigma^2}{\eta_\phi^4 \phi_\star^4} \exp[4N(\eta_\phi - \eta_\sigma)] 
\left(\frac{H_\star}{2\pi}\right)^4 \ln(kL) \nonumber \\
&=& \frac{\eta_\sigma^2}{\eta_\phi^4} \exp[4N(\eta_\phi - \eta_\sigma)] \left(\frac{m_P}{\phi_\ast}\right)^4 \left(\frac{r 
\mathcal{P}_\zeta}{8}\right)^2 \ln(kL) \,,
\eea
which reduces to
\be
\left(\frac{\phi_\star}{m_P}\right)^4 \gg \left(\frac{r}{8}\right)^2 \mathcal{P}_\zeta \frac{\eta_\sigma^2}{\eta_\phi^4} 
\exp[4N(\eta_\phi - \eta_\sigma)] \ln(kL)\,, \label{norml}
\ee
where $\mathcal{P}_\zeta$ must be replaced by the observed value $\mathcal{P}_\zeta^{1/2} = (4.957 \pm 0.094) \times 10^{-5}$ 
\cite{bunn}.

\subsubsection{$\zeta$ not generated during inflation and $\mathcal{P}_\zeta$ dominated by the tree-level term}
The equation \ref{normt} tells us that in this case the constraint to satisfy is
\be\label{normtd}
\left(\frac{\phi_\star}{m_P}\right)^2 \gg \frac{1}{\eta_\phi^2} \frac{r}{8} \,.
\ee

%%%%%%%%%%%%%%%%%%%%%%%%%%%%%%%%%%%%%%%%%%%%%%%%%%%%%%%%%%%%%%
\section{Non-Gaussianity: $\tau_{NL}$} \label{taonl}	%%%%%%
%%%%%%%%%%%%%%%%%%%%%%%%%%%%%%%%%%%%%%%%%%%%%%%%%%%%%%%%%%%%%%
In this section we will calculate the level of non-gaussianity represented in the parameter $\tau_{NL}$.
Since the contributions to the trispectrum $T_\zeta$, calculated in Eqs. (\ref{tt2}) and (\ref{tl2}), and coming from Figs.
\ref{tf}a and \ref{tf}b respectively, present a wavevector dependence as that of the first line in Eq. (\ref{tfp}), we conclude
that for the specific subclass of inflationary models we are considering, the non-gaussianity in $T_\zeta$ is parameterized in
terms of $\tau_{NL}$. This automatically leads to $g_{NL} \ll \tau_{NL}$. In view of this, $\tau_{NL}$ is
given in this case as \cite{bl}:
\be
\frac{1}{2} \tau_{NL} = \frac{T_\zeta}{8\pi^6 \left[\frac{1}{k_2^3 k_4^3 |{\bf k}_3 + {\bf k}_4|^3} + 23 \;\; {\rm
permutations}\right] \mathcal{P}_\zeta^3} \,. \label{tau}
\ee
%%%%%%%%%%%%%%%%%%%%%%%%%%%%%%%%%%%%%%%%%%%%%%%%%%%%%%%%%%%
\subsection{The intermediate $\phi_\star$ $T$-region}
%%%%%%%%%%%%%%%%%%%%%%%%%%%%%%%%%%%%%%%%%%%%%%%%%%%%%%%%%%%
The level of non-gaussianity $\tau_{NL}$ according to Eqs. (\ref{pt2}), (\ref{tl2}) and (\ref{tau}), is given by
\bea
\frac{1}{2} \tau_{NL} &=& \frac{T^{1-loop}_\zeta}{8\pi^6 \left[\frac{1}{k_2^3 k_4^3 |{\bf k}_3 + {\bf k}_4|^3} + 23 \;\; {\rm
permutations}\right] (\mathcal{P}^{tree}_\zeta)^3} \nonumber \\
&=&\frac{2\eta_\sigma^4}{\eta_\phi^2\phi_\star^2}\exp[8N(|\eta_\sigma|-|\eta_\phi|)] \left(\frac{H_\star}
{2\pi}\right)^2\ln(kL)\nonumber\\
&=&\frac{2 \eta_\sigma^4}{\eta_\phi^2}\exp[8N(|\eta_\sigma|-|\eta_\phi|)]\left(\frac{m_P}{\phi_\star}\right)^2\frac{r{\mathcal
P}_\zeta}{8}\ln(kL)\nonumber\\
&=&2 \eta_\sigma^4\exp[8N(|\eta_\sigma|-|\eta_\phi|)]{\mathcal P}_\zeta\ln(kL)\nonumber\\
\Rightarrow\;\;\frac{1}{2}\tau_{NL}&\simeq& 4.91\times 10^{-9}|\eta_\sigma|^4\exp[400\ln(5.657\times10^{-2}r^{-1/2})(|
\eta_\sigma|-0.020)]\,,
\eea
where in the last line we have used the expressions in Eqs. (\ref{normt}) and (\ref{amount}).

\begin{figure*} [t]
\begin{center}
\includegraphics[width=15cm,height=10cm]{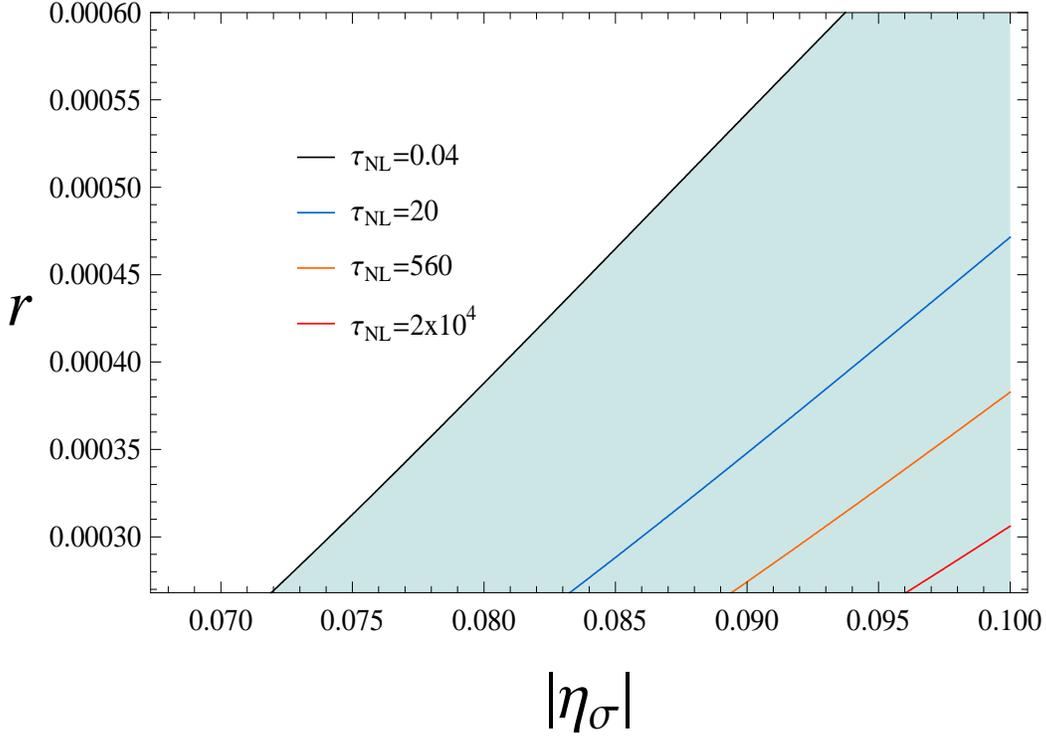}
\end{center}
\caption[Contours of $\tau_{NL}$ in the $r$ vs $|\eta_\sigma|$ plot]{Contours of $\tau_{NL}$ in the $r$ vs $|\eta_\sigma|$
plot. The intermediate (high) $\phi_\star$ $T$-region corresponds to the shaded (white) region. The observationally expected
$2\sigma$ range of values, for WMAP, PLANCK, and even the 21 cm background anisotropies, and for positive $\tau_{NL}$,
$\tau_{NL} > 20$ are completely inside the intermediate $\phi_\star$ $T$-region.  Notice that the boundary line between the
high and the intermediate $\phi_\star$ $T$-regions matches almost exactly the $\tau_{NL} = 0.04$ line.}
\label{fig1}
\end{figure*}

Now, by implementing the spectral tilt constraint in \eq{tiltt} in the spectrum normalisation constraint in Eq. (\ref{normt})
and the amount of inflation constraint in Eq. (\ref{amountc}), we conclude that the tensor to scalar ratio $r$ is bounded from
below:  $r \gsim 2.680 \times 10^{-4}$.

In the $r$ vs $|\eta_\sigma|$ plot in figure \ref{fig1}, we show lines of constant $\tau_{NL}$ corresponding to the values
$\tau_{NL} = 20, 560, 2 \times 10^4$.  We also show the high (in white) and intermediate (shaded) $\phi_\star$ $T$-regions in
agreement with the constraint in Eq. (\ref{intct}):
\be
\frac{r \mathcal{P}_\zeta}{8} \frac{\eta_\sigma^2}{\eta_\phi^2} \exp[4N(|\eta_\sigma|-|\eta_\phi|)] \ll \left(\frac{\phi_\star}
{m_P}\right)^2 \ll \frac{r \mathcal{P}_\zeta}{2} \frac{\eta_\sigma^4}{\eta_\phi^4} \exp[8N(|\eta_\sigma|-|\eta_\phi|)] \,,
\nonumber
\ee
\be
\Rightarrow\;\;4.070\times10^4\ll|\eta_\sigma|^4\exp[400\ln(5.657\times10^{-2}r^{-1/2})(|
\eta_\sigma|-0.020)]\ll1.656\times10^{17} \,.
\ee
As is evident from the plot, the observationally expected $2\sigma$ range of values for WMAP, $|\tau_{NL}| \gsim 2 \times 10^4$
\cite{kogo}, PLANCK, $|\tau_{NL}| \gsim 560$ \cite{kogo}, and even the 21 cm background anisotropies, $|\tau_{NL}| \gsim 20$
\cite{cooray2}, and for positive $\tau_{NL}$, are completely inside the intermediate $\phi_\star$ $T$-region as required.
Higher values for $\tau_{NL}$, up to $\tau_{NL} = 1.7 \times 10^5$ are consistent within our framework for the intermediate
$\phi_\star$ $T$-region.

\begin{figure*} [t]
\begin{center}
\includegraphics[width=15cm,height=10cm]{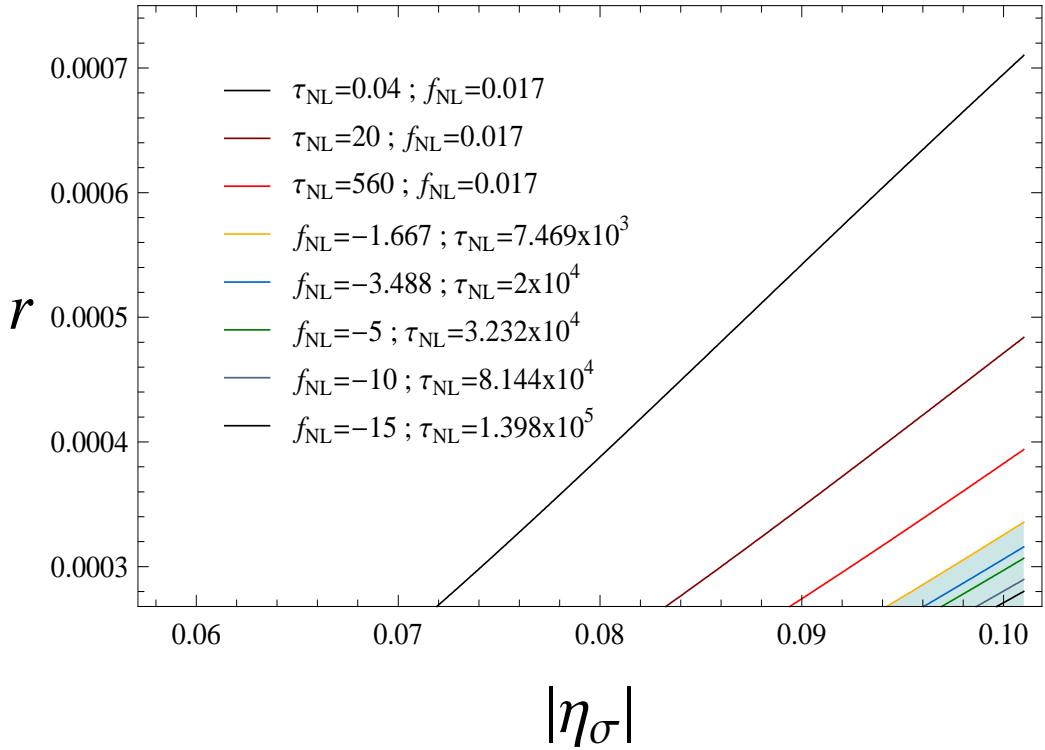}
\end{center}
\caption[Contours of both $f_{NL}$ and $\tau_{NL}$ in the $r$ vs $|\eta_\sigma|$ plot]{Contours of both $f_{NL}$ and
$\tau_{NL}$ in the $r$ vs
$|\eta_\sigma|$ plot. The intermediate (high) $\phi_\star$ region corresponds to the shaded (white) region. Lines for constant
$\tau_{NL}$ almost exactly matches lines for constant $f_{NL}$. According to this figure, and to the observational status, {\it non-gaussianity is more likely to be detected through the trispectrum than through the bispectrum}, for the
inflationary model studied in this chapter with concave downward potential, and from the WMAP, PLANCK, and even the 21 cm
background anisotropies observations. These lines also show some {\it consistency relations between the values of $f_{NL}$ and
$\tau_{NL}$} that will be useful at testing the inflationary model considered with concave downward potential against
observations.}
\label{fig3}
\end{figure*}

In subsection \ref{fnl}, we studied $f_{NL}$ for the case when $\zeta$ is generated during inflation, $B_\zeta$ is
dominated by the one-loop correction, and $P_\zeta$ is dominated by the tree-level term. Fig. \ref{fig32} shows the results
found.  The WMAP \cite{wmap} (and also PLANCK \cite{komatsu}) observationally allowed $2\sigma$ range of values for negative
$f_{NL}$, $-9 < f_{NL}$, is completely inside the intermediate $\phi_\star$ region\footnote{The intermediate $\phi_\star$ $T$-
region (where $T_\zeta$ is dominated by the one-loop correction and $P_\zeta$ is dominated by the tree-level term) encloses the
intermediate $\phi_\star$ region (where $B_\zeta$ is dominated by the one-loop correction and $P_\zeta$ is dominated by the
tree-level term).}. Fig. \ref{fig3} shows both Figs. \ref{fig32} and \ref{fig1} in the same plot. Incidentally, for the
available parameter window, lines for constant $\tau_{NL}$ almost exactly matches lines for constant $f_{NL}$. Thus, it is
possible to see that, according the observational status presented in the Section \ref{observational}, {\it
non-gaussianity is more likely to be detected through the trispectrum than through the bispectrum}, for the inflationary model
studied in this chapter with concave downward potential, and from the WMAP, PLANCK, and even the 21 cm background anisotropies
observations. Fig. \ref{fig3} also shows some {\it consistency relations between the values of $f_{NL}$ and $\tau_{NL}$} that
will be useful at testing the inflationary model considered with concave downward potential against observations.  For
instance, if WMAP detected non-gaussianity through the trispectrum with $\tau_{NL} \geq 8 \times 10^4$ at the $2\sigma$ level,
the slow-roll inflationary model with concave downward potential considered in this chapter would be ruled out since the
predicted $f_{NL}$ would be outside the current observational interval.

Similarly to the $f_{NL}$ case, it is interesting to see a slow-roll inflationary model with canonical kinetic terms where
large, {\it and observable}, values for $\tau_{NL}$ may be obtained (in contrast to the expected $\tau_{NL} \sim \mathcal{O}
(r)$ from the tree-level calculation \cite{seery3,ssv}). So we conclude that {\it if} $T_\zeta$ {\it is dominated by the one-
loop correction but} $P_\zeta$ {\it is dominated by the tree-level term, sizeable non-gaussianity is generated even if} $\zeta$
{\it is generated during inflation}.
\subsection{The high $\phi_\star$ $T$-region}
According to the expressions in Eqs. (\ref{asidef}), (\ref{tiltt}), (\ref{tt2}), (\ref{pt2}) and (\ref{tau}) the value of
$\tau_{NL}$ is in this case
\be
\frac{1}{2}\tau_{NL} = \frac{T^{tree}_\zeta}{8\pi^6 \left[\frac{1}{k_2^3 k_4^3 |{\bf k}_3 + {\bf k}_4|^3} + 23 \;\; {\rm
permutations}\right] (\mathcal{P}^{tree}_\zeta)^3}=\frac{1}{2}\eta_\phi^2 = 2\times10^{-4}\label{taonlslowroll}\,,
\ee
in agreement with the general expectations of Ref. \cite{bsw1} for slow-roll inflationary models with canonical kinetic terms
where only the tree-level contributions are considered and the field perturbations are assumed to be gaussian. This result is
no observational interest because the generated non-gaussianity is too small to be observable.
%%%%%%%%%%%%%%%%%%%%%%%%%%%%%%%%%%%%%%%%%%%%%%%%%%%%%%%%%%%%%%%%%%%%%%%%%%%%
\section{$\zeta$ not generated during inflation}\label{afterinfla}	%%%%%%%%
%%%%%%%%%%%%%%%%%%%%%%%%%%%%%%%%%%%%%%%%%%%%%%%%%%%%%%%%%%%%%%%%%%%%%%%%%%%%
We will assume in this Section that the fields driving inflation have nothing to do with the generation of $\zeta$;
nevertheless, they will generate the primordial non-gaussianity (see for instance Refs.
\cite{bl,hikage,kawakami,kawasaki3,kawasaki1,kawasaki2,langlois,suyama}). To this end, the post-inflationary evolution,
particularly the generation of $\zeta$, will be assumed not to generate significative levels of non-gaussianity in comparison
with those generated during inflation.
%%%%%%%%%%%%%%%%%%%%%%%%%%%%%%%%%%%%%%%%%%%%%%%%%%%%
\subsection{$\tnl$}                         %%%%%%%%
%%%%%%%%%%%%%%%%%%%%%%%%%%%%%%%%%%%%%%%%%%%%%%%%%%%%
\subsubsection{The low $\phi_\star$ $T$-region}\label{nolow}
It is possible, in principle, that $P_\zeta$ is dominated by the one-loop correction as long as $\zeta$ is not generated during
inflation.  Thus, the observed spectral index constraint is no longer required and, therefore, the low $\phi_\star$ $T$-region
is in principle viable.

Combining the conditions in Eqs. (\ref{ptloop}) and (\ref{norml}) with the expression for the number of e-folds in Eq.
(\ref{amount}), we get:
\be
1\lsim \frac{rn^2}{16} {\mathcal P}_\zeta \exp[N|\eta_\sigma|(4-2/n)]\,,\label{no1}
\ee
and
\be
1\gsim10^6\left(\frac{rn}{16}\right)^2{\mathcal P}_\zeta\exp(4N|\eta_\sigma|)\,,\label{no2}
\ee
where we have defined the parameter $n$ as the ratio between the two $\eta$ parameters: $n=\eta_\sigma/\eta_\phi$. These two
expressions lead to
\be
rn\lsim1.6\times10^{-5}|\eta_\sigma|\exp(-2N|\eta_\sigma|/n) \,,
\ee
as a necessary but not sufficient condition to satisfy both Eqs. (\ref{no1}) and (\ref{no2}). However, by introducing such a
condition in Eq. (\ref{no1}), we see that the latter translate into the following constraint:
\be
1\lsim 10^{-6}|\eta_\sigma|^2{\mathcal P}_\zeta\exp[4N|\eta_\sigma|(1-1/n)] \,.
\ee
The previous expression is impossible to satisfy because the highest value  the right hand side may take is for $n\rightarrow
\infty$ and, of course, $\eta_\sigma=0.1$ and $N=62$. Such a value, $1.45\times10^{-6}$, is much less than one. We conclude
that this case is of no interest because it is imposible to satisfy the normalisation spectrum condition in Eq. (\ref{norml}).

\subsubsection{The intermediate $\phi_\star$ $T$-region} \label{iphitregion}
The level of non-gaussianity $\tau_{NL}$ in this case is given by
\bea
\frac{1}{2}\tau_{NL} &=& \frac{T_\zeta^{1-loop}}{8\pi^6\left[\frac{1}{k_2^3 k_4^3 |{\bf k}_3 + {\bf k}_4|^3} + 23 \;\; {\rm
permutations}\right] {\mathcal{P}_\zeta^3}} \nonumber\\
&=& \frac{2\eta_{\sigma}^4}{\eta_{\phi}^8\phi_\star^8} \exp[8N(|\eta_\sigma|-|\eta_\phi|)]\left(\frac{H_\star}{2\pi}\right)
{\mathcal P}_\zeta^{-3} \ln(kL)\nonumber\\
&=&\frac{2\eta_{\sigma}^4}{\eta_{\phi}^8} \exp[8N(|\eta_\sigma|-|\eta_\phi|)]\left(\frac{m_P}
{\phi_\star}\right)^8\left(\frac{r}{8}\right)^4{\mathcal P}_\zeta \ln(kL) \nonumber\\
&=&\frac{2\eta_{\sigma}^4}{\eta_{\phi}^4}\left(\frac{1}{2\times10^{-2}}\right)^4 \exp(8N|\eta_\sigma|)\left(\frac{r}
{8}\right)^4{\mathcal P}_\zeta \ln(kL) \nonumber\\
&\simeq&2.60\times10^{16} \ (nr)^4 \,, \label{taunonzeta}
\eea
where in the last line we have introduced again the ratio $n$ defined in previous subsubsection, and chosen for simplicity $|
\eta_\sigma| = 0.1$ and $N=62$ so that the non-gaussianity is maximized.

From Eqs. (\ref{intct}), (\ref{normtd}) and (\ref{amount}), and those coming from the $\zeta$ series convergence constraints in
Eqs. (\ref{converg1}) and (\ref{converg2}), with $|\eta_\sigma| = 0.1$ and $N = 62$, lead to the following conditions that
reduce the available parameter space:
\begin{itemize}
\item The perturbative regime constraint $|x| \ll 1$:
\begin{equation}
r \lsim 6.51 \times 10^4 \ n \exp\left[-\frac{12.4}{n}\right] \,.
\end{equation}
\item The perturbative regime constraint $|y| \ll 1$:
\begin{equation}
r \lsim \frac{2.68 \times 10^{-1}}{n^2} \,.
\end{equation}
\item The $P_\zeta$ dominated by the tree-level term constraint:
\begin{equation}
r \lsim \frac{1.10 \times 10^{-4}}{n} \exp\left[\frac{12.4}{n}\right] \,.
\end{equation}
\item The $T_\zeta$ dominated by the one-loop correction constraint:
\begin{equation}
r \gsim \frac{4.68 \times 10^{-12}}{n^3} \exp\left[\frac{37.2}{n}\right] \,. \label{plnon1}
\end{equation}
\item The spectrum normalisation constraint:
\begin{equation}
r \lsim \frac{1.6 \times 10^{-4}}{n} \exp\left[-\frac{12.4}{n}\right] \,. \label{sncnonzeta}
\end{equation}
\end{itemize}

\begin{figure*} [t]
\begin{center}
\includegraphics[width=15cm,height=10cm]{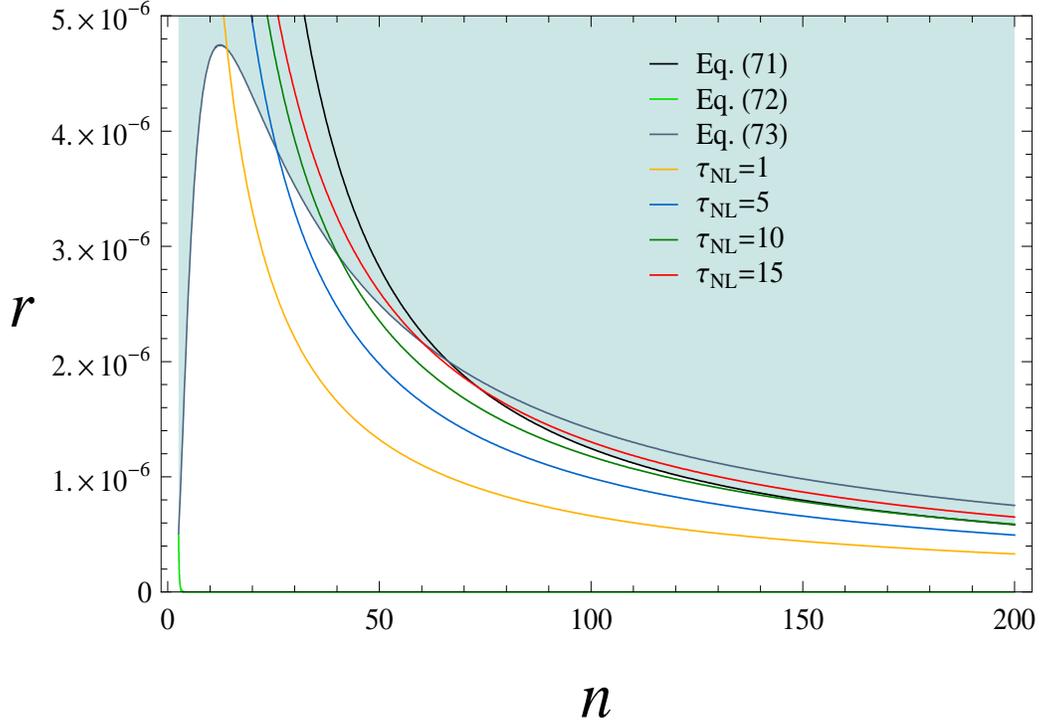}
\end{center}
\caption[Contours of $\tau_{NL}$ in the $r$ vs $n$ plot, for $2.58 \leq n \leq 200$, when $\zeta$ is not generated during
inflation]{Contours of $\tau_{NL}$ in the $r$ vs $n$ plot, for $2.58 \leq n \leq 200$, when $\zeta$ is not generated during
inflation. The allowed parameter space corresponds to the white region. The constraint in Eq. (\ref{plnon1}) almost matches
(visually) the horizontal axis. The largest possible value $\tau_{NL}$ may take in this range is 15.}
\label{fig4}
\end{figure*}

\begin{figure*} [t]
\begin{center}
\includegraphics[width=15cm,height=10cm]{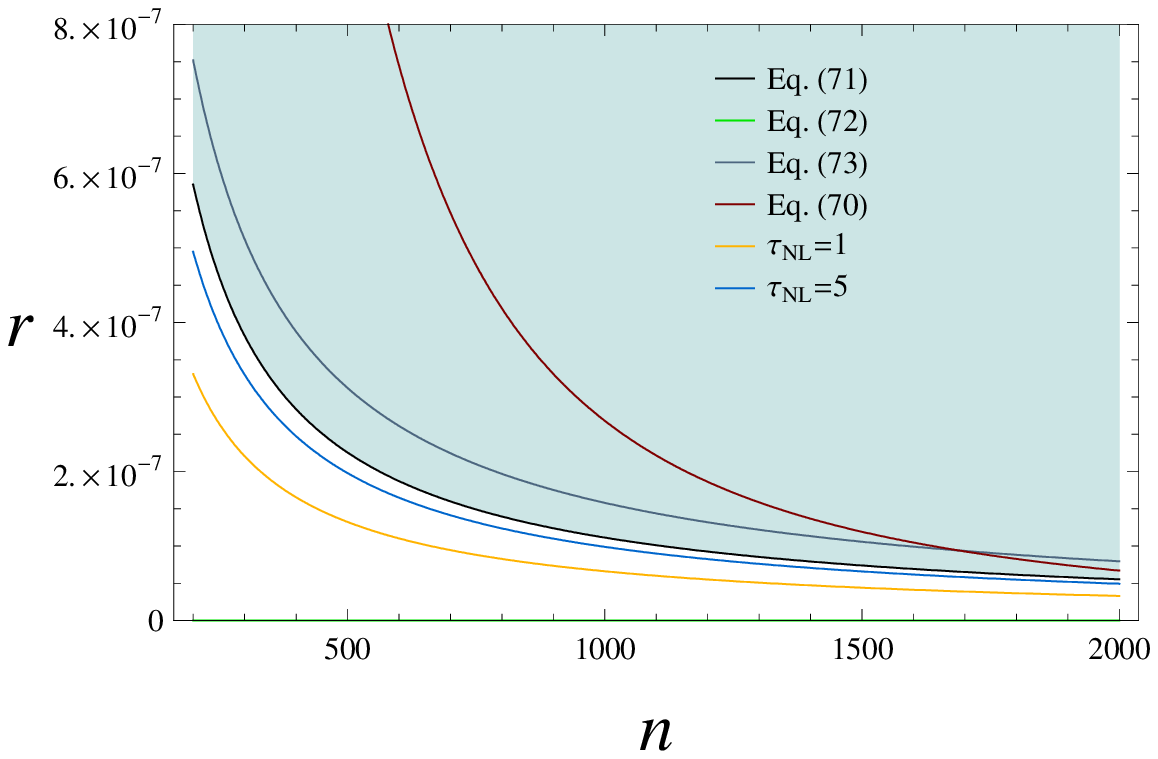}
\end{center}
\caption[Contours of $\tau_{NL}$ in the $r$ vs $n$ plot, for $200 \leq n \leq 2000$, when $\zeta$ is not generated during
inflation]{Contours of $\tau_{NL}$ in the $r$ vs $n$ plot, for $200 \leq n \leq 2000$, when $\zeta$ is not generated during
inflation. The allowed parameter space corresponds to the white region. The constraint in Eq. (\ref{plnon1}) matches (visually)
the horizontal axis. The largest possible value $\tau_{NL}$ may take in this range is a bit higher than 5.}
\label{fig5}
\end{figure*}

\begin{figure*} [t]
\begin{center}
\includegraphics[width=15cm,height=10cm]{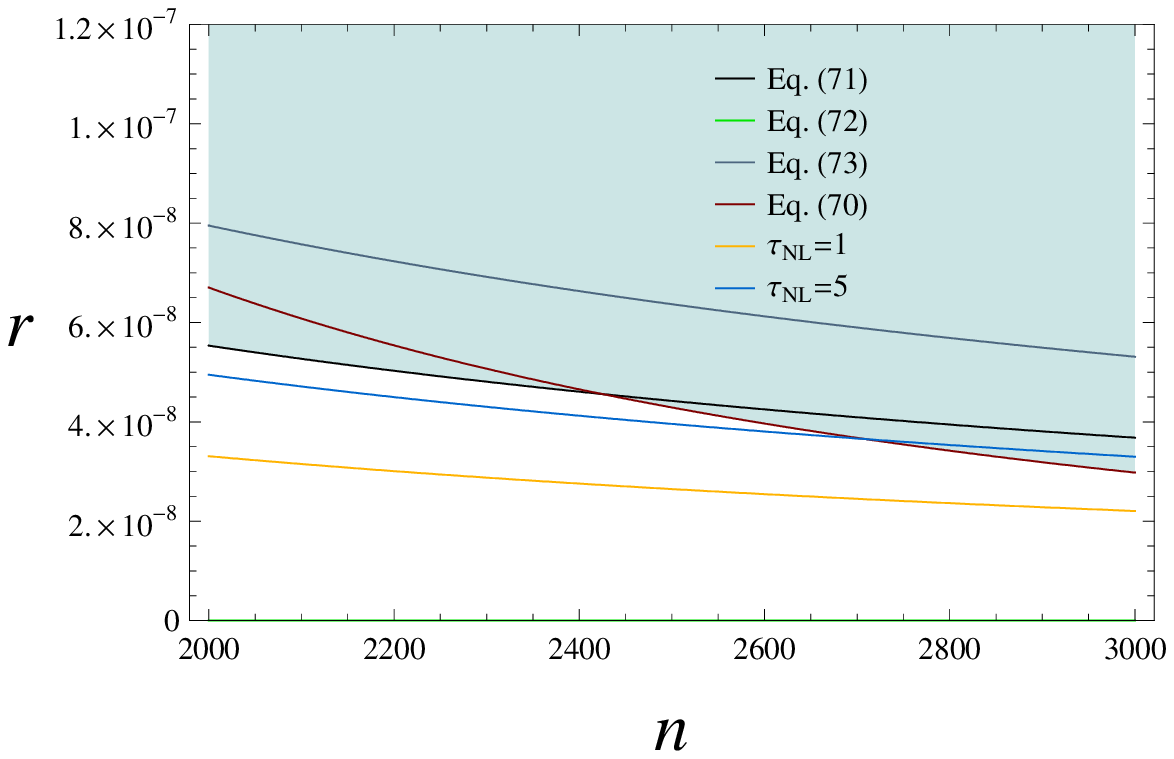}
\end{center}
\caption[Contours of $\tau_{NL}$ in the $r$ vs $n$ plot, for $2000 \leq n \leq 3000$, when $\zeta$ is not generated during
inflation.]{Contours of $\tau_{NL}$ in the $r$ vs $n$ plot, for $2000 \leq n \leq 3000$, when $\zeta$ is not generated during
inflation. The allowed parameter space corresponds to the white region. The constraint in Eq. (\ref{plnon1}) matches (visually)
the horizontal axis. The largest possible value $\tau_{NL}$ may take in this range is a bit higher than 5.}
\label{fig6}
\end{figure*}

Analysing these expressions, we conclude that the constraint in the first item is automatically satisfied once the constraint
in the fifth item is satisfied. Moreover, from the constraint in the fifth item, we see that the highest possible value $r$ may
take is $4.75 \times 10^{-6}$. And finally, to make the constraint in the fourth item consistent with the constraints in the
second, third, and fifth items, the lower bound $n \gsim 2.58$ is required. The resultant available parameter window, together
with the lines for constant values of $\tau_{NL}$, $\tau_{NL} = 1,5,10,15$, is presented in Fig. \ref{fig4} for $2.58 \leq n
\leq 200$. Fig. \ref{fig5} shows the range $200 \leq n \leq 2000$ with the lines $\tau_{NL} = 1,5$, while Fig. \ref{fig6} shows
the range $2000 \leq n \leq 3000$ also with the lines $\tau_{NL} = 1,5$. As the figures reveal, {\it when $T_\zeta$ is
dominated by the one-loop correction and $P_\zeta$ is dominated by the tree-level term, large values for $\tau_{NL}$ are
obtained although not so large as in the case where $\zeta$ is generated during inflation}.  Indeed, an upper bound on
$\tau_{NL}$, according to Eqs. (\ref{taunonzeta}) and (\ref{sncnonzeta}), is $34.078$ when $n \rightarrow \infty$.

We conclude that, even if $\zeta$ is not generated during inflation, we may find {\it observable} values for $\tau_{NL}$.
However, such observable values could only be observed by the 21 cm background anisotropies at the $1\sigma$ level according to
the observational status presented in Section \ref{observational}. We also conclude that, {\it for non-
gaussianity to be observable, primordial gravitational waves must be undetectable}.

\subsubsection{The high $\phi_\star$ $T$-region}
This case is of no interest because the generated non-gaussianity is too small to be observable:
\bea
\frac{1}{2}\tau_{NL} &=& \frac{T_\zeta^{tree}}{8\pi^6\left[\frac{1}{k_2^3 k_4^3 |{\bf k}_3 + {\bf k}_4|^3} + 23 \;\; {\rm
permutations}\right] {\mathcal{P}_\zeta^3}} \nonumber\\
&=&  \frac{T_\zeta^{tree}}{8\pi^6\left[\frac{1}{k_2^3 k_4^3 |{\bf k}_3 + {\bf k}_4|^3} + 23 \;\; {\rm permutations}\right]
\left(\mathcal{P}_\zeta^{tree}\right)^3} \frac{\left(\mathcal{P}_\zeta^{tree}\right)^3}{\mathcal{P}_\zeta^3}\;=\; \frac{1}{2} |
\eta_\phi|^2 \frac{\left(\mathcal{P}_\zeta^{tree}\right)^3}{\mathcal{P}_\zeta^3}\nonumber\\
&\Rightarrow& \tau_{NL}\;\ll\;|\eta_\phi|^2 \,,
\eea
according to the expressions in Eqs. (\ref{pt2}) and (\ref{tt2}).

%%%%%%%%%%%%%%%%%%%%%%%%%%%%%%%%%%%%%%%%%%%%%
\subsection{$f_{NL}$} \label{fnlafter}
%%%%%%%%%%%%%%%%%%%%%%%%%%%%%%%%%%%%%%%%%%%%
\subsubsection{The low $\phi_\star$ region}
From Eqs. (\ref{lowphi}) and (\ref{ptloop}) we see that the low $\phi_\star$ region and the low $\phi_\star$ $T$-region are
exactly the same, being only constrained by the fact that $P_\zeta$ is dominated by the one-loop correction.  Thus, the
obtained conclusions in Subsubsection \ref{nolow} equally apply. Therefore, this case is of no interest because it is
impossible to satisfy the normalisation spectrum condition in Eq. (\ref{norml}) for the low $\phi_\star$ region.

\subsubsection{The intermediate $\phi_\star$ region}

The level of non-gaussianity is in this case given by
\bea
\frac{6}{5} f_{NL} &=& \frac{B_\zeta^{1-loop}}{4\pi^4 \frac{\sum_i k_i^3}{\prod_i k_i^3} \mathcal{P}_\zeta^2} =
\frac{\eta_\sigma^3}{\eta_\phi^6 \phi_\star^6} \exp[6N(|\eta_\sigma|-|\eta_\phi|)] \left(\frac{H_\star}{2\pi}\right)^6
\mathcal{P}_\zeta^{-2} \ln(kL) \nonumber \\
&=& \frac{\eta_\sigma^3}{\eta_\phi^6} \exp[6N(|\eta_\sigma|-|\eta_\phi|)] \left(\frac{m_P}{\phi_\star}\right)^6 \left(\frac{r}
{8}\right)^3 \mathcal{P}_\zeta \ln(kL) \nonumber \\
&=& -\frac{\eta_\sigma^3}{\eta_\phi^3} \left(\frac{1}{2\times10^{-2}}\right)^3 \exp(6N |\eta_\sigma|) \left(\frac{r}
{8}\right)^3 \mathcal{P}_\zeta \ln(kL) \nonumber \\
&\approx& -8.59 \times 10^{9} (nr)^3 \,, \label{fnonzeta}
\eea
where
in the last line we have chosen again for simplicity $|\eta_\sigma| = 0.1$ and $N = 62$ so that the non-gaussianity is
maximized.

Since the spectrum normalisation constraint in Eq. (\ref{sncnonzeta}) equally applies to this case, we conclude from it and
from Eq. (\ref{fnonzeta}) that an upper bound on $|f_{NL}|$ is $2.93 \times 10^{-2}$ when $n \rightarrow \infty$. $f_{NL}$ is,
of course, unobservable.  We conclude that {\it when $\zeta$ is not generated during inflation, but the primordial non-
gaussianity is, it is impossible to detect non-gaussianity through the bispectrum}. However, in view of Subsubsection
\ref{iphitregion}, {\it it is possible to detect it through the trispectrum}.

\subsubsection{The high $\phi_\star$ region}

This case is of no interest because the generated non-gaussianity is too small to be observable:
\be
\frac{6}{5} f_{NL} = \frac{B_\zeta^{tree}}{4\pi^4 \frac{\sum_i k_i^3}{\prod_i k_i^3} \mathcal{P}_\zeta^2} =
\frac{B_\zeta^{tree}}{4\pi^4 \frac{\sum_i k_i^3}{\prod_i k_i^3} \left(\mathcal{P}_\zeta^{tree}\right)^2}
\frac{\left(\mathcal{P}_\zeta^{tree}\right)^2}{\mathcal{P}_\zeta^2} = |\eta_\phi|
\frac{\left(\mathcal{P}_\zeta^{tree}\right)^2}{\mathcal{P}_\zeta^2} \ll |\eta_\phi| \,,
\ee
according to the expressions in Eqs. (\ref{pt}) and (\ref{bt}).
%%%%%%%%%%%%%%%%%%%%%%%%%%%%%%%%%%%%%%%%%%%%%%
\section{Conclusions} \label{conclusca2}	%%%%%%
%%%%%%%%%%%%%%%%%%%%%%%%%%%%%%%%%%%%%%%%%%%%%%

In this chapter we extended the analysis given in the previous one, but this time we calculated the trispectrum $T_\zeta$ 
of the primordial curvature perturbation $\zeta$, generated during a {\it slow-roll} inflationary epoch and considering a two-
field quadratic model of inflation with canonical kinetic terms. In order to obtain a large level of non-gaussianity, we consider loop 
contributions as well as tree level terms, and show that it is possible to attain very high, including observable, values for 
the level of non-gaussianity $\tau_{NL}$ if $T_\zeta$ is dominated by the one-loop contribution and $\pz$ is dominated by
tree level term. The statement presented in Ref. \cite{bch2} about the suppression of the loop corrections
against the tree-level terms when considering classicality was analyzed in this chapter and
argued to be too strongly stated leading to non-general conclusions. The probability that a
typical observer sees a non-gaussian distribution in the model considered in this thesis
was investigated and found to be non-negligible.
%%%%%%%%%%%%%%%%%%%%%%%%%%%%%%%%%%%%%%%%%%%%%%%%%%%%%%%%%%%%%%%%%%%%%%%%%%%%%%%
%%%%%%%%%%%%%%%%%%%%%%%%%%%%%%%%%%%%%%%%%%%%%%%%%%%%%%%%%%%%%%%%%%%%%%%%%%%%%%%%%
\chapter{NON-GAUSSIANITY FROM VECTOR FIELD PERTURBATIONS}\label{chaptvec}     %%%
%%%%%%%%%%%%%%%%%%%%%%%%%%%%%%%%%%%%%%%%%%%%%%%%%%%%%%%%%%%%%%%%%%%%%%%%%%%%%%%%%
%%%%%%%%%%%%%%%%%%%%%%%%%%%%%%%%%%%%%%%%%%%%%%%%%%%%%%%%%%%%%%%%%%%%%%%%%%%%%%%%%

%%%%%%%%%%%%%%%%%%%%%%%%%%%%%%%%%%%%%
\section{Introduction}      %%%%%%%%%
%%%%%%%%%%%%%%%%%%%%%%%%%%%%%%%%%%%%%

The anisotropies in the temperature of the cosmic microwave background (CMB) radiation, which have strong connections with the
origin of the large-scale structure in the observable Universe, is one of hottest topics in modern cosmology. The properties of
the CMB temperature anisotropies are described in terms of the spectral functions, like the spectrum, bispectrum, trispectrum,
etc., of the primordial curvature perturbation $\zeta$ \cite{cogollo}. In most of the cosmological models the $n$-point
correlators of $\zeta$ are supposed to be translationally and rotationally invariants. However, violations of such invariances
entail modifications of the usual definitions for the spectral functions in terms of the statistical descriptors
\cite{acw,armendariz,carroll}. These violations may be consequences either of the presence of vector field perturbations
\cite{armendariz,bdmr,vc,vc2,RA2,dklr,dkw,dkw2,go,gmv2,gmv,gvnm,himmetoglu3,himmetoglu,himmetoglu2,himmetoglu4,kksy,dkl,koh,ys},
spinor field perturbations \cite{bohmer,shan}, or p-form perturbations \cite{germani,germani2,kobayashi,koivisto,koivisto2},
contributing significantly to $\zeta$, of anisotropic expansion
\cite{bamba,bohmer,dechant,gcp,himmetoglu4,kksy,koivisto,ppu1,ppu2,watanabe} or of an
inhomogeneous background \cite{armendariz,carroll,dklr}.
Violation of the statistical isotropy (i.e. violation of the rotational invariance in the
$n$-point correlators of $\zeta$) seems to be present in the data \cite{app,ge,hl,samal} and, although its statistical
significance is still low, the continuous presence of anomalies in every CMB data analysis (see for instance Refs.
\cite{bunn1,dvorkin,dipole2,dipole1,hansen,dipole3,hoftuft,hou,land1,land2,oliveira,schwarz,tegmark}) suggests the evidence
might be decisive in the forthcoming years. Since the statistical anisotropy is observationally low, it entails a big problem
when vector fields are present during inflation, because they generically lead to a high amount of statistical anisotropy,
higher than that coming from observations \cite{dklr,gmv,kksy}. To solve this problem, people use different mechanisms in order
to make those models consistent with observation, for example using a triad of orthogonal vectors \cite{armendariz,bento}, a
large number of identical randomly oriented vectors fields \cite{gmv}, or assuming that the contribution of
vector fields to the total energy density is negligible \cite{dklr,kksy}.

The amount of statistical anisotropy is quantified throught the parameter $g_\zeta$, usually called the level of statistical
anisotropy in the spectrum. \eq{astadef} gives us the
primordial power spectrum that takes into account the leading effects of violations of statistical isotropy by
the presence of some vector field in the inflationary era. As we could see in Section \ref{observational} the $\gz$ parameter
has observational bounds and works, together with the non-gaussianity parameters $\fnl$, $\tnl$, $\gnl$, etc., as
statistical descriptors for $\zeta$. Therefore, it could be a crucial tool to discriminate between some of the more usual 
cosmological models.

Recent works point out the possibility that a vector field causes part of the primordial curvature perturbation and show that
the particular presence of vector fields in the inflationary dynamics may generate sizeable levels of non-gaussianity described
by $\fnl$ \cite{bdmr1,dkl,vrl} and $\tnl$ \cite{bdmr2,vr}. In such works the authors included both vector and scalar field
perturbations, and asummed that the contributions to the spectrum from vector field perturbations were smaller than those
coming from scalar fields and in an opposite way for bispectrum and trispectrum.

In this chapter we use the $\dn$ formalism to calculate the tree-level and one-loop contributions to
the bispectrum $\bz$ and trispectrum $\tz$ of $\zeta$, including vector and scalar field perturbations. We then
calculate the order of magnitude of the levels of non-gaussianity in $\bz$ and $\tz$ including the
one-loop contributions and write down formulas that relate the order of magnitude of the levels of non-gaussianity $\fnl$
and $\tnl$ with the amount of statistical anisotropy in the spectrum $\gz$. Finally, comparison with the
expected observational bound from WMAP is done.

%%%%%%%%%%%%%%%%%%%%%%%%%%%%%%%%%%%%%%%%%%%%%%%%%%%%%%%%%%%%%%%%%%%%%%%%%%
\section{Statistical descriptors from vector field perturbations}       %%
%%%%%%%%%%%%%%%%%%%%%%%%%%%%%%%%%%%%%%%%%%%%%%%%%%%%%%%%%%%%%%%%%%%%%%%%%%

As we saw in Chapter \ref{chaptgen}, the $\delta N$ formalism \cite{dklr,lms,lr,st,starobinsky,ss}
is a powerful tool to calculate the primordial curvature perturbation and all its statistical descriptor to any
desired order. In the simplest case where $\zeta$ is generated by one scalar field and one vector field and assuming that the anisotropy in the expansion of the Universe is negligible, it can be
calculated up to quadratic terms by means of the following truncated expansion \cite{dklr}:
\be
\zeta(\bfx)\equiv\delta N (\phi(\bfx),A_i(\bfx),t)=N_\phi \delta\phi + N_A^i\delta A_i+\frac{1}{2}N_{\phi\phi}(\delta\phi)^2+
N_{\phi A}^i\delta\phi\delta A_i+\frac{1}{2}N_{AA}^{ij}\delta A_i \delta A_j \,,\label{deltan}
\ee
where
\be
N_\phi\equiv\frac{\partial N}{\partial \phi}\,,\quad
N_A^{i}\equiv\frac{\partial N}{\partial A_i}\,,\quad
N_{\phi\phi} \equiv\frac{\partial^2 N}{\partial \phi^2}\,,\quad
N_{AA}^{ij}\equiv\frac{\partial^2 N}{\partial A_i\partial A_j}\,,
\quad N_{\phi A}^i\equiv\frac{\partial^2 N}{\partial A_i\partial\phi}\,,
\ee
$\phi$ being the scalar field and ${\bf A}$ the vector field, with $i$ denoting the spatial
indices running from 1 to 3. In the Section \ref{descriptors} we defined the power spectrum $\pz$, the
bispectrum $\bz$ and trispectrum $\tz$ for the primordial curvature perturbation, through the Fourier modes of $\zeta$ as:
{\small\bea
\langle\zeta(\bfk)\zeta(\bfk')\rangle&\equiv&(2\pi)^3\delta(\bfk+\bfk')P_\zeta(\bfk)
\;\;\equiv \;\;(2\pi)^3\delta(\bfk+\bfk')\frac{2\pi^2}{k^3}\calpz(\bfk) \,,
\label{spdef}\\
\langle\zeta(\bfk)\zeta(\bfk')\zeta(\bfk'')\rangle&\equiv&(2\pi)^3\delta(\bfk+\bfk'+\bfk'')\bz(\bfk,\bfk',\bfk'')
\;\;  \no\\&\equiv& \;\;(2\pi)^3\delta(\bfk+\bfk'+\bfk'')\frac{4\pi^4}{k^3k'^3}\calbz(\bfk,\bfk',\bfk'') \,.
\label{bsdef}\\
\langle\zeta(\bfk_1)\zeta(\bfk_2)\zeta(\bfk_3)\zeta(\bfk_4)\rangle&\equiv&(2\pi)^4\delta(\bfk_1+\bfk_2+\bfk_3+
\bfk_4)\tz(\bfk_1,\bfk_2,\bfk_3,\bfk_4)\no\\
&\equiv& (2\pi)^3\delta(\bfk_1+\bfk_2+\bfk_3+\bfk_4)\frac{\(2\pi^2\)^3}{k_1^3k_2^3|\bfk_2+\bfk_3|^3}\caltz(\bfk_1,\bfk_2,\bfk_3,\bfk_4) \,.
\label{tsdef}
\eea}
Using \eq{deltan} and the definitions given in Eqs. (\ref{spdef}) and (\ref{bsdef}), it was
found in Ref. \cite{dklr} that the tree-level contribution to the spectrum is of the form shown in \eq{curvquad}, that is
\be
\pz(\bfk) = P_\zeta\su{iso}(k) \( 1 + g_\zeta
(\hat{\bfd}\cdot \hat \bfk)^2 \) \,.
\label{curvquad1} \ee
In addition, an analogous form for the contribution to $\fnl$ was given in Ref. \cite{dkl}, showing that both, $\pz$ and
$\fnl$ have anisotropic contributions coming from the vector field perturbation. The one-loop
correction to the spectrum was also given in Ref. \cite{dklr}, however they kept it in an integral form. In this chapter we
give the tree-level and one-loop contributions to the bispectrum and to the trispectrum. We also 
estimate the integrals coming from loop corections in order to get an order of magnitude for $\fnl$ and $\tnl$.

Using the Fourier fmodes for Eqs. \eq{deltan}, (\ref{spdef}), (\ref{bsdef}) and (\ref{tsdef}) and 
considering contributions up to one-loop order, the expressions for $\calpz$, $\calbz$ and $\caltz$ , are\footnote{These 
expresions can be calculated using the diagrammatic tool that we will present in a forthcoming paper \cite{val}.}:
{\small\bea
\calp_\zeta^{\rm tree}(\bfk) &=& N_\phi^2 \calp_{\delta \phi}(k) +
N_A^iN_A^j\calt_{ij}(\bfk)  \nonumber\\
&=&
 N_\phi^2 \calp_{\delta \phi}(k) +
N_A^2 \calp_+(k) + ({\bf N}_A\cdot\hat\bfk )^2 \calp_+(k)  \( r\sub{long} - 1  \)\label{Pzetat}
\,, \eea
\bea
\calp_\zeta^{\rm 1-loop} (\bfk) &=& \int
\frac{d^3p k^3}{4\pi|\bfk + \bfp|^3 p^3}
\[\frac{1}{2} N_{\phi \phi}^2  \calp_{\delta \phi} (|\bfk + \bfp|)
\calp_{\delta \phi}(p) +
N_{\phi A}^i N_{\phi A}^j \calp_{\delta \phi} (|\bfk + \bfp|)
\calt_{ij}(\bfp)\right. \nonumber \\&&\left. + \frac{1}{2} N_{AA}^{ij} N_{AA}^{kl}\calt_{ik}(\bfk+\bfp)\calt_{jl}(\bfp) \] \,,
\label{Pzetal}
\eea
\bea
\calb_\zeta^{\rm tree} (\bfk,\bfk',\bfk'') &=& N_\phi^2 N_{\phi \phi}
[\calp_{\delta \phi} (k) \calp_{\delta \phi} (k') + {\rm c. \ p.}]  + N_A^i N_A^k N_{AA}^{mn}
\Big[ \calt_{im}(\bfk)\calt_{kn}(\bfk') + {\rm c. \ p.}\Big] \no\\
 &+& N_\phi N_A^i N_{\phi A}^j \Big[\calp_{\delta \phi} (k)
\calt_{ij}(\bfk') + 5 \ {\rm perm.} \Big] \,, \label{bzt}
\eea
\bea
\calb_\zeta^{\rm 1-loop}(\bfk,\bfk',\bfk'')&=&N_{\phi \phi}^6\int\frac{d^3pk^3k'^3}{4\pi p^3|\bfk+\bfp|^3|\bfk'-\bfp|
^3}\calp_{\delta\phi}(p)\calp_{\delta\phi} (|\bfk+\bfp|)
\calp_{\delta \phi}(|\bfk'-\bfp|)\no\\
&+& N_{AA}^{ij}N_{AA}^{kl}N_{AA}^{mn}\int\frac{d^3pk^3k'^3}{4\pi p^3|\bfk+\bfp|^3|\bfk'-\bfp|^3}\calt_{il}(\bfp)\calt_{kn}
(\bfk+\bfp)\calt_{jm}(\bfk''-\bfp)\no\\
&+&N_{\phi \phi}N_{\phi A}^{i}N_{\phi A}^{j}\int\frac{d^3pk^3k'^3}{4\pi p^3|\bfk''+\bfp|^3|\bfk'-\bfp|
^3}\bigg\{\calp_{\delta\phi}(p)\calp_{\delta\phi} (|\bfk''+\bfp|)\calt_{ij}(\bfk'-\bfp)\no\\
&+&\calp_{\delta\phi}(p)\calp_{\delta\phi}(|\bfk'-\bfp|)\calt_{ij}(\bfk''+\bfp)+\calp_{\delta\phi}(|
\bfk'-\bfp|)\calp_{\delta\phi}(|\bfk''+\bfp|)\calt_{ij}(\bfp)\bigg\}\no\\
&+&N_{\phi A}^{i}N_{\phi A}^{j}N_{AA}^{kl}\int\frac{d^3pk^3k'^3}{4\pi p^3|\bfk''+\bfp|^3|\bfk'-\bfp|
^3}\bigg\{\calp_{\delta\phi}(p)\calt_{ik}(\bfk'-\bfp)\calt_{jl} (\bfk''+\bfp)\no\\
&+&\calp_{\delta\phi}(|\bfk''+\bfp|)\calt_{ik}(\bfp)\calt_{jl}(\bfk'-\bfp)+\calp_{\delta\phi}(|
\bfk'-\bfp|)\calt_{ik}(\bfp)\calt_{jl}(\bfk''+\bfp)\bigg\}\label{bsl} \,,
\eea
\bea
\calt_\zeta^{\rm tree} (\bfk_1,\bfk_2,\bfk_3,\bfk_4) &=& N_\phi^2 N_{\phi \phi}^2
[\calp_{\delta \phi} (k_2) \calp_{\delta \phi} (k_4)  \calp_{\delta \phi} (|\bfk_1+\bfk_2|)+ {\rm 11 \ perm.}] \no\\
&+& N_A^i N_A^j N_{AA}^{kl} N_{AA}^{mn}\Big[ \calt_{ik}(\bfk_2)\calt_{jm}(\bfk_4)\calt_{ln}(\bfk_1+\bfk_2) + {\rm 11 \ perm.}\Big] \no\\
 &+& N_\phi^2 N_{A \phi}^i N_{A \phi}^j \Big[\calp_{\delta \phi} (k_2) \calp_{\delta \phi} (k_4)
\calt_{ij}(\bfk_1+\bfk_2) + 11 \ {\rm perm.} \Big] \no\\
&+& N_A^i N_A^j N_{A \phi}^k N_{A \phi}^l \Big[\calt_{ik}(\bfk_2)\calt_{jl}(\bfk_4)\calp_{\delta \phi}(|\bfk_1+\bfk_2|) + 11 \
  {\rm perm.} \Big] \no\\
&+& N_\phi N_{\phi \phi}N_A^i N_{A\phi}^{j}\Big[\calp_{\delta \phi}(k_2)\calt_{ij}(\bfk_4)\calp_{\delta \phi}(|\bfk_1+\bfk_2|) + 23 \
  {\rm perm.} \Big] \no\\
&+& N_\phi N_A^iN_{A\phi}^{j} N_{AA}^{kl} \Big[\calp_{\delta \phi}(k_2)\calt_{ik}(\bfk_4)\calt_{jl}(\bfk_1+\bfk_2) + 23 \ {\rm perm.} \Big]
\,, \label{tst}
\eea
\bea
\calt_{\zeta A}^{\rm 1-loop}(\bfk_1,\bfk_2,\bfk_3,
\bfk_4)&=&N_{AA}^{ij}N_{AA}^{kl}N_{AA}^{mn}N_{AA}^{op}\int\frac{d^3p \ k_1^3k_3^3|\bfk_3+\bfk_4|^3}{4\pi p^3|
\bfk_1-\bfp|^3|\bfk_3+\bfp|^3|\bfk_3+\bfk_4+\bfp|^3}\times\no\\
&&\times\calt_{im}(\bfp)\calt_{jk}(\bfk_1-\bfp)\calt_{np}(\bfk_3+\bfp)\calt_{lo}(\bfk_3+\bfk_4+\bfp)\label{tsl} \,,
\eea}
where
\be\label{def1}
\calt_{ij}(\bfk)\equiv T_{ij}^{\rm even}(\bfk)\calp_+(k)+iT_{ij}^{\rm odd}(\bfk)\calp_-(k)+T\su{long}_{ij}(\bfk)\calp_{\rm
long}(k) \,,
\ee
and
\be
T_{ij}\su{even} (\bfk) \equiv \delta_{ij} -  \hat k_i \hat k_j \,,\qquad
T\su{odd}_{ij} (\bfk) \equiv \epsilon_{ijk}\hat k_k \,,\qquad
T\su{long}_{ij} (\bfk) \equiv \hat k_i \hat k_j\,.
\ee
\eq{Pzetat} was writen in the form of \eq{curvquad1} with $\hat{\bfd}=\hat{\bfn}_A$, $\bfn_A$
being a vector
with magnitude  $N_A\equiv\sqrt{N^i_AN^i_A}$, and $r\sub{long}\equiv\calp\sub{long}/\calp_+$, where $\calp\sub{long}$ is the
power spectrum for the longitudinal component, and $\calp_+$ and $\calp_-$ are the parity conserving and violating
power spectra defined by
\be
\calp_{\pm}\equiv\frac{1}{2}\(\calp_R\pm\calp_L\) \,,
\ee
with $\calp_R$ and $\calp_L$ denoting the power spectra for the transverse components with right-handed and
left-handed polarisations \cite{dklr}.

The above expressions can be further separated into different terms: one due to perturbations in the scalar
field, another due to the vector field perturbations, and the other due to the mixed terms:
{\bea
\calpz^{\rm tree}(\bfk)&=&\calpz_\phi^{\rm tree}(k)+\calpz_ A^{\rm tree}(\bfk)\label{sst},\\
\calp_\zeta^{\rm 1-loop}(\bfk)&=&\calpz_{\phi}^{\rm 1-loop}(k)+\calpz_{A}^{\rm 1-loop}(\bfk)+\calpz_{\phi A}^{\rm
1-loop}(\bfk)\label{ssl} \,,
\eea
\bea
\calb_\zeta^{\rm tree} (\bfk,\bfk',\bfk'') &=& \calbz_\phi^{\rm tree} (\bfk,\bfk',\bfk'')+\calbz_A^{\rm tree}
(\bfk,\bfk',\bfk'')+\calbz_{\phi A}^{\rm tree} (\bfk,\bfk',\bfk'')\label{bsst} \,,\\
\calb_\zeta^{\rm 1-loop}(\bfk,\bfk',\bfk'')&=&\calbz_\phi^{\rm 1-loop} (\bfk,\bfk',\bfk'')+\calbz_A^{\rm 1-loop}
(\bfk,\bfk',\bfk'')\no\\&+&\calbz_{\phi A}^{\rm 1-loop} (\bfk,\bfk',\bfk'')\label{bssl}\,,
\eea
\bea
\calt_\zeta^{\rm tree} (\bfk_1,\bfk_2,\bfk_3,\bfk_4) &=&\caltz_\phi^{\rm tree} (\bfk_1,\bfk_2,\bfk_3,\bfk_4)+\caltz_A^{\rm 
tree}(\bfk_1,\bfk_2,\bfk_3,\bfk_4) \no\\&+&\caltz_{\phi A}^{\rm tree} (\bfk_1,\bfk_2,\bfk_3,\bfk_4)\label{tsst} \,,\\
\calt_\zeta^{\rm 1-loop}(\bfk_1,\bfk_2,\bfk_3,\bfk_4)&=&\caltz_\phi^{\rm 1-loop} (\bfk_1,\bfk_2,\bfk_3,\bfk_4)+
\caltz_A^{\rm 1-loop} (\bfk_1,\bfk_2,\bfk_3,\bfk_4)\no\\&+&\caltz_{\phi A}^{\rm 1-loop} (\bfk_1,\bfk_2,\bfk_3,
\bfk_4)\label{tssl} \,,
\eea
Observational analysis tell us that the statistical anisotropy in CMB temperature perturbation
could be observable in a future through current experiments like WMAP or PLANCK. \eq{curvquad1} combined with recent studies
\cite{ge} tells us that the level of statistical anisotropies $g_\zeta$ has an
upper bound and in the best case ($99\%$ confidence level) this is $\gz\lsim 0.383$ \cite{gawe}. During our analysis  
we will adopt an upper bound for $g_\zeta$: $\gz\lsim 0.1$. In order to satisfy the latter observational constraint over the 
spectrum, we must be sure that the contributions coming from vector fields in Eqs. (\ref{Pzetat}) and (\ref{Pzetal}) are
smaller than those coming from scalar fields. That means that the first term in  \eq{sst} dominates over all
the other terms, even those coming from one-loop contributions. With the previous conclusion in mind we feel free to make
assumptions over the other contributions, specially for those coming from vector field perturbations.
%%%%%%%%%%%%%%%%%%%%%%%%%%%%%%%%%%%%%%%%%%%%%%%%%%%%%%%%%%%%%%%%%%%%%%%%%%%%%%%%%%
\section{Vector field contributions to the statistical descriptors}		   %%%%%%%
%%%%%%%%%%%%%%%%%%%%%%%%%%%%%%%%%%%%%%%%%%%%%%%%%%%%%%%%%%%%%%%%%%%%%%%%%%%%%%%%%%
As we explain in the previous section, our unique restriction from observation is related to
the amount of statistical anisotropy present in the spectrum, so we need to be sure that the first term in \eq{sst} always
dominates. In our study we will assume that the terms coming only from the vector field dominate over those coming from
the mixed terms and from the scalar fields only, except for the case of the tree-level spectrum\footnote{For an actual
realisation of this scenario, we need to show that such constraints are fully satisfied.}. Based on the assumption made, Eqs.
(\ref{sst}) - (\ref{bssl}) take the form:
\bea
\calpz^{\rm tree}(\bfk)&=&\calpz_\phi^{\rm tree}(k)+\calpz_ A^{\rm tree}(\bfk)\label{sst1} \,,\\
\calp_\zeta^{\rm 1-loop}(\bfk)&=&\calpz_{A}^{\rm 1-loop}(\bfk)\label{ssl1} \,,\\
\calb_\zeta^{\rm tree} (\bfk,\bfk',\bfk'') &=&\calbz_A^{\rm tree} (\bfk,\bfk',\bfk'')\label{bsst1} \,,\\
\calb_\zeta^{\rm 1-loop}(\bfk,\bfk',\bfk'')&=&\calbz_A^{\rm 1-loop} (\bfk,\bfk',\bfk'')\label{bssl1} \, \\
\calt_\zeta^{\rm tree} (\bfk_1,\bfk_2,\bfk_3,\bfk_4) &=&\caltz_A^{\rm tree} (\bfk_1,\bfk_2,\bfk_3,\bfk_4)\label{tsst1} \,,\\
\calt_\zeta^{\rm 1-loop}(\bfk_1,\bfk_2,\bfk_3,\bfk_4)&=&\caltz_A^{\rm 1-loop} (\bfk_1,\bfk_2,\bfk_3,\bfk_4)\label{tssl1} \,,
\eea
The above expressions lead us to eight different ways that allow us to study and probably get a high
level of non-
gaussianity\footnote{Our assumption is inspired in the one given in Ref. \cite{bl}.
In that work the authors use two scalar fields instead of one scalar and one vector field as
in this chapter. A realisation of such a scenario can be found in Chapter \ref{chaptsca} (see also Refs. \cite{cogollo,valenzuela,leblond}).}
\begin{itemize}
\item Vector field spectrum ($\calp_{\zeta_A}$) and bispectrum ($\calb_{\zeta_A}$) dominated
by the tree-level terms \cite{dkl}.
\item Vector field spectrum ($\calp_{\zeta_A}$) and bispectrum ($\calb_{\zeta_A}$) dominated
by the 1-loop contributions.
\item Vector field spectrum ($\calp_{\zeta_A}$) dominated by the tree-level terms and
bispectrum ($\calb_{\zeta_A}$) dominated by the 1-loop contributions.
\item Vector field spectrum ($\calp_{\zeta_A}$) dominated by the 1-loop contributions and
bispectrum ($\calb_{\zeta_A}$) dominated by the tree-level terms.
\item Vector field spectrum ($\calp_{\zeta_A}$) and trispectrum ($\calt_{\zeta_A}$) dominated
by the tree-level terms.
\item Vector field spectrum ($\calp_{\zeta_A}$) and trispectrum ($\calt_{\zeta_A}$) dominated
by the one-loop contributions.
\item Vector field spectrum ($\calp_{\zeta_A}$) dominated by the tree-level terms and
trispectrum ($\calt_{\zeta_A}$) dominated by the 1-loop contributions.
\item Vector field spectrum ($\calp_{\zeta_A}$) dominated by the 1-loop contributions and
trispectrum ($\calt_{\zeta_A}$) dominated by the tree-level terms.
\end{itemize}
In order to study these possibilities, we first need to estimate the integrals coming from loop
contributions. From Eqs. (\ref{Pzetal}), (\ref{bsl}), (\ref{tsl}), (\ref{ssl1}), (\ref{bssl1}) and (\ref{tssl1}) the integrals to solve are:
{\small\bea
\calpz^{\rm 1-loop}(\bfk)&=& \frac{1}{2} N_{AA}^{ij} N_{AA}^{kl} \int \frac{d^3p k^3}{4\pi p^3|\bfk + \bfp|^3}
\calt_{ik}(\bfk+\bfp)\calt_{jl}(\bfp) \,, \label{intsl}\\
\calbz^{\rm 1-loop}(\bfk,\bfk',\bfk'')&=&N_{AA}^{ij}N_{AA}^{kl}N_{AA}^{mn}\int\frac{d^3pk^3k'^3}{4\pi p^3|
\bfk+\bfp|^3|\bfk'-\bfp|^3}\times\no\\&&\times\calt_{il}(\bfp)\calt_{kn}(\bfk+\bfp)\calt_{jm}(\bfk'-\bfp) \, \label{intbsl}\\
\calt_{\zeta A}^{\rm 1-loop}(\bfk_1,\bfk_2,\bfk_3,
\bfk_4)&=&N_{AA}^{ij}N_{AA}^{kl}N_{AA}^{mn}N_{AA}^{op}\int\frac{d^3p \ k_1^3k_3^3|\bfk_3+\bfk_4|^3}{4\pi p^3|
\bfk_1-\bfp|^3|\bfk_3+\bfp|^3|\bfk_3+\bfk_4+\bfp|^3}\times\no\\
&&\times \calt_{im}(\bfp)\calt_{jk}(\bfk_1-\bfp)\calt_{np}(\bfk_3+\bfp)\calt_{lo}(\bfk_3+\bfk_4+\bfp)\,. \label{inttsl}.
\eea}
The above integrals cannot be done analytically, but they can be estimated in the same way as that presented in
Refs.\cite{bl,lythaxions,lythbox}. In Appendix \ref{Integrals} we show that the integrals are proportional to $\ln(kL)$
where $L$ is the box size. To evaluate them we take the spectrum to be scale-invariant, which will be a good approximation if
both scalar field $\phi$ and vector field  ${\bf A}$ are sufficiently light during inflation. The integrals are logarithmically
divergent at the zeros of the denominator and in each direction, but there is a cutoff at $k\sim L^{-1}$. We found
that in our case the integrals are also proportional to $\ln(kL)$ and that each singularity gives equal
contributions to the overall result.  We find from Eqs. (\ref{intsl}) and (\ref{intbsl}):
\bea
\calp_{\zeta A}^{\rm 1-loop} (\bfk)&=&\frac{1}{2}N_{AA}^{ij}N_{AA}^{kl}(2\calp_++\calp_{long})\delta_{ik}\calt_{jl}
(\bfk)\ln(kL) \,, \label{sploop}\\
\calb_\zeta^{\rm 1-loop}(\bfk,\bfk',
\bfk'')&=&N_{AA}^{ij}N_{AA}^{kl}N_{AA}^{mn}\ln(kL)\big(2\calp_++\calp_{long})\delta_{il}\big[\calt_{kn}(\bfk)\calt_{jm}
(\bfk')\big] \, \label{bsploop}\\
\calt_{\zeta A}^{\rm 1-loop}(\bfk_1,\bfk_2,\bfk_3,\bfk_4)
&=&N_{AA}^{ij}N_{AA}^{kl}N_{AA}^{mn}N_{AA}^{op}\ln(kL)\big(2\calp_++\calp_{long})\delta_{im}\calt_{jk}(\bfk_1)\,\no\\
&&\times\,\calt_{np}(\bfk_3)\calt_{lo}(\bfk_4+\bfk_3).\label{tsploop}
\eea
%%%%%%%%%%%%%%%%%%%%%%%%%%%%%%%%%%%%%%%%%%%%%%%%%%%%%%%%%%%%%%%%%%%%%%%%%%%%%%
\section{Calculation of the non-gaussianity parameter $\fnl$}   %%%%%%%%%%%%%%
%%%%%%%%%%%%%%%%%%%%%%%%%%%%%%%%%%%%%%%%%%%%%%%%%%%%%%%%%%%%%%%%%%%%%%%%%%%%%%
The non-gaussianity parameter in the biespectrum $\bz$ is defined by \cite{maldacena,komatsu}\footnote{We employ the WMAP
sign convention.}:
\be
\fnl=\frac{5}{6}\frac{\calbz(\bfk,\bfk',\bfk'')}{\big[\calpz(k)\calpz(k')+{\rm cyc.\  perm.}\big]}\label{fnl1} \,.
\ee
Since the isotropic contribution to the curvature perturbation
is always dominant compared to the anisotropic one, we can write in the above expression only the
isotropic part of the spectrum $\calpz^{\rm iso}(k)$:
\be
\fnl=\frac{5}{6}\frac{\calbz(\bfk,\bfk',\bfk'')}{\big[\calpz^{\rm iso}(k)\calpz^{\rm iso}(k')+{\rm cyc.\  perm.}\big]} \,.
\label{fnl2}
\ee
Keeping in mind the above expression, we will estimate the possible amount of non-gaussianity $\fnl$ in the bispectrum $\bz$, generated by the
anisotropic part of the primordial curvature perturbation. To do it we take into account the different possibilities
mentioned in the previous section, where the non-gaussianity is produced solely by vector field perturbations.
%%%%%%%%%%%%%%%%%%%%%%%%%%%%%%%%%%%%%%%%%%%%%%%%%%%%%%%%%%%%%%%%%%%%%%%%%%%%%%%%%%%%%%%%%%%
\subsection{Vector field spectrum ($\calp_{\zeta_A}$) and bispectrum ($\calb_{\zeta_A}$)
dominated by the tree-level terms}
%%%%%%%%%%%%%%%%%%%%%%%%%%%%%%%%%%%%%%%%%%%%%%%%%%%%%%%%%%%%%%%%%%%%%%%%%%%%%%%%%%%%%%%%%%%
We start our analysis by considering the case studied in Ref. \cite{dkl}, where the authors assume that
the bispectrum is dominated by vector fields perturbations and that the higher order
contributions from the vector field are always sub-dominant, i.e $N_A^i\delta A_i\gg N_{AA}^{ij}\delta A_i \delta A_j$.
This means that both the spectrum and the bispectrum are dominated by the tree level terms, i.e. $\calpz_A^{\rm tree} \gg
\calpz_A^{\rm 1-loop}$ and $\calbz_A^{\rm tree} \gg \calbz_A^{\rm 1-loop}$, so that
the level of non-gaussianity $f_{NL}$ is given by:
\be
\fnl\:=\: \frac{5}{6} \frac{\calb_\zeta^{\rm tree} (\bfk,\bfk',\bfk'')}{\big[\calpz^{\rm iso}(k)\calpz^{\rm iso}(k')+{\rm cyc.\
perm.}\big]}\:\simeq\:\frac{5}{6}\frac{\calbz_A^{\rm tree} (\bfk,\bfk',\bfk'')}{\big[\calpz^{\rm iso}(k)\calpz^{\rm iso}
(k')+{\rm cyc.\  perm.}\big]} \,.
\label{fnlm1}
\ee
Since the anisotropic contribution to the curvature perturbation is subdominant, we can take
$\calpz\sim\calpz^{\rm iso}$, so we may write:
\be
\fnl\:\simeq\:\frac{N_A^i N_A^k N_{AA}^{mn}\big[\calt_{im}(\bfk)\calt_{kn}(\bfk')+{\rm cyc.\  perm.}\big]}
{\big[\calpz(k)\calpz(k')+{\rm cyc.\  perm.}
\big]} \,.
\label{fnlm2}
\ee
Assuming that $\calp_{\rm long}$, $\calp_+$, and $\calp_-$ are all of the same order of
magnitude, and that the spectrum is scale invariant, we
may write the above equation as:
\be
\fnl\:\simeq\:\frac{\calp_A^2 N_A^2 N_{AA}}{\calpz^2} \,,
\label{fnlm3}
\ee
where $\calp_A=2\calp_++\calp_{\rm long}$. Taking as a typical value for the vector field perturbation $\delta
A=\sqrt{\calp_A}$ and $ N_A\delta A > N_{AA}\delta A^2$, the contribution of the vector field to $\zeta$  is given by
$\zeta_A\sim \sqrt{\calpz_A}\sim N_A \sqrt{\calp_A}$. %where $\calpz_A$ is the spectrum of the anisotropic curvature
%perturbation.
Thus, we may write an upper bound for $\fnl$:
\be
\fnl\:\lsim\:\frac{\calpz_A\threehalf}{\calpz^2} \,.
\label{fnlm4}
\ee
Since the level of statistical anisotropy in the power spectrum is of order $\gz\sim\calpz_A/\calpz$,
 and since $\calpz\half \simeq 5\times 10 ^{-5}$ \cite{wmap5}, \eq{fnlm3} yields \cite{dkl}:
\be
\fnl\:\lsim\:10^3\(\frac{\gz}{0.1}\)\threehalf \,.
\label{fnlm5}
\ee
The above expression gives an upper bound for the level of non-gaussianity $\fnl$ in terms of
the level of statistical anisotropy in the power spectrum $\gz$ when the former is generated
by the anisotropic contribution to the curvature perturbation. As we may see, the recent
observational bounds on $\fnl$: $-9 < f_{NL} <111$ \cite{wmap5}\footnote{The bispectrum (trispectrum) in this scenario might be 
either of the local, equilateral, or orthogonal type. We are not interested in this thesis on the shape of the non-gaussianity
but on its order of magnitude. Being that the case, comparing with the expected bound on the
{\it local} $\fnl$ \cite{wmap5} ($\tnl$ \cite{kogo}) makes no sensible difference under the assumption that the
expected bounds on the equilateral and orthogonal $\fnl$ ($\tnl$) are of the same order of magnitude,
as analogously happens in the $\fnl$ case for single-field inflation \cite{smz}.}, may easily be exceeded.

As an example of this model, we apply the previous results to a specific model, e.g. the vector curvaton
scenario \cite{vc,vc2,RA2}, where the $N$-derivatives are \cite{dkl}:
\bea
N_{A}&=&\frac{2}{3A}r \,,\label{navc}\\
N_{AA}&=&\frac{2}{A^2}r\label{naavc} \,,
\eea
where $A\equiv|\bfA|$ is the value of vector field just before the vector curvaton field decays
and the parameter $r$ is the ratio between the energy density of the vector curvaton field and
the total energy density of the Universe just
before the vector curvaton decay. We begin exploring the conditions under which the vector field
spectrum and bispectrum are always dominated by the tree-level terms. From Eqs. (\ref{Pzetat}), (\ref{bzt}),
(\ref{sploop}) and (\ref{bsploop}) our constraint leads to:
\bea
\calp_A N_A^2 &\gg & \calp_A^2 N_{AA}^2 \,,\\
\calp_A^2 N_A^2 N_{AA}&\gg &\calp_A^3 N_{AA}^3 \,.
\eea
Thus, it follows that:
\be
\calp_A \ll \(\frac{N_A}{N_{AA}}\)^2 \,.
\ee
We have to remember that in the present case the contribution of the vector field to $\zeta$  is given by
$\zeta_A\sim \sqrt{\calpz_A}\sim N_A \sqrt{\calp_A}$. Then, the above equation combined with Eqs. (\ref{navc}) and
(\ref{naavc}) leads to:
\be
r \gg 2.25\times 10^{-4}\gz\half \,.\label{rbt}
\ee
This is a lower bound on the $r$ parameter we have to consider when building a realistic
particle physics model of the vector curvaton scenario.

Finally, from \eq{fnlm3}, the $f_{NL}$ parameter in this scenario is given by:
\be
\fnl\simeq \frac{4.5\times 10^{-2}}{r}\(\frac{\gz}{0.1}\)^2\label{fnlgr}.
\ee
This is a consistency relation between $f_{NL}$, $g_\zeta$, and $r$ which will help when
confronting the specific vector curvaton realisation against observation.
%%%%%%%%%%%%%%%%%%%%%%%%%%%%%%%%%%%%%%%%%%%%%%%%%%%%%%%%%%%%%%%%%%%%%%%%%%%%%%%%%%%%%%%%%%%
\subsection{Vector field spectrum ($\calp_{\zeta_A}$) and bispectrum ($\calb_{\zeta_A}$)
dominated by the 1-loop contributions}
%%%%%%%%%%%%%%%%%%%%%%%%%%%%%%%%%%%%%%%%%%%%%%%%%%%%%%%%%%%%%%%%%%%%%%%%%%%%%%%%%%%%%%%%%%%
Since the bispectrum is dominated by 1-{\rm loop} contributions and is given by \eq{bsploop}, we may
write \eq{fnl2} as:
\be
\fnl\:\simeq\:\frac{N_{AA}^{ij}N_{AA}^{kl}N_{AA}^{mn}\ln(kL)\big(2\calp_++\calp_{long})\delta_{il}\big[\calt_{kn}
(\bfk)\calt_{jm}(\bfk')+{\rm cyc.\  perm.}\big]}{\big[\calpz(k)\calpz(k')+{\rm cyc.\  perm.}\big]} \,.
\label{fnl11}
\ee
Assuming again that $\calp_{\rm long}$, $\calp_+$, and $\calp_-$ are all of the same order of
magnitude, and that the spectrum is scale invariant, the above equation leads to:
\be
\fnl\:\simeq\:\frac{\calp_A^3 N_{AA}^3}{\calpz^2} \,.
\label{fnl12}
\ee
Since the vector field spectrum is dominated by the 1-{\rm loop} contribution,
$\zeta_A\sim \sqrt{\calpz_A}\sim N_{AA}
\calp_A$. Thus, and taking into account that $\gz\sim\calpz_A/\calpz$ and $\calpz\half\simeq
5\times 10 ^{-5}$ \cite{wmap5}, we
find:
\be
\fnl\:\sim\:\frac{1}{\sqrt{\calpz}}\(\frac{\calpz_A}{\calpz}\)\threehalf\:\sim\:10^3\(\frac{\gz}{0.1}\)\threehalf.
\label{fnl13}
\ee
The biggest difference between the result found in Ref. \cite{dkl}, given by \eq{fnlm5}, and the result given by
\eq{fnl13}, is that the latter gives an equality relation between the non-gaussianity parameter $\fnl$ and the
level of statistical anisotropy in the power spectrum $\gz$. Following the recent bounds for
$\fnl$: $-9 < f_{NL} <111$ \cite{wmap5}, this scenario predicts an upper bound for the $\gz$ parameter:
\be
\gz<\,0.02 \,.\label{gz}
\ee
This bound is stronger than that obtained from direct observations in Ref. \cite{ge}.

Again we apply our result to the vector curvaton scenario. Since we are assuming that the vector field
spectrum and bispectrum are dominated by 1-loop contributions, we get from Eqs. (\ref{Pzetat}), (\ref{bzt}),
(\ref{sploop}), and (\ref{bsploop}):
\be
\calp_A > \(\frac{N_A}{N_{AA}}\)^2 \,,
\ee
which for the vector curvaton scenario becomes:
\be
r <2.25\times 10^{-4}\gz\half \,. \label{rbl}
\ee
This is an upper bound on the $r$ parameter we have to consider when building a realistic
particle physics model of the vector curvaton scenario.
%%%%%%%%%%%%%%%%%%%%%%%%%%%%%%%%%%%%%%%%%%%%%%%%%%%%%%%%%%%%%%%%%%%%%%%%%%%%%%%%%%%%%%%%%%%
\subsection{Vector field spectrum ($\calp_{\zeta_A}$) dominated by the tree-level terms and
bispectrum ($\calb_{\zeta_A}$) dominated by the 1-loop contributions}\label{ptb1l}
%%%%%%%%%%%%%%%%%%%%%%%%%%%%%%%%%%%%%%%%%%%%%%%%%%%%%%%%%%%%%%%%%%%%%%%%%%%%%%%%%%%%%%%%%%%
In order to check the viability of this case, we start studying the implications of the restrictions over the
spectrum and the bispectrum, i.e. what happens when we assume that the vector field spectrum
is dominated by the tree-level terms and the bispectrum is dominated by the 1-loop contributions.
From Eqs. (\ref{Pzetat}), (\ref{bzt}),
(\ref{sploop}), and (\ref{bsploop}) it follows that:
\bea
\calp_A N_A^2 \gg  \calp_A^2 N_{AA}^2 &\Rightarrow & \calp_A\ll\frac{N_A^2}{N_{AA}^2} \,,
\label {fc}\\
\calp_A^2 N_A^2 N_{AA}\ll  \calp_A^3 N_{AA}^3 &\Rightarrow & \calp_A\gg\frac{N_A^2}{N_{AA}^2} \,.
\label{sc}
\eea
As we may see, it is impossible to satisfy simultaneously Eqs. (\ref{fc}) and (\ref{sc}).
This is perhaps related to the fact that we have
taken into account only one vector field. Such a conclusion may be relaxed if we take into
account more than one vector field, as analogously happens in the scalar multi-field case
\cite{cogollo,valenzuela}.
%%%%%%%%%%%%%%%%%%%%%%%%%%%%%%%%%%%%%%%%%%%%%%%%%%%%%%%%%%%%%%%%%%%%%%%%%%%%%%%%%%%%%%%%%%%
\subsection{Vector field spectrum ($\calp_{\zeta_A}$) dominated by the 1-loop contributions
and bispectrum ($\calb_{\zeta_A}$) dominated by the tree-level terms}\label{p1lbt}
%%%%%%%%%%%%%%%%%%%%%%%%%%%%%%%%%%%%%%%%%%%%%%%%%%%%%%%%%%%%%%%%%%%%%%%%%%%%%%%%%%%%%%%%%%%
As in the previous case, it is impossible to satisfy the conditions under which
the spectrum is always dominated by the 1-loop contributions and the bispectrum is always dominated
by the tree-level terms:
%we arrived to an inconsistency. In this particular case the conditions also are:
\bea
\calp_A &\gg& \frac{N_A^2}{N_{AA}^2} \,, \\
\calp_A &\ll& \frac{N_A^2}{N_{AA}^2} \,.
\eea
%it the bispectrum is to be dominated by the tree-level terms.
Again, the conclusion may be relaxed if we take into account more than one vector field.

%%%%%%%%%%%%%%%%%%%%%%%%%%%%%%%%%%%%%%%%%%%%%%%%%%%%%%%%%%%%%%%%%%%%%%%%%%%%%%
\section{Calculation of the non-gaussianity parameter $\tnl$}   %%%%%%%%%%%%%%
%%%%%%%%%%%%%%%%%%%%%%%%%%%%%%%%%%%%%%%%%%%%%%%%%%%%%%%%%%%%%%%%%%%%%%%%%%%%%%
The non-gaussianity parameter $\tnl$ in the trispectrum $\tz$ is defined by \cite{bl}:
\be
\tnl=\frac{2\,\caltz(\bfk_1,\bfk_2,\bfk_3,\bfk_4)}{\big[\calpz(\bfk_1)\calpz(\bfk_2)\calpz(\bfk_1+\bfk_4)+{\rm 23\  perm.}\big]}\label{tnl1} \,.
\ee
Remember that the isotropic contribution in the \eq{curvquad1} is always dominant compared to the anisotropic one
so that we may write in the above expression only the isotropic part of the spectrum $\calpz^{\rm iso}(k)$:
\be
\tnl=\frac{2\caltz(\bfk_1,\bfk_2,\bfk_3,\bfk_4)}{\big[\calpz^{\rm iso}(k_1)\calpz^{\rm iso}(k_2)\calpz^{\rm iso}(|
\bfk_1+\bfk_4|)+{\rm 23\  perm.}\big]} \,.
\label{tnl2}
\ee
Using the above expression, we will estimate the possible amount of non-gaussianity generated by the
anisotropic part of the primordial curvature perturbation, taking into account different possibilities
and assuming that the non-gaussianity is produced solely by vector field perturbations.
%%%%%%%%%%%%%%%%%%%%%%%%%%%%%%%%%%%%%%%%%%%%%%%%%%%%%%%%%%%%%%%%%%%%%%%%%%%%%%%%%%%%%%%%%%%%%
\subsection{Vector field spectrum ($\calp_{\zeta_A}$) and trispectrum ($\calt_{\zeta_A}$)
dominated by the tree-level terms}
%%%%%%%%%%%%%%%%%%%%%%%%%%%%%%%%%%%%%%%%%%%%%%%%%%%%%%%%%%%%%%%%%%%%%%%%%%%%%%%%%%%%%%%%%%%%%%
In this first case, we assume that the trispectrum is dominated by vector field perturbations
and that the higher order terms in the $\dn$ expansion in Eq. (\ref{deltan}) involving the
vector field are sub-dominant against the first-order term:
$N_A^i\delta A_i\gg N_{AA}^{ij}\delta A_i \delta A_j$. The latter implies that both the
spectrum and the trispectrum are dominated by the tree-level terms, i.e.
$\calpz_A^{\rm tree} \gg \calpz_A^{\rm 1-loop}$ and $\caltz_A^{\rm tree} \gg \caltz_A^{\rm 1-loop}$.
Thus, we have from \eq{tnl2}:
\be
\tnl=\frac{2\caltz_A^{\rm tree}(\bfk_1,\bfk_2,\bfk_3,\bfk_4)}{\big[\calpz^{\rm iso}(k_1)\calpz^{\rm iso}(k_2)\calpz^{\rm iso}(|
\bfk_1+\bfk_4|)+{\rm 23\  perm.}\big]} \,,
\label{tnl11}
\ee
which, in view of Eqs. (\ref{tst}) and (\ref{tsst1}), looks like:

\be
\tnl\:\simeq\:\frac{2N_A^i N_A^j N_{AA}^{kl} N_{AA}^{mn}\Big[ \calt_{ik}(\bfk_2)\calt_{jm}(\bfk_4)\calt_{ln}(\bfk_1+\bfk_2) + 
{\rm 11 \ perm.}\Big]}{\big[\calpz^{\rm iso}(k_1)\calpz^{\rm iso}(k_2)\calpz^{\rm iso}(|\bfk_1+\bfk_4|)+{\rm 23\  perm.}\big]} 
\,.
\label{tnl12}
\ee

We will just consider here the order of magnitude of $\tnl$.  Therefore, we will ignore the
specific ${\bf k}$ dependence of $\calt_{ij}$.  Instead, we will assume that
$\calp_{\rm long}$, $\calp_+$, and $\calp_-$ are all of the same order of
magnitude, which is a good approximation for some specific actions (see for instance Ref.
\cite{dklr}), and take advantage of the fact that the spectrum is almost scale invariant
\cite{wmap5}. Thus, after getting rid of all the ${\bf k}$ dependences, the order of magnitude
of $\tnl$ looks like:
\be
\tnl\:\simeq\:\frac{\calp_A^3 N_A^2 N_{AA}^2}{(\calpz^{\rm iso})^3} \,,
\label{tnl13}
\ee
where $\calp_A=2\calp_++\calp_{\rm long}$. Employing our assumption that $N_A\delta A > N_{AA}\delta A^2$,
and since the root mean squared value for the vector field perturbation $\delta
A$ is $\sqrt{\calp_A}$, the contribution of the vector field to $\zeta$  is given
by $\zeta_A\sim \sqrt{\calpz_A}\sim N_A \sqrt{\calp_A}$. An upper bound for $\tnl$ is therefore
given by:
\be
\tnl\:\lsim\:\frac{\calpz_A^2}{(\calpz^{\rm iso})^3} \,.
\label{tnl14}
\ee
Since the order of magnitude of $\gz$ is $\calpz_A/\calpz^{\rm iso}$, under the assumptions made above
we get:
\be
\tnl\:\lsim\:8\times10^6\(\frac{\gz}{0.1}\)^2 \,,
\label{tnl15}
\ee
where $(\calpz^{\rm iso})\half \simeq 5\times 10 ^{-5}$ \cite{wmap5} % \eq{tnl13} yields
has been used.
Eq. (\ref{tnl15}) gives an upper bound for the level of non-gaussianity $\tnl$ in terms of
the level of statistical anisotropy in the power spectrum $\gz$ when the former is generated
by the anisotropic contribution to the curvature perturbation. Comparing with the expected
observational limit on $\tnl$ coming from future WMAP data releases, $\tnl \sim2\times 10^4$
\cite{kogo}, we conclude that in this scenario a large level of non-gaussianity in the
trispectrum $T_\zeta$ of $\zeta$ is possible, leaving some room for ruling out this scenario
if the current expected observational limit is overtaken.

As an example of this scenario, we apply the previous results to a specific model, e.g. the vector curvaton
model \cite{vc,vc2,RA2}, where the $N$-derivatives are \cite{dkl}:
\bea
N_{A}&=&\frac{2}{3A}r \,,\label{navc}\\
N_{AA}&=&\frac{2}{A^2}r\label{naavc} \,,
\eea
where $A\equiv|\bfA|$ is the value of vector field just before the vector curvaton field decays
and the parameter $r$ is the ratio between the energy density of the vector curvaton field and
the total energy density of the Universe just
before the vector curvaton decay.

First, we check if the conditions under which the vector field
spectrum and trispectrum are always dominated by the tree-level terms are fully satisfied.
From
Eqs. (\ref{Pzetat}), (\ref{tst}), (\ref{sploop}) and (\ref{tsploop}) our constraint leads to:
\bea
\calp_A N_A^2 &\gg & \calp_A^2 N_{AA}^2 \,, \label{teq} \\
\calp_A^3 N_A^2 N_{AA}^2&\gg &\calp_A^4 N_{AA}^4 \,,
\eea
which mean that the if the vector field spectrum is dominated by the tree-level terms so is
the vector field trispectrum.  An analogous situation happens when the vector field spectrum
is dominated by the one-loop terms:  the vector field trispectrum is also dominated by this
kind of terms. As a result, it is impossible that simultaneously the vector field spectrum
is dominated by the tree-level (one-loop) terms and the vector field trispectrum is dominated
by the one-loop (tree-level) terms.
Following Eq. (\ref{teq}), we get:
\be
\calp_A \ll \(\frac{N_A}{N_{AA}}\)^2 \,,
\ee
which, in view of
$\zeta_A\sim \sqrt{\calpz_A}\sim N_A \sqrt{\calp_A}$ and 
Eqs. (\ref{navc}) and (\ref{naavc}), reduces to:
\be
r \gg 2.25\times 10^{-4}\gz\half \,.\label{rbt}
\ee
This lower bound on the $r$ parameter has to be considered when building a realistic
particle physics model of the vector curvaton scenario.

Second, looking at \eq{tnl13}, we obtain the level of non-gaussianity $\tnl$ for this scenario:
\be
\tnl\simeq \frac{2\times 10^{-2}}{r^2}\(\frac{\gz}{0.1}\)^3 \,.
\ee
This is a consistency relation between $\tnl$, $g_\zeta$, and $r$ which will help when
confronting the specific vector curvaton realisation against observation.  Indeed, a similar
consistency relation between $\fnl$ and $g_\zeta$ was derived for this scenario in \ref{fnlgr}
\cite{vrl}:
\be
\fnl\simeq \frac{4.5\times 10^{-2}}{r}\(\frac{\gz}{0.1}\)^2 \,.
\ee
Thus, in the framework of the vector curvaton scenario, the levels of non-gaussianity $\fnl$
and $\tnl$ are related to each other via the $r$ parameter in this way:
\be
\tnl \simeq \frac{2.1}{r^{1/2}} \fnl^{3/2} \,,
\ee
in contrast to the standard result
\begin{equation}
\tnl = \frac{36}{25} \fnl^2 \,, \label{relbyrnes}
\end{equation}
for the scalar field case (including the scalar curvaton scenario) found in Ref. \cite{bsw1}.
%%%%%%%%%%%%%%%%%%%%%%%%%%%%%%%%%%%%%%%%%%%%%%%%%%%%%%%%%%%%%%%%%%%%%%%%%%%%%%%%%%%%%%%%%%%
\subsection{Vector field spectrum ($\calp_{\zeta_A}$) and trispectrum ($\calt_{\zeta_A}$)
dominated by the one-loop contributions}
%%%%%%%%%%%%%%%%%%%%%%%%%%%%%%%%%%%%%%%%%%%%%%%%%%%%%%%%%%%%%%%%%%%%%%%%%%%%%%%%%%%%%%%%%%%
From Eqs. (\ref{tsploop}) and (\ref{tnl2}) we get
\be
\tnl\:\simeq\:\frac{N_{AA}^{ij}N_{AA}^{kl}N_{AA}^{mn}N_{AA}^{op}\ln(kL)\big(2\calp_++\calp_{long})\delta_{im}\big[\calt_{jk}
(\bfk_1)\calt_{np}(\bfk_3)\calt_{lo}(|\bfk_4+\bfk_3|)\big]}{\big[\calpz^{\rm iso}(k_1)\calpz^{\rm iso}(k_2)\calpz^{\rm iso}(|
\bfk_1+\bfk_4|)+{\rm 23\  perm.}\big]} \,.
\label{tnl21}
\ee

Assuming again that $\calp_{\rm long}$, $\calp_+$, and $\calp_-$ are all of the same order of
magnitude, and that the spectrum is scale invariant, we end up with:
\be
\tnl\:\simeq\:\frac{\calp_A^4 N_{AA}^4}{(\calpz^{\rm iso})^3} \,.
\label{tnl22}
\ee
Performing a similar analysis as done in the previous subsection, but this time taking into
account that the vector field spectrum is dominated by the one-{\rm loop} contribution and
therefore $\zeta_A\sim \sqrt{\calpz_A}\sim
N_{AA} \calp_A$, %Now, remembering that $\gz\sim\calpz_A/\calpz$ and $\calpz\half\simeq
%5\times 10 ^{-5}$ \cite{wmap5},
we arrive at:
\be
\tnl\:\sim\:\frac{\calpz_A^2}{(\calpz^{\rm iso})^3}\:\sim\:8\times10^6\(\frac{\gz}{0.1}\)^2.
\label{tnl23}
\ee
The above result gives a relation between the non-gaussianity parameter $\tnl$ and the
level of statistical anisotropy in the power spectrum $\gz$.

Now, we call a similar result that we found for the
non-gaussianity parameter $\fnl$ in \eq{fnl13}, that is:
\be
\fnl\:\sim\:10^3\(\frac{\gz}{0.1}\)\threehalf.
\label{fnlv}
\ee
By combining Eqs. (\ref{tnl23}) and (\ref{fnlv}) we get:
\be
\tnl\:\sim\:8\times10^2\fnl^{4/3},\label{tnlfnl}
\ee
which gives a consistency relation between the non-gaussianity parameters $\fnl$ and $\tnl$ for this
particular scenario. The consistency relations in Eqs. (\ref{tnl23}), (\ref{fnlv}), and
(\ref{tnlfnl}) will put under test this scenario against future observations. In particular,
the consistency relation in Eq. (\ref{tnlfnl}) differs significantly from those obtained when
$\zeta$ is generated only by scalar fields (see e.g. Eq. (\ref{relbyrnes}) and Ref. \cite{bsw1}).

Again when we apply our result to the vector curvaton scenario, we get from Eqs. (\ref{Pzetat}), (\ref{tst}),
(\ref{sploop}), (\ref{tsploop}), (\ref{navc}) and (\ref{naavc}) :
\be
r <2.25\times 10^{-4}\gz\half \,, \label{rbl}
\ee
which is an upper bound on the $r$ parameter that must be considered when building a realistic
particle physics model of the vector curvaton scenario.

As happens in Sections \ref{ptb1l} and \ref{p1lbt}, the other two possibilities are not viable because it is impossible to 
satisfy simultaneously that the vector field spectrum ($\calp_{\zeta_A}$) is dominated by the tree-level terms and
the trispectrum ($\calt_{\zeta_A}$) is dominated by the one-loop contributions, or the vector field spectrum
($\calp_{\zeta_A}$) is dominated by the one-loop contributions and the trispectrum ($\calt_{\zeta_A}$) is
dominated by the tree-level terms.
%%%%%%%%%%%%%%%%%%%%%%%%%%%%%%%%%%%%%%%%%
\section{Conclusions}	    %%%%%%%%%%%%%
%%%%%%%%%%%%%%%%%%%%%%%%%%%%%%%%%%%%%%%%%
We have studied in this chapter the order of magnitude of the levels of non-gaussianity $\fnl$ and $\tnl$ in the bispectrum 
$\bz$ and  in the trispectrum $\tz$, respectively, when statistical anisotropy is generated by the presence of one
vector field. Particularly, we have shown that it is possible to get an upper bound on the order of magnitude of
$\fnl$ if we assume that tree level contributions on $\calpz_A$ and $\calbz_A$ domiante over all other terms, this result
given in the \eq{fnlm5}. On the other hand if we assume that the 1-loop contributions dominate over the tree-level terms in 
both the vector field spectrum ($\calp_{\zeta_A}$) and the bispectrum ($\calb_{\zeta_A}$) a high level of non-gaussianity 
$\fnl$ is obtained. $\fnl$ is given in this case by \eq{fnl13}, where we may see that 
there is a consistency  relation between $\fnl$ and the amount of statistical anisotropy in the spectrum $\gz$. We also have 
shown that it is possible to get an upper bound on the order of magnitude of $\tnl$ if we assume that the tree-level 
contributions dominate over all higher order terms in both the vector field spectrum ($\calp_{\zeta_A}$) and the trispectrum
($\calt_{\zeta_A}$); this bound is given in \eq{tnl15}. We also found that it is possible to
get a high level of non-gaussianity $\tnl$, easyly exceeding the expected observational
bound from WMAP, if we assume that the one-loop contributions dominate over the tree-level
terms in both the vector field spectrum ($\calp_{\zeta_A}$) and the trispectrum
($\calt_{\zeta_A}$). $\tnl$ is given in this case by \eq{tnl23}, where we may see that there
is a consistency relation between the order of magnitude of $\tnl$ and the amount of
statistical anisotropy in the spectrum $\gz$. Two other consistency relations are given by
Eqs. (\ref{fnlv}) and (\ref{tnlfnl}), this time relating the order of magnitude of the
non-gaussianity parameter $\fnl$ in the bispectrum $\bz$ with the amount of statistical
anisotropy $g_\zeta$ and the order of magnitude of the level of non-gaussianity $\tnl$ in the
trispectrum $\tz$.  Such consistency relations let us fix two of the three parameters by
knowing about the other one, i.e. if the non-gaussianity in the bispectrum (or trispectrum)
is detected and our scenario is appropriate, the amount of statistical anisotropy in the
power spectrum and the order of magnitude of the non-gaussianity parameter $\tnl$ (or $\fnl$)
must have specific values, which are given by Eqs. (\ref{fnlv}) (or (\ref{tnl23})) and
(\ref{tnlfnl}). A similar conclusion is reached if the statistical anisotropy in the
power spectrum is detected before the non-gaussianity in the bispectrum or the trispectrum is.
%%%%%%%%%%%%%%%%%%%%%%%%%%%%%%%%%%%%%%%%%%%%%%%%%
%%%%%%%%%%%%%%%%%%%%%%%%%%%%%%%%%%%%%%%%%%%%%%%%%
\chapter{CONCLUSIONS}\label{chaptconclu}      %%%
%%%%%%%%%%%%%%%%%%%%%%%%%%%%%%%%%%%%%%%%%%%%%%%%%
%%%%%%%%%%%%%%%%%%%%%%%%%%%%%%%%%%%%%%%%%%%%%%%%%
Observational cosmology is in its golden age: current satellite and balloon experiments are working
extremely well \cite{hinshaw,wmap}, dramatically improving the quality of data \cite{wmap5}.
Moreover, foreseen experiments \cite{planck,planck1} will take the field to a state of unprecedent precission
where theoretical models will be subjected to the most demanding tests. Given such a state of affairs, it is
essential to study the higher order statistical descriptors for cosmological quantities such as the primordial
curvature perturbation $\zeta$, which give us information about the non-gaussianity and about the possible violations
of statistical isotropy in their corresponding probability distribution functions.

The slow-roll class of inflationary models with canonical kinetic terms
are the most popular and studied to date. Inflationary models of the slow-roll variety predict very well
the spectral index in the spectrum $P_\zeta$ of $\zeta$ but, if the kinetic terms are canonical, they seem to
generate unobservable levels of non-gaussianity in the bispectrum $B_\zeta$ and the trispectrum $T_\zeta$ of
$\zeta$ making them impossible to test against the astonishing forthcoming data. Where does this conclusion
come from? The answer relies on careful calculations of the levels of non-gaussianity $f_{NL}$ and $\tau_{NL}$
by making use of the $\delta N$ formalism \cite{battefeld,seery3,vernizzi,yokoyama1}. In this framework, $\zeta$
is given in terms of the perturbation $\delta N$ in the amount of expansion from the time the cosmologically
relevant scales exit the horizon until the time at which one wishes to calculate $\zeta$.

Due to the functional dependence of the amount of expansion, $\zeta$ is usually Taylor-expanded (see Eq. (\ref{Nexp}))
and truncated up to some desired order so that $f_{NL}$ and $\tau_{NL}$ are easily calculated (see for instance Eq.
(\ref{fdnf})). Two key questions arise when noting that it is impossible to extract general and useful information
from the $\zeta$ series expansion in Eq. (\ref{Nexp}) until one chooses a definite inflationary model and calculates
 explicitly the $N$ derivatives. First of all, when writing a general expression for $f_{NL}$ or $\tau_{NL}$ in terms
of the $N$ derivatives, how do we know that such an expression is correct if the series convergence has not been examined?
Moreover, if the convergence radius of the $\zeta$ series is already known, why is each term is the $\zeta$ series
supposed to be smaller than the previous one so that cutting the series at any desired order is thought to be enough
to keep the leading terms? Nobody seems to have formulated these questions before and, by following a naive line of
thinking, $f_{NL}$ and $\tau_{NL}$ were calculated for slow-roll inflationary models with canonical kinetic terms
without checking the $\zeta$ series convergence and keeping only the presumably leading tree-level terms
\cite{battefeld,maldacena,seery3,seery7,vernizzi,yokoyama1}.

These two questions have been addressed in this thesis (see Chapt. 3 and 4) by paying attention to a particular quadratic 
small-field slow-roll model of inflation with two components and canonical kinetic terms (see Eq. (\ref{pot})). Although
the non-diagrammatic approach followed in Section \ref{seccou} to find the necessary condition for the convergence
of the $\zeta$ series in our model might not be applicable to all the cases, we have been able to show that not
being careful enough when choosing the right available parameter space could make the $\zeta$ series, and therefore
the calculation of $f_{NL}$ and $\tau_{NL}$ from the truncated series (e.g. Eq. (\ref{fdnf})), meaningless. We also
have been able to show in our model that the one-loop terms in the spectrum $P_\zeta$, the bispectrum $B_\zeta$
and trispectrum $\tz$ of $\zeta$ could be bigger or lower than the corresponding tree-level terms, but are always much bigger 
than the corresponding terms whose order is higher than the one-loop order. If $B_\zeta$ is dominated by the one-loop
correction but $P_\zeta$ is dominated by the tree-level term, {\it sizeable and observable values for} $f_{NL}$
{\it are generated}, so they can be tested against present and forthcoming observational data, a similar 
conclusion was reached when the trispectrum is dominated by one-loop corrections and the $P_\zeta$ is dominated by the tree-
level term. Finally, if both $P_\zeta$ and $B_\zeta$ or $\tz$ are dominated by the tree-level terms, $f_{NL}$ or $\tnl$ {\it 
are slow-roll suppressed} (see Eqs. \ref{fnlslowroll} and \ref{taonlslowroll}) as was originally predicted in Refs. 
\cite{battefeld,vernizzi,yokoyama1}. What these results teach us is that the issue of the $\zeta$ series convergence and loop 
corrections is essential for making correct predictions about the statistical descriptors of $\zeta$ in the framework of the 
$\delta N$ formalism, and promising for finding high levels of non-gaussianity that can be compared with observations. 

The above disccusion about $\zeta$ was made assuming that the $n$-point correlators of $\zeta$ are tranlationally and 
rotationally invariant. However as we could see in the section \ref{obsvec}, violations of the translational (rotational) 
invariance (i.e. violations of the statistical homogeneity (isotropy)) seem to be present in the data
\cite{dipole2,dipole1,hansen,dipole3,hoftuft,hou} (\cite{app,gawe,ge,hl,samal}); therefore it is pertinent to study theoretical 
models 
that include those violations. This is the reason why in the chapter \ref{chaptvec} we studied the statistical descriptors for 
$\zeta$ for models with vector field perturbations, which are responsible of violations of statistcal isotropy. We 
studied in that chapter the order of magnitude of the levels of non-gaussianity $\fnl$ and $\tnl$ in the bispectrum $\bz$ 
and in the trispectrum $\tz$, when statistical anisotropy is generated by the presence of one massive vector field. 
We have shown that it is possible to get an upper bound on the order of magnitude of $\fnl$ (see \eq{fnlm5}) and $\tnl$ (see 
\eq{tnl15}) if we assume that the tree-level contributions dominate over all higher order terms in both the vector field 
spectrum ($\calp_{\zeta_A}$), the bispectrum ($\calbz_A$) and trispectrum ($\calt_{\zeta_A}$). We also show that it is possible 
to get high levels of non-gaussianity $\fnl$ and $\tnl$, easily exceeding the expected observational
bounds from WMAP, if we assume that the one-loop contributions dominate over the tree-level
terms in both the vector field spectrum ($\calp_{\zeta_A}$) and the bispectrum
($\calb_{\zeta_A}$) or in both the vector field spectrum ($\calp_{\zeta_A}$) and the trispectrum
($\calt_{\zeta_A}$). We could see that there are a consistency 
relations between the order of magnitude of $\fnl$ and 
the amount of statistical anisotropy in the spectrum $\gz$ [\eq{fnl13}] and between the order of magnitude $\tnl$ and $\gz$ 
[\eq{tnl23}]. Two other 
consistency relations are given by Eqs. (\ref{fnlv}) and (\ref{tnlfnl}), this time relating the order of magnitude of the
non-gaussianity parameter $\fnl$ in the bispectrum $\bz$ with the amount of statistical
anisotropy $g_\zeta$ and the order of magnitude of the level of non-gaussianity $\tnl$ in the
trispectrum $\tz$.  Such consistency relations let us fix two of the three parameters by
knowing about the other one, i.e. if the non-gaussianity in the bispectrum (or trispectrum)
is detected and our scenario is appropriate, the amount of statistical anisotropy in the
power spectrum and the order of magnitude of the non-gaussianity parameter $\tnl$ (or $\fnl$)
must have specific values, which are given by Eqs. (\ref{fnlv}) (or (\ref{tnl23})) and
(\ref{tnlfnl}). A similar conclusion is reached if the statistical anisotropy in the
power spectrum is detected before the non-gaussianity in the bispectrum or the trispectrum is.

\appendix
%%%%%%%%%%%%%%%%%%%%%%%%%%%%%%%%%%%%%%%%%%%%%%%%%%%%%%%%%%%%%%%%%%%%%%%%%%%%%%%%%%%%%%%%%%%%%%%%%%%%%%%%%%%%%%%%%%%%%%%%%
%%%%%%%%%%%%%%%%%%%%%%%%%%%%%%%%%%%%%%%%%%%%%%%%%%%%%%%%%%%%%%%%%%%%%%%%%%%%%%%%%%%%%%%%%%%%%%%%%%%%%%%%%%%%%%%%%%%%%%%%%
\chapter{TREE-LEVEL AND ONE-LOOP DIAGRAMS FOR $P_\zeta$, $B_\zeta$ AND $\tz$ : SCALAR FIELDS} \label{app}           %%%%%
%%%%%%%%%%%%%%%%%%%%%%%%%%%%%%%%%%%%%%%%%%%%%%%%%%%%%%%%%%%%%%%%%%%%%%%%%%%%%%%%%%%%%%%%%%%%%%%%%%%%%%%%%%%%%%%%%%%%%%%%%
%%%%%%%%%%%%%%%%%%%%%%%%%%%%%%%%%%%%%%%%%%%%%%%%%%%%%%%%%%%%%%%%%%%%%%%%%%%%%%%%%%%%%%%%%%%%%%%%%%%%%%%%%%%%%%%%%%%%%%%%%
We show in this appendix the mathematical expressions for the tree-level and one-loop Feynman-like diagrams associated with the
spectrum $P_\zeta$, the bispectrum $B_\zeta$ an trispectrum of $\zeta$, following the set of rules presented in Ref. 
\cite{byrnes1}. To this end we have taken into account the $N$ derivatives for our small-field slow-roll model given in Eqs. 
(\ref{1d}), (\ref{2d}), and (\ref{3d}).  After presenting the mathematical expressions, we will estimate the order of magnitude 
of each diagram in order to determine the respective leading terms at tree-level and one-loop for both $P_\zeta$ and $B_\zeta$.
%%%%%%%%%%%%%%%%%%%%%%%%%%%%%%%%%%%%%%%%%%%%%%%%%%%%%%%%%%%%%%%%%%%%%%%%%%%%%%%%%%
\section{Tree-level diagram for $P_\zeta$} \label{diagramsP}			%%%%%%%%%%
%%%%%%%%%%%%%%%%%%%%%%%%%%%%%%%%%%%%%%%%%%%%%%%%%%%%%%%%%%%%%%%%%%%%%%%%%%%%%%%%%%
\begin{figure}
\begin{center}
\includegraphics[width=7cm]{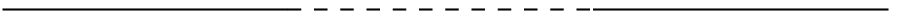}
\end{center}
\caption[Tree-level Feynman-like diagram for $P_\zeta$]{Tree-level Feynman-like diagram for $P_\zeta$. The internal dashed line corresponds to a
two-point correlator of field perturbations.} \label{tlpd}
\end{figure}

Looking at Fig. \ref{tlpd}, we see that $P_\zeta^{tree}$ is given by
\begin{eqnarray}
P_\zeta^{tree} &=& N_\phi^2 \ P_{\delta \phi} (k) \nonumber \\
&=& \frac{2\pi^2}{k^3} \frac{1}{\eta_\phi^2 \phi_\star^2} \left(\frac{H_\star}{2\pi}\right)^2 \,. \label{pt1}
\end{eqnarray}
Of course, there is only one tree-level diagram for $P_\zeta$ and therefore Eq. (\ref{pt1}) is the associated leading tree-
level term.

Our calculation in this appendix goes up to the one-loop diagrams so, in order to have complete consistency in the calculation
\cite{seery2}, we should also take into account the one-loop correction to the two-point correlator in the field perturbations
when calculating the diagram in Fig. \ref{tlpd}. Such a correction has been studied in Refs.
\cite{seery1,sloth1,sloth2,weinberg1,weinberg2} where the most general result for single-field slow-roll inflation with
$N_{total}$ not very much bigger than 62 is \cite{seery1}
\begin{equation}
P_{\delta \phi}^{1-loop} = \frac{2\pi^2}{k^3} \left( \frac{H_\star}{2\pi}\right)^2 \left \{1 + \left(\frac{H_\star}{2\pi
m_P}\right)^2 \left[\frac{35}{6} \ln(kL) + \beta \right] \right \} \,, \label{q1l}
\end{equation}
where $L$ is the infrared cutoff for a minimal box \cite{bernardeu4,lythbox}, and $\beta$ is a renormalisation scheme-dependent
constant that is expected to be negligible on large scales compared to $\ln (kL) \sim \mathcal{O}(1)$.  The one-loop correction
to the field perturbation spectrum in Eq. (\ref{q1l}) is, therefore, negligible compared to the tree-level contribution
$P_{\delta \phi}^{tree} = (2\pi^2/k^3) (H_\star/2\pi)^2$ if $H_\star \ll m_P$ as usually required.  In our model $H_\star \ll
m_P$ is indeed given but, since we are dealing with a two-component model, the result in Eq. (\ref{q1l}) may not be applicable.
Anyway, we feel quite confident that the (up to now unknown) extension of Eq. (\ref{q1l}) to the multiple-field case will yield
similar results, so we will keep the expression in Eq. (\ref{pt1}) as the leading tree-level contribution to $P_\zeta$.
%%%%%%%%%%%%%%%%%%%%%%%%%%%%%%%%%%%%%%%%%%%%%%%%%%%%%%%%%%%%%%%%%%%%%%%%%%%%%%%%%%
\section{One-loop diagrams for $P_\zeta$} \label{Ldis}				%%%%%%%%%%%
%%%%%%%%%%%%%%%%%%%%%%%%%%%%%%%%%%%%%%%%%%%%%%%%%%%%%%%%%%%%%%%%%%%%%%%%%%%%%%%%%%
\begin{figure}
\begin{center}
\begin{tabular}{cccc}
\includegraphics[width=7cm,height=1cm]{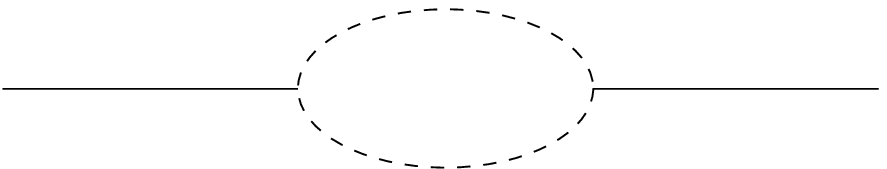} & & & \includegraphics[width=7cm,height=1cm]{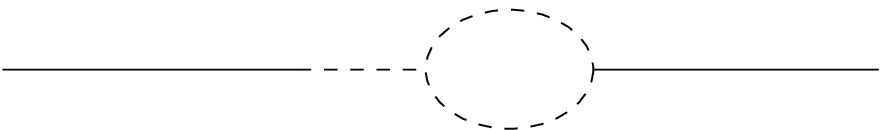} \\
(a) & & & (b)
\end{tabular}
\end{center}
\caption[One-loop Feynman-like diagrams for $P_\zeta$]{One-loop Feynman-like diagrams for $P_\zeta$. (a). The two internal dashed lines
correspond to two-point correlators of field perturbations. (b). The internal dashed lines correspond to a three-point correlator of field
perturbations.}
\label{olpd}
\end{figure}

Looking at Figs. \ref{olpd}a and \ref{olpd}b, we see that $P_\zeta^{1-loop}$ is given by two contributions $P_\zeta^{1-loop \;
a}$ and $P_\zeta^{1-loop \; b}$:
{\small\begin{eqnarray}
P_\zeta^{1-loop \; a} &=& \frac{1}{2} \left[N_{\phi \phi}^2 + N_{\sigma \sigma}^2\right] \int \frac{d^3 q}{(2\pi)^3} P_{\delta
\phi}(q) P_{\delta \phi} (|{\bf k} + {\bf q}|) \nonumber \\
&=& \frac{1}{2} \left[\frac{1}{\eta_\phi^2 \phi_\star^4} + \frac{\eta_\sigma^2}{\eta_\phi^4 \phi_\star^4}
\exp\left[4N(\eta_\phi - \eta_\sigma)\right]\right] \frac{4\pi^2}{k^3} \ln(kL) \left(\frac{H_\star}{2\pi}\right)^4 \,,
\label{1la} \\
P_\zeta^{1-loop \; b} &=& N_\phi N_{\phi \phi} \int \frac{d^3 q}{(2\pi)^3} B_{\delta \phi \ \delta \phi \ \delta \phi} (k,q,|
{\bf k} + {\bf q}|) + \nonumber \\
&& + N_\phi N_{\sigma \sigma} \int \frac{d^3 q}{(2\pi)^3} B_{\delta \phi \ \delta \sigma \ \delta \sigma} (k,q,|{\bf k} + {\bf
q}|) \nonumber \\
&=& -\frac{1}{\eta_\phi^2 \phi_\star^3} \left[\int \frac{d^3 q}{(2\pi)^3} 4\pi^4 \sum_{perm} \left(\frac{H_\star}
{2\pi}\right)^4 \frac{\epsilon_\phi^{1/2}}{2\sqrt{2} m_P} \frac{\mathcal{M}_3(k,q,|{\bf k} + {\bf q}|)}{k^3 q^3 |{\bf k} + {\bf
q}|^3} \right] + \nonumber \\
&& + \frac{\eta_\sigma}{\eta_\phi^3 \phi_\star^3} \exp\left[2N(\eta_\phi - \eta_\sigma)\right] \left[\int \frac{d^3 q}
{(2\pi)^3} 4\pi^4 \sum_{perm. \ l2a.} \left(\frac{H_\star}{2\pi}\right)^4 \frac{\epsilon_\phi^{1/2}}{2\sqrt{2} m_P}
\frac{\mathcal{M}_3(k,q,|{\bf k} + {\bf q}|)}{k^3 q^3 |{\bf k} + {\bf q}|^3} \right] \,, \nonumber \\
&& \label{1lb}
\end{eqnarray}}
where the $\ln(kL) \sim \mathcal{O}(1)$ factor comes from the evaluation of the momentum integrals in a minimal box
\cite{bernardeu4,klv,lythbox}, the $\mathcal{M}_3 (k_1,k_2,k_3)$ function is defined by \cite{seery5}
\begin{equation}
\mathcal{M}_3 (k_1,k_2,k_3) = -k_1 k_2^2 - 4 \frac{k_2^3 k_3^3}{k_t} + \frac{1}{2} k_1^3 + \frac{k_2^2 k_3^2}{k_t^2} (k_2-k_3)
\,,
\end{equation}
with $k_t = k_1+k_2+k_3$,
and the subindex $perm. \ l2a.$ means a permutation over the last two arguments in $\mathcal{M}_3$.

A quick glance reveals that the first term in Eq. (\ref{1la}) is subleading with respect to the second one because $|
\eta_\sigma| > |\eta_\phi|$ and $\exp[4N(\eta_\phi - \eta_\sigma)] \gg 1$. The same is true for Eq. (\ref{1lb}) where
$\exp[2N(\eta_\phi - \eta_\sigma)] \gg 1$. Now, by comparing the orders of magnitude of the leading terms in Eqs. (\ref{1la})
and (\ref{1lb}), we conclude that:
\begin{eqnarray}
\frac{P_\zeta^{1-loop \; a}}{P_\zeta^{1-loop \; b}} &\sim& \frac{\frac{\eta_\sigma^2}{\eta_\phi^4 \phi_\star^4} \exp \left[4N
(\eta_\phi - \eta_\sigma)\right] \left(\frac{H_\star}{2\pi}\right)^4 \frac{2\pi^2}{k^3}}{\frac{\eta_\sigma}{\eta_\phi^3
\phi_\star^3} \exp \left[2N (\eta_\phi - \eta_\sigma)\right] \left(\frac{H_\star}{2\pi}\right)^4 \frac{\epsilon_\phi^{1/2}}
{m_P} \frac{2\pi^2}{k^3}} \nonumber \\
&=& \frac{\eta_\sigma}{\eta_\phi} \frac{m_P}{\phi_\star} \exp \left[2N(\eta_\phi - \eta_\sigma)\right] \frac{1}
{\epsilon_\phi^{1/2}} \gg 1 \,,
\end{eqnarray}
where $m_P \gg \phi_\star$ and $\epsilon_\phi \ll 1$. Thus, the one-loop leading term for $P_\zeta$ in our model is given by
\begin{equation}
P_\zeta^{1-loop} =\frac{2\pi^2}{k^3}  \frac{\eta_\sigma^2}{\eta_\phi^4 \phi_\star^4} \exp\left[4N(\eta_\phi -
\eta_\sigma)\right]  \left(\frac{H_\star}{2\pi}\right)^4 \ln(kL) \,. \label{1lfpd}
\end{equation}

Having presented the leading tree-level and one-loop contributions to $P_\zeta$ in Eqs. (\ref{pt1}) and (\ref{1lfpd}), a
consistency issue to think about is the dependence of the expression in Eq. (\ref{q1l}) on the infrared cutoff $L$. This
quantity is in principle an artefact of the series expansion, and the final series result should in principle be independent on
the chosen value for $L$ (see for instance Ref. \cite{riotto}).  In fact, by assuming that this is the case, Refs.
\cite{bartolo,enqvist2,lythbox} have shown that there is a running on the $N$ derivatives with respect to $L$ so that changes 
in the $\ln (kL)$ factors are compensated by the running of the $N$ derivatives. This is similar to what happens in Quantum 
Field
Theory where physical results independent on the energy scale must be independent of the chosen value for the renormalisation
scale $Q$.  Changing $Q$ only modifies the relative weight of the tree-level and loop contributions, usually making the tree-
level terms dominate over the loop corrections if $Q$ is chosen around the relevant energy scale of the process studied.
Nevertheless, we see that the $\ln (kL)$ term in Eq. (\ref{q1l}) does not compensate for the $\ln(kL)$ term in Eq.
(\ref{1lfpd}), %and similarly for the bispectrum,
which is a real concern as we could expect since $\zeta$ and its spectral functions are a set of observables. The solution to
this paradox relies on the fact that the observed $\zeta$ depends on $L$ as the stochastic properties of the distributions
depend on the size of the available region in which we are actually able to perform observations. In this regard $\zeta$ is
analogous to for instance the fine structure constant in Quantum Field Theory which, being an observable, depends on the energy
scale for which experiments are done and, therefore, on $Q$. Likewise, $\zeta$ and its spectral functions, though being
observables, depend on the size of the regions where observations are done and, therefore, on $L$. Having this in mind it is
essential to work in a minimal box \cite{bartolo}, i.e. with $L$ a bit bigger than $H_0^{-1}$ (with the subscript 0 meaning
today), so that $\ln (kL) \sim \mathcal{O} (1)$ as has been done throughout this thesis.
%%%%%%%%%%%%%%%%%%%%%%%%%%%%%%%%%%%%%%%%%%%%%%%%%%%%%%%%%%%%%%%%%%%%%%%%%%%%%%%%%%
\section{Tree-level diagrams for $B_\zeta$}                         %%%%%%%%%%%%%%
%%%%%%%%%%%%%%%%%%%%%%%%%%%%%%%%%%%%%%%%%%%%%%%%%%%%%%%%%%%%%%%%%%%%%%%%%%%%%%%%%%
\begin{figure}
\begin{center}
\begin{tabular}{cccc}
\includegraphics[width=7cm,height=2cm]{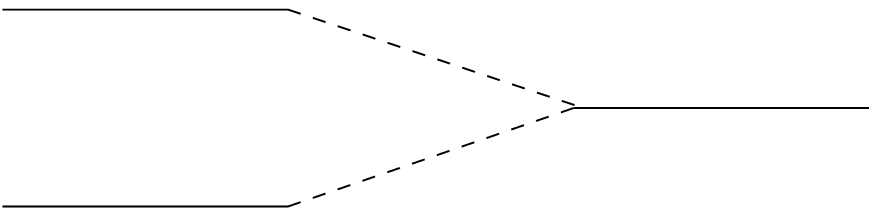} & & & \includegraphics[width=7cm,height=2cm]{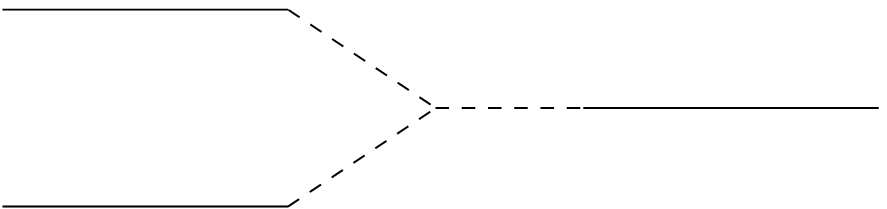} \\
(a) & & & (b)
\end{tabular}
\end{center}
\caption[Tree-level Feynman-like diagrams for $B_\zeta$]{Tree-level Feynman-like diagrams for $B_\zeta$. (a). The two internal dashed lines
correspond to two-point correlators of field perturbations. (b). The internal dashed lines correspond to a three-point correlator of field
perturbations.}
\label{tlbd}
\end{figure}

Looking at Figs. \ref{tlbd}a and \ref{tlbd}b, we see that $B_\zeta^{tree}$ is given by two contributions $B_\zeta^{tree \; a}$
and $B_\zeta^{tree \; b}$:
\begin{eqnarray}
B_\zeta^{tree \; a} &=& N_\phi^2 N_{\phi \phi} \left[P_{\delta \phi} (k_1) \ P_{\delta \phi} (k_2) + 2 \ {\rm permutations}
\right] \nonumber \\
&=& - \frac{1}{\eta_\phi^3 \phi_\star^4} \left(\frac{\sum_i k_i^3}{\prod_i k_i^3}\right) 4\pi^4 \left(\frac{H_\star}
{2\pi}\right)^4 \,. \label{tlba} \\
B_\zeta^{tree \; b} &=& N_\phi^3 B_{\delta \phi \ \delta \phi \ \delta \phi} (k_1,k_2,k_3) \nonumber \\
&=& \frac{1}{\eta_\phi^3 \phi_\star^3} 4\pi^4 \sum_{perm} \left(\frac{H_\star}{2\pi}\right)^4 \frac{\epsilon_\phi^{1/2}}
{2\sqrt{2}m_P} \frac{\mathcal{M}_3(k_1,k_2,k_3)}{\prod_i k_i^3} \,. \label{tlbb}
\end{eqnarray}

Now, from comparing the order of magnitude of the expressions in Eqs. (\ref{tlba}) and (\ref{tlbb}), we conclude that:
\begin{eqnarray}
\frac{B_\zeta^{tree \; a}}{B_\zeta^{tree \; b}} &\sim& \frac{\frac{1}{\eta_\phi^3 \phi_\star^4} \left(\frac{\sum_i k_i^3}
{\prod_i k_i^3}\right) 4\pi^4 \left(\frac{H_\star}{2\pi}\right)^4}{\frac{1}{\eta_\phi^3 \phi_\star^3} 4\pi^4 \sum_{perm}
\left(\frac{H_\star}{2\pi}\right)^4 \frac{\epsilon_\phi^{1/2}}{m_P} \left(\frac{\sum_i k_i^3}{\prod_i k_i^3}\right)} \nonumber
\\
&=& \frac{m_P}{\phi_\star} \frac{1}{\epsilon_\phi^{1/2}} \gg 1 \,,
\end{eqnarray}
which in fact is usual as demonstrated in Refs. \cite{lyth7,vernizzi}. Thus, the tree-level leading term for $B_\zeta$ in our
model is given by:
\begin{equation}
B_\zeta^{tree} = - \frac{1}{\eta_\phi^3 \phi_\star^4} \left(\frac{H_\star}{2\pi}\right)^4  4\pi^4 \left(\frac{\sum_i k_i^3}
{\prod_i k_i^3}\right)\,. \label{tlbd1}
\end{equation}

As was done for $P_\zeta$ in Subsection \ref{diagramsP}, %being our calculation at one loop, we must be consistent and take
% into account
the one-loop correction to the spectrum of the field perturbations must be taken into account for the sake of consistency when
calculating the contribution associated to the diagram in Fig. \ref{tlbd}a.  The discussion about the relevance of this quantum
one-loop correction is actually the same as in Subsection \ref{diagramsP} and, therefore, we may conclude with some confidence
that the expression in Eq. (\ref{tlba}) is reliable.  As regards the diagram in Fig. \ref{tlbd}b, it is necessary to include
the one-loop correction the three-point correlator of the field perturbations in Eq. (\ref{tlbb}), which in fact nobody has
calculated yet even for the single-field case.  Nevertheless we might conjecture that, analogously to that for the $P_\zeta$
case, such a correction is negligible compared to the tree-level contribution to $B_\zeta$ and, therefore, the expression in
Eq. (\ref{tlbb}) will also be reliable.
%%%%%%%%%%%%%%%%%%%%%%%%%%%%%%%%%%%%%%%%%%%%%%%%%%%%%%%%%%%%%%%%%%%%%%%%%%%%%%%%%%
\section{One-loop diagrams for $B_\zeta$}                          %%%%%%%%%%%%%%%
%%%%%%%%%%%%%%%%%%%%%%%%%%%%%%%%%%%%%%%%%%%%%%%%%%%%%%%%%%%%%%%%%%%%%%%%%%%%%%%%%%
\begin{figure}
\begin{center}
\begin{tabular}{cccc}
\includegraphics[width=7cm,height=2cm]{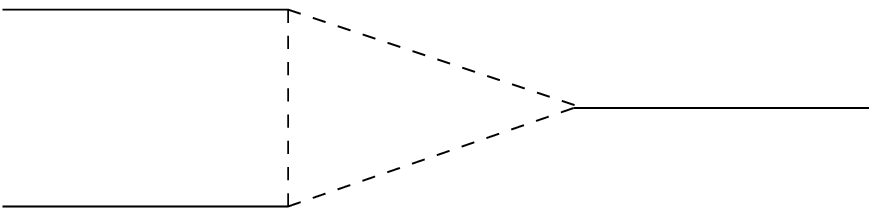} & & & \includegraphics[width=7cm,height=2cm]{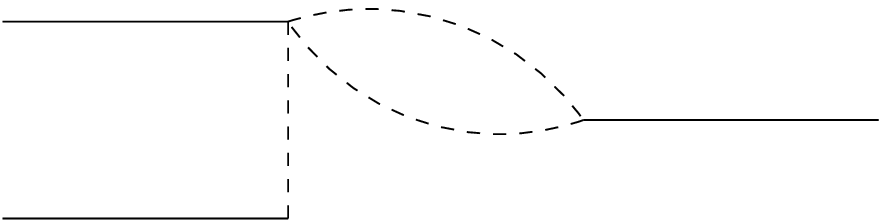} \\
(a) & & & (b) \\
& & & \\
& & & \\
%\vspace{3mm}
\includegraphics[width=7cm,height=2cm]{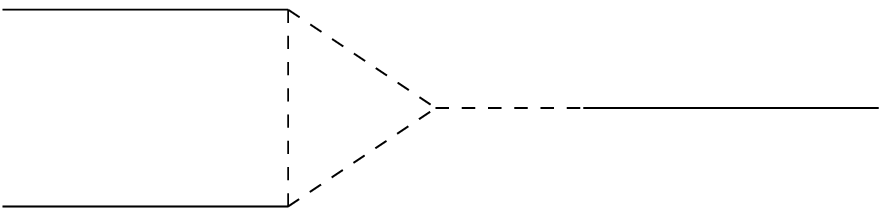} & & & \includegraphics[width=7cm,height=2cm]{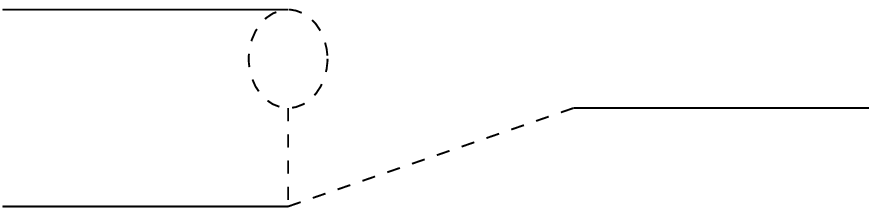} \\
(c) & & & (d) \\
%\vspace{3mm}
& & & \\
& & & \\
\includegraphics[width=7cm,height=2cm]{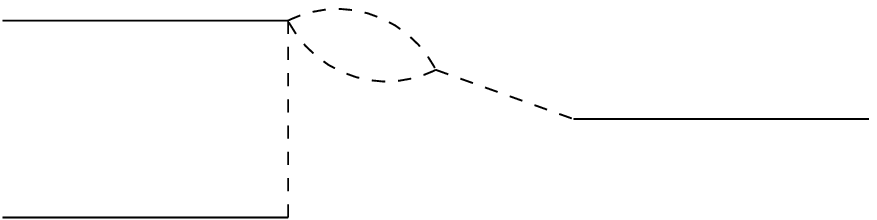} & & & \includegraphics[width=7cm,height=2cm]{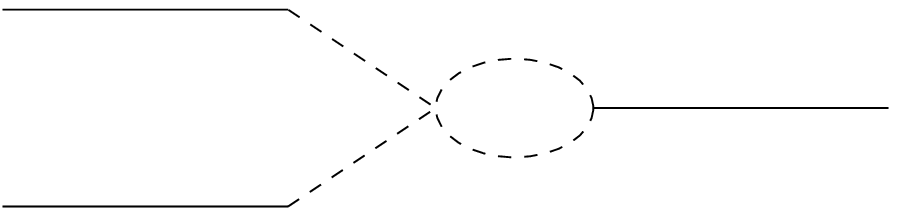} \\
(e) & & & (f)
\end{tabular}
\end{center}
\caption[One-loop Feynman-like diagrams for $B_\zeta$]{One-loop Feynman-like diagrams for $B_\zeta$. (a) and (b). The three internal dashed
lines correspond to two-point correlators of field perturbations. (c), (d), and (e). The internal dashed lines correspond to a two-point and a
three-point correlator of field perturbations. (f). The internal dashed lines correspond to a four-point correlator of field
perturbations.} \label{olbd}
\end{figure}

Looking at Figs. \ref{olbd}a, \ref{olbd}b, \ref{olbd}c, \ref{olbd}d, \ref{olbd}e, and \ref{olbd}f, we see that $B_\zeta^{tree}$
is given by six contributions $B_\zeta^{1-loop \; a}$, $B_\zeta^{1-loop \; b}$, $B_\zeta^{1-loop \; c}$, $B_\zeta^{1-loop \;
d}$, $B_\zeta^{1-loop \; e}$, and $B_\zeta^{1-loop \; f}$:
\bea
B_\zeta^{1-loop \; a} &=& \left[N_{\phi \phi}^3 + N_{\sigma \sigma}^3\right] \int \frac{d^3 q}{(2\pi)^3} P_{\delta \phi}(q)
P_{\delta \phi} (|{\bf k_1} + {\bf q}|) P_{\delta \phi} (|{\bf k_3} - {\bf q}|) \nonumber \\
&=& \left[-\frac{1}{\eta_\phi^3 \phi_\star^6} + \frac{\eta_\sigma^3}{\eta_\phi^6 \phi_\star^6} \exp\left[6N (\eta_\phi -
\eta_\sigma)\right] \right] \left(\frac{\sum_i k_i^3}{\prod_i k_i^3}\right) \ln(kL) \left(\frac{H_\star}{2\pi}\right)^6 4\pi^4
\,.  \label{poskomb} \\
B_\zeta^{1-loop \; b} &=& \frac{1}{2} \left[N_\phi N_{\phi \phi} N_{\phi \phi \phi} + N_\phi N_{\sigma \sigma} N_{\sigma \sigma
\phi} \right] \times \nonumber \\
&&\times \left[\int \frac{d^3 q}{(2\pi)^3} P_{\delta \phi}(q) P_{\delta \phi}(|{\bf k_3} - {\bf q}|) P_{\delta \phi} (k_2) + 5
\ {\rm permutations}\right] \nonumber \\
&=& \frac{1}{2} \left[-\frac{2}{\eta_\phi^3 \phi_\star^6} - \frac{2\eta_\sigma^3}{\eta_\phi^6 \phi_\star^6} \exp \left[4N
(\eta_\phi - \eta_\sigma)\right] \right] 16\pi^4 \left(\frac{\sum_i k_i^3}{\prod_i k_i^3}\right) \ln(kL) \left(\frac{H_\star}
{2\pi}\right)^6 \,. \\
B_\zeta^{1-loop \; c} &=& N_\phi N_{\phi \phi}^2 \int \frac{d^3 q}{(2\pi)^3} \left[B_{\delta \phi \ \delta \phi \ \delta \phi}
(q, |{\bf k_3} + {\bf q}|, k_3) P_{\delta \phi} (|{\bf k_1} - {\bf q}|) + 2 \ {\rm permutations} \right] + \nonumber \\
&& + N_\phi N_{\sigma \sigma}^2 \int \frac{d^3 q}{(2\pi)^3} \left[B_{\delta \sigma \ \delta \sigma \ \delta \phi} (q, |{\bf
k_3} + {\bf q}|, k_3) P_{\delta \phi} (|{\bf k_1} - {\bf q}|) + 2 \ {\rm permutations} \right] \nonumber \\
&=& \frac{1}{\eta_\phi^3 \phi_\star^5} \Big[\int \frac{d^3 q}{(2\pi)^3} 8\pi^6 \sum_{perm} \left(\frac{H_\star}{2\pi}\right)^6
\frac{\epsilon_\phi^{1/2}}{2\sqrt{2}m_P} \frac{\mathcal{M}_3(q,|{\bf k_3} + {\bf q}|,k_3)}{q^3 |{\bf k_3} + {\bf q}|^3 k_3^3}
\frac{1}{|{\bf k_1} - {\bf q}|^3} + \nonumber \\
&&+ 2 \ {\rm permutations} \Big] + \nonumber \\
&& + \frac{\eta_\sigma^2}{\eta_\phi^5 \phi_\star^5} \exp \left[4N (\eta_\phi - \eta_\sigma)\right] \times \nonumber \\
&& \times \Big[\int \frac{d^3 q}{(2\pi)^3} 8\pi^6 \sum_{perm. \ l2a.} \left(\frac{H_\star}{2\pi}\right)^6
\frac{\epsilon_\phi^{1/2}}{2\sqrt{2}m_P} \frac{\mathcal{M}_3(k_3,q,|{\bf k_3} + {\bf q}|)}{k_3^3 q^3 |{\bf k_3} + {\bf q}|^3}
\frac{1}{|{\bf k_1} - {\bf q}|^3} + \nonumber \\
&& + 2 \ {\rm permutations} \Big] \,.
\eea
\bea
B_\zeta^{1-loop \; d} &=& \frac{1}{2} N_\phi N_{\phi \phi}^2 \int \frac{d^3 q}{(2\pi)^3} \left[B_{\delta \phi \ \delta \phi \
\delta \phi} (k_3, q, |{\bf k_3} - {\bf q}|) P_{\delta \phi} (k_2) + 5 \ {\rm permutations} \right] + \nonumber \\
&& + \frac{1}{2} N_\phi N_{\phi \phi} N_{\sigma \sigma} \int \frac{d^3 q}{(2\pi)^3} \left[B_{\delta \phi \ \delta \sigma \
\delta \sigma} (k_3, q, |{\bf k_3} - {\bf q}|) P_{\delta \phi} (k_2) + 5 \ {\rm permutations} \right] \nonumber \\
&=& \frac{1}{2\eta_\phi^3 \phi_\star^5} \Big[\int \frac{d^3 q}{(2\pi)^3} 8\pi^6 \sum_{perm} \left(\frac{H_\star}{2\pi}\right)^6
\frac{\epsilon_\phi^{1/2}}{2\sqrt{2}m_P} \frac{\mathcal{M}_3(k_3,q,|{\bf k_3} - {\bf q}|)}{k_3^3 q^3 |{\bf k_3} - {\bf q}|^3}
\frac{1}{k_2^3} + \nonumber \\
&& + 5 \ {\rm permutations} \Big] - \nonumber \\
&& - \frac{\eta_\sigma}{2\eta_\phi^4 \phi_\star^5} \exp \left[2N (\eta_\phi - \eta_\sigma)\right] \times \nonumber \\
&& \times \Big[\int \frac{d^3 q}{(2\pi)^3} 8\pi^6 \sum_{perm. \ l2a.} \left(\frac{H_\star}{2\pi}\right)^6
\frac{\epsilon_\phi^{1/2}}{2\sqrt{2}m_P} \frac{\mathcal{M}_3(k_3,q,|{\bf k_3} - {\bf q}|)}{k_3^3 q^3 |{\bf k_3} - {\bf q}|^3}
\frac{1}{k_2^3} + \nonumber \\
&& + 5 \ {\rm permutations} \Big] \,. \\
B_\zeta^{1-loop \; e} &=& \frac{1}{2} N_\phi^2 N_{\phi \phi \phi} \int \frac{d^3 q}{(2\pi)^3} \left[B_{\delta \phi \ \delta
\phi \ \delta \phi} (k_1, q, |{\bf k_1} + {\bf q}|) P_{\delta \phi} (k_2) + 5 \ {\rm permutations} \right] + \nonumber \\
&& + \frac{1}{2} N_\phi^2 N_{\sigma \sigma \phi} \int \frac{d^3 q}{(2\pi)^3} \left[B_{\delta \phi \ \delta \sigma \ \delta
\sigma} (k_1, q, |{\bf k_1} + {\bf q}|) P_{\delta \phi} (k_2) + 5 \ {\rm permutations} \right] \nonumber \\
&=& \frac{1}{\eta_\phi^3 \phi_\star^5} \Big[\int \frac{d^3 q}{(2\pi)^3} 8\pi^6 \sum_{perm} \left(\frac{H_\star}{2\pi}\right)^6
\frac{\epsilon_\phi^{1/2}}{2\sqrt{2}m_P} \frac{\mathcal{M}_3(k_1,q,|{\bf k_1} + {\bf q}|)}{k_1^3 q^3 |{\bf k_1} + {\bf q}|^3}
\frac{1}{k_2^3} + \nonumber \\
&&+ 5 \ {\rm permutations} \Big] - \nonumber \\
&& - \frac{\eta_\sigma^2}{\eta_\phi^5 \phi_\star^5} \exp \left[2N (\eta_\phi - \eta_\sigma)\right] \times \nonumber \\
&& \times \Big[\int \frac{d^3 q}{(2\pi)^3} 8\pi^6 \sum_{perm. \ l2a.} \left(\frac{H_\star}{2\pi}\right)^6
\frac{\epsilon_\phi^{1/2}}{2\sqrt{2}m_P} \frac{\mathcal{M}_3(k_1,q,|{\bf k_1} + {\bf q}|)}{k_1^3 q^3 |{\bf k_1} + {\bf q}|^3}
\frac{1}{k_2^3} + \nonumber \\
&& + 5 \ {\rm permutations} \Big] \,.
\eea
\bea
B_\zeta^{1-loop \; f} &=& \frac{1}{2} N_\phi^2 N_{\phi \phi} \int \frac{d^3 q}{(2\pi)^3} \left[T_{\delta \phi \ \delta \phi \
\delta \phi \ \delta \phi} ({\bf k_1}, {\bf q}, {\bf k_3} - {\bf q}, {\bf k_2}) + 2 \ {\rm permutations} \right] + \nonumber \\
&& + \frac{1}{2} N_\phi^2 N_{\sigma \sigma} \int \frac{d^3 q}{(2\pi)^3} \left[T_{\delta \phi \ \delta \sigma \ \delta \sigma \
\delta \phi} ({\bf k_1}, {\bf q}, {\bf k_3} - {\bf q}, {\bf k_2}) + 2 \ {\rm permutations} \right] \nonumber \\
&=& - \frac{1}{2\eta_\phi^3 \phi_\star^4} \Big[\int \frac{d^3 q}{(2\pi)^3} 8\pi^6 \sum_{perm} \left(\frac{H_\star}
{2\pi}\right)^6 \frac{\mathcal{M}_4({\bf k_1},{\bf q},{\bf k_3} - {\bf q}, {\bf k_2})}{k_1^3 q^3 |{\bf k_3} - {\bf q}|^3 k_2^3}
\frac{1}{m_P^2} + \nonumber \\
&&+ 2 \ {\rm permutations} \Big] + \nonumber \\
&& + \frac{\eta_\sigma}{2\eta_\phi^4 \phi_\star^4} \exp \left[2N (\eta_\phi - \eta_\sigma)\right] \times \nonumber \\
&& \times \Big[\int \frac{d^3 q}{(2\pi)^3} 8\pi^6 \sum_{perm. \ f2a. \ l2a.} \left(\frac{H_\star}{2\pi}\right)^6
\frac{\mathcal{M}_4({\bf k_1},{\bf k_2},{\bf q},{\bf k_3} - {\bf q})}{k_1^3 k_2^3 q^3 |{\bf k_3} - {\bf q}|^3} \frac{1}{m_P^2}
+ \nonumber \\
&& + 2 \ {\rm permutations} \Big] \,,
\eea
where the subindex $perm. \ f2a. \ l2a.$ means a permutation over the first two arguments and simultaneously over the last two
arguments in $\mathcal{M}_4 ({\bf k_1},{\bf k_2},{\bf k_3},{\bf k_4})$ defined by \cite{seery4}
\begin{eqnarray}
\mathcal{M}_4 ({\bf k_1},{\bf k_2},{\bf k_3},{\bf k_4}) &=& -2 \frac{k_1^2 k_3^2}{k_{12}^2 k_{34}^2} \frac{W_{24}}{k_t}
\left[\frac{{\bf Z}_{12}\cdot{\bf Z}_{34}}{k_{34}^2} + 2{\bf k}_2 \cdot {\bf Z}_{34} + \frac{3}{4} \sigma_{12} \sigma_{34}
\right] \nonumber \\
&& -\frac{1}{2} \frac{k_3^2}{k_{34}^2} \sigma_{34} \left[\frac{{\bf k}_1 \cdot {\bf k}_2}{k_t} W_{124} + \frac{k_1^2 k_2^2}
{k_t^3} \left(2 + 6\frac{k_4}{k_t}\right) \right] \,,
\end{eqnarray}
with ${\bf k}_{ij} = {\bf k}_i + {\bf k}_j$, $k_t = k_1 + k_2 + k_3 + k_4$, and
\begin{eqnarray}
&&\sigma_{ij} = {\bf k}_i \cdot {\bf k}_j + k_j^2 \,, \\
&&{\bf Z}_{ij} = \sigma_{ij} {\bf k}_i - \sigma_{ji}{\bf k}_j \,, \\
&&W_{ij} = 1 + \frac{k_i + k_j}{k_t} + \frac{2k_ik_j}{k_t^2} \,, \\
&&W_{lmn} = 1 + \frac{k_l + k_m + k_n}{k_t} + \frac{2(k_lk_m + k_lk_n + k_mk_n)}{k_t^2} + \frac{6k_lk_mk_n}{k_t^3} \,.
\end{eqnarray}

Following the same kind of analysis as we carried out for the one-loop diagrams of $P_\zeta$ and the tree-level terms for
$B_\zeta$ we conclude the following:
\begin{eqnarray}
\frac{B_\zeta^{1-loop \; a}}{B_\zeta^{1-loop \; b}} &\sim& \frac{\frac{\eta_\sigma^3}{\eta_\phi^6 \phi_\star^6} \exp\left[6N
(\eta_\phi - \eta_\sigma)\right] \left(\frac{\sum_i k_i^3}{\prod_i k_i^3}\right) \left(\frac{H_\star}{2\pi}\right)^6 4\pi^4}
{\frac{\eta_\sigma^3}{\eta_\phi^6 \phi_\star^6} \exp \left[4N (\eta_\phi - \eta_\sigma)\right] 4\pi^4 \left(\frac{\sum_i k_i^3}
{\prod_i k_i^3}\right) \left(\frac{H_\star}{2\pi}\right)^6} \nonumber \\
&=& \exp \left[2N(\eta_\phi - \eta_\sigma)\right] \gg 1 \,,
\end{eqnarray}
\begin{eqnarray}
\frac{B_\zeta^{1-loop \; a}}{B_\zeta^{1-loop \; c}} &\sim& \frac{\frac{\eta_\sigma^3}{\eta_\phi^6 \phi_\star^6} \exp\left[6N
(\eta_\phi - \eta_\sigma)\right] \left(\frac{\sum_i k_i^3}{\prod_i k_i^3}\right) \left(\frac{H_\star}{2\pi}\right)^6 4\pi^4}
{\frac{\eta_\sigma^2}{\eta_\phi^5 \phi_\star^5} \exp \left[4N (\eta_\phi - \eta_\sigma)\right] 4\pi^4 \left(\frac{H_\star}
{2\pi}\right)^6 \frac{\epsilon_\phi^{1/2}}{m_P} \left(\frac{\sum_i k_i^3}{\prod_i k_i^3}\right)} \nonumber \\
&=& \frac{\eta_\sigma}{\eta_\phi} \frac{m_P}{\phi_\star} \exp \left[2N(\eta_\phi - \eta_\sigma)\right] \frac{1}
{\epsilon_\phi^{1/2}} \gg 1 \,,
\end{eqnarray}
\begin{eqnarray}
\frac{B_\zeta^{1-loop \; a}}{B_\zeta^{1-loop \; d}} &\sim& \frac{\frac{\eta_\sigma^3}{\eta_\phi^6 \phi_\star^6} \exp\left[6N
(\eta_\phi - \eta_\sigma)\right] \left(\frac{\sum_i k_i^3}{\prod_i k_i^3}\right) \left(\frac{H_\star}{2\pi}\right)^6 4\pi^4}
{\frac{\eta_\sigma}{\eta_\phi^4 \phi_\star^5} \exp \left[2N (\eta_\phi - \eta_\sigma)\right] 4\pi^4 \left(\frac{H_\star}
{2\pi}\right)^6 \frac{\epsilon_\phi^{1/2}}{m_P} \left(\frac{\sum_i k_i^3}{\prod_i k_i^3}\right)} \nonumber \\
&=& \left(\frac{\eta_\sigma}{\eta_\phi}\right)^2 \frac{m_P}{\phi_\star} \exp \left[4N(\eta_\phi - \eta_\sigma)\right] \frac{1}
{\epsilon_\phi^{1/2}} \gg 1 \,,
\end{eqnarray}
\begin{eqnarray}
\frac{B_\zeta^{1-loop \; a}}{B_\zeta^{1-loop \; e}} &\sim& \frac{\frac{\eta_\sigma^3}{\eta_\phi^6 \phi_\star^6} \exp\left[6N
(\eta_\phi - \eta_\sigma)\right] \left(\frac{\sum_i k_i^3}{\prod_i k_i^3}\right) \left(\frac{H_\star}{2\pi}\right)^6 4\pi^4}
{\frac{\eta_\sigma^2}{\eta_\phi^5 \phi_\star^5} \exp \left[2N (\eta_\phi - \eta_\sigma)\right] 4\pi^4 \left(\frac{H_\star}
{2\pi}\right)^6 \frac{\epsilon_\phi^{1/2}}{m_P} \left(\frac{\sum_i k_i^3}{\prod_i k_i^3}\right)} \nonumber \\
&=& \frac{\eta_\sigma}{\eta_\phi} \frac{m_P}{\phi_\star} \exp \left[4N(\eta_\phi - \eta_\sigma)\right] \frac{1}
{\epsilon_\phi^{1/2}} \gg 1 \,,
\end{eqnarray}
\begin{eqnarray}
\frac{B_\zeta^{1-loop \; a}}{B_\zeta^{1-loop \; f}} &\sim& \frac{\frac{\eta_\sigma^3}{\eta_\phi^6 \phi_\star^6} \exp\left[6N
(\eta_\phi - \eta_\sigma)\right] \left(\frac{\sum_i k_i^3}{\prod_i k_i^3}\right) \left(\frac{H_\star}{2\pi}\right)^6 4\pi^4}
{\frac{\eta_\sigma}{\eta_\phi^4 \phi_\star^4} \exp \left[2N (\eta_\phi - \eta_\sigma)\right] 4\pi^4 \left(\frac{H_\star}
{2\pi}\right)^6 \frac{1}{m_P^2} \left(\frac{\sum_i k_i^3}{\prod_i k_i^3}\right)} \nonumber \\
&=& \left(\frac{\eta_\sigma}{\eta_\phi}\right)^2 \left(\frac{m_P}{\phi_\star}\right)^2 \exp \left[4N(\eta_\phi -
\eta_\sigma)\right] \gg 1 \,.
\end{eqnarray}
Thus, the one-loop leading term for $B_\zeta$ in our model is given by:
\begin{equation}
B_\zeta^{1-loop} = \frac{\eta_\sigma^3}{\eta_\phi^6 \phi_\star^6} \exp\left[6N (\eta_\phi - \eta_\sigma)\right]
\left(\frac{H_\star}{2\pi}\right)^6 \ln(kL) 4\pi^4 \left(\frac{\sum_i k_i^3}{\prod_i k_i^3}\right) \,. \label{1lfbd}
\end{equation}

Once again, the $\ln (kL)$ dependence in Eq. (\ref{1lfbd}) does not look like that obtained from introducing Eq. (\ref{q1l})
into Eq. (\ref{tlbd1}).  However the situation here is the same as that discussed at the end of Subsection \ref{Ldis}, leading
us to identical conclusions.

%%%%%%%%%%%%%%%%%%%%%%%%%%%%%%%%%%%%%%%%%%%%%%%%%%%%%%%%%%%%%%%%%%%%%%%%%%%%%%%%%%
\section{Tree-level and one-loop diagrams for $\tz$}                   %%%%%%%%%%%
%%%%%%%%%%%%%%%%%%%%%%%%%%%%%%%%%%%%%%%%%%%%%%%%%%%%%%%%%%%%%%%%%%%%%%%%%%%%%%%%%%
Taking into account the last two Sections, the existence of a perturbative regime and
the truncation of the series in \eq{dntrun} at second-order, we can see
that just one Feynman-like diagram per spectral function of $\zeta$ is
necessary to study the tree or loop corrections to these spectral functions. In theses cases, just one leading diagram
for the one-loop correction to $P_\zeta$ Fig. \ref{olpd}a, as well as one leading
diagram for the one-loop correction to $B_\zeta$ Fig. \ref{olbd}a are necessary. When applied to $T_\zeta$,
this analysis shows that the only diagrams to consider are the one in Fig. \ref{tf}a for the tree-level terms, and the one in 
Fig. \ref{tf}b for the loop corrections. Such diagrams lead to for $T_\zeta^{tree}$ and $T_\zeta^{ 1-loop}$.
\begin{figure}
\begin{center}
\begin{tabular}{cccc}
\includegraphics[width=7cm,height=2cm]{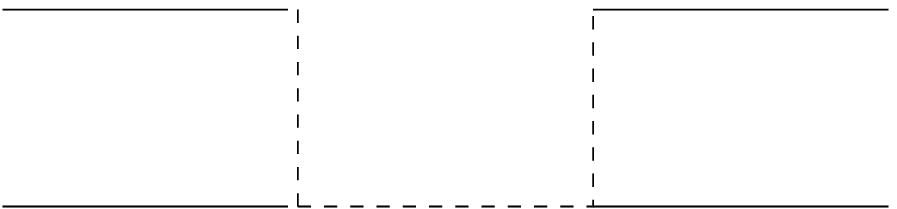} & & & \includegraphics[width=7cm,height=2cm]{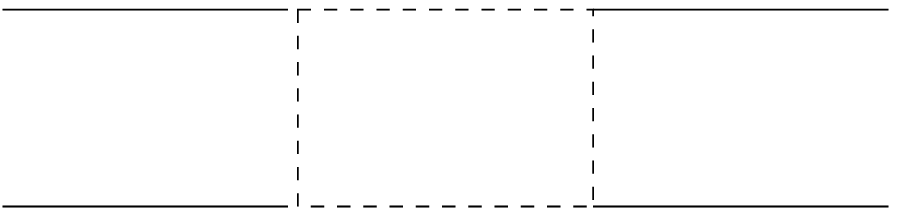} \\
(a) & & & (b)
\end{tabular}
\end{center}
\caption[Tree-level and one-loop Feynman-like diagrams for $T_\zeta$]{(a). Tree-level Feynman-like diagram for $T_\zeta$. (b). 
One-loop Feynman-like diagram for $T_\zeta$. The internal dashed lines correspond to two-point correlators of field 
perturbations.} \label{tf}
\end{figure}
\bea
\tz^{tree}&=& N_{\phi\phi}^2N_\phi^2\[P_{\delta \phi} (k_2) \ P_{\delta \phi} (k_4) P_{\delta \phi} (|\bfk_3+\bfk_4|)+ 11 \
{\rm permutations}\]\no\\
&=&\frac{1}{\eta_\phi^4\phi_\star^6}\left(\frac{H_\star}{2 \pi}\right)^6\left[\frac{2\pi^2}{k_2^3}\frac{2\pi^2}
{k_4^3}\frac{2\pi^2}{|{\bf k}_3+
{\bf k}_4|^3} + 11 \ {\rm permutations}\right] \,.\label{ttap}
\eea

\bea\label{t1loop}
\tz^{1-loop}&=& \left[N_{\phi \phi}^4 + N_{\sigma \sigma}^4\right] \int \frac{d^3 q}{(2\pi)^3} \[P_{\delta \phi}(q)
P_{\delta \phi} (|{\bf k_1} - {\bf q}|) P_{\delta \phi} (|{\bf k_3} + {\bf q}|) P_{\delta \phi} (|{\bf k_1}+\bfk_2 + {\bf
q}|)\right.\no\\
&+&\left. 11 \ {\rm permutations}\]\no\\
&=&\left[\frac{1}{\eta_\phi^4 \phi_\star^8} + \frac{\eta_\sigma^4}{\eta_\phi^8 \phi_\star^8} \exp\left[8N (\eta_\phi -
\eta_\sigma)\right] \right]\left(\frac{H_\star}{2 \pi}\right)^8\ln(kL) \ 4
\Big[ \frac{2\pi^2}{k_2^3}\frac{2\pi^2}{k_4^3}\frac{2\pi^2}{|{\bf k}_3+{\bf k}_4|^3} + \nonumber \\
&&+11 \ {\rm permutations} \Big]\no \\
&=&\frac{\eta_\sigma^4}{\eta_\phi^8\phi_\star^8}\exp[8N(\eta_\phi-\eta_\sigma)] \left(\frac{H_\star}{2 \pi}\right)^8\ln(kL) \ 4
\Big[ \frac{2\pi^2}{k_2^3}\frac{2\pi^2}{k_4^3}\frac{2\pi^2}{|{\bf k}_3+{\bf k}_4|^3} + \nonumber \\
&&+11 \ {\rm permutations} \Big] \,.\label{tlap}
\eea
%%%%%%%%%%%%%%%%%%%%%%%%%%%%%%%%%%%%%%%%%%%%%%%%%%%%%%%%%%%%%%%%%%%%%%%%%%%%%%%%%%%%%%%%%%%%%%%%%
%%%%%%%%%%%%%%%%%%%%%%%%%%%%%%%%%%%%%%%%%%%%%%%%%%%%%%%%%%%%%%%%%%%%%%%%%%%%%%%%%%%%%%%%%%%%%%%%%
\chapter{THE ONE-LOOP INTEGRAL FOR $\pz$} \label{Integrals}    %%%%%%%%
%%%%%%%%%%%%%%%%%%%%%%%%%%%%%%%%%%%%%%%%%%%%%%%%%%%%%%%%%%%%%%%%%%%%%%%%%%%%%%%%%%%%%%%%%%%%%%%%%
%%%%%%%%%%%%%%%%%%%%%%%%%%%%%%%%%%%%%%%%%%%%%%%%%%%%%%%%%%%%%%%%%%%%%%%%%%%%%%%%%%%%%%%%%%%%%%%%%
We sketch in this appendix the mathematical procedure to estimate the integrals that appear when we consider the loop
corrections. We only work one integral since the other ones are estimated in a similar way.

The one-loop contribution to the spectrum is:
\bea
\calp_\zeta^{\rm 1-loop} (\bfk) &=& \int
\frac{d^3p \ k^3}{4\pi|\bfk + \bfp|^3 p^3}
\[\frac{1}{2} N_{\phi \phi}^2  \calp_{\delta \phi} (|\bfk + \bfp|)
\calp_{\delta \phi}(p) +
N_{\phi A}^i N_{\phi A}^j \calp_{\delta \phi} (|\bfk + \bfp|)
\calt_{ij}(\bfp)\right. \nonumber \\&&\left. + \frac{1}{2} N_{AA}^{ij} N_{AA}^{kl}\calt_{ik}(\bfk+\bfp)\calt_{jl}(\bfp) \] \,.
\eea
As we can see, the total contribution to $\calp_\zeta^{\rm 1-loop}$ corresponds to three integrals, each one
having two singularities: one in $\bfp=0$ and the other one in $\bfp=-\bfk$. If the fields spectra
are scale invariant, the first integral may be written as:
\be
\calp_\zeta^{\rm 1-loop (a)} (\bfk) = \frac{1}{8\pi}\calp_{\delta \phi}^2 N_{\phi \phi}^2
\int\frac{d^3p \ k^3}{4\pi|\bfk + \bfp|^3 p^3} \,,
\ee
so the actual integral to estimate is:
\be
I=\int_{L^{-1}} \frac{d^3p \ k^3}{|\bfk + \bfp |^3 p^3} \,. \label{dint}
\ee
This integral is logaritmically divergent at the zeros in the denominator,
but there is a cutoff at $k=L^{-1}$. The subscript $L^{-1}$ indicates that the integrand is
set equal to zero
in a sphere of radius $L^{-1}$ around each singularity, and the discussion makes sense only for $L^{-1}\ll k\ll k_{max}$.
If we consider the infrared divergences, that means $\bfp\ll\bfk$, we may write:
\be
I=\int_{L^{-1}}^k\frac{d^3p}{p^3}\sim4\pi\ln(kL).
\ee
To calculate the contribution coming from the other singularity we can make the  substitution $\bfq=\bfk+\bfp$.
After evaluating this latter integral, we find that the contibution is  again $4\pi\ln(kL)$.
The integral in Eq. (\ref{dint}) may be finally estimated by adding the contibutions of the two
singularities:
\be
I=\int\frac{d^3p \ k^3}{|\bfk + \bfp |^3 p^3}=8\pi\ln(kL).
\ee
More details to evaluate these integrals may be found in Refs. \cite{bl,lythaxions,lythbox}.

The technique to evaluate this kind of integrals when considering vector fields is the same, although
the procedure is algebraically more tedious. Nevertheless, one can finally arrive to the same
conclusion. A more detailed discussion about this issue will be found in a forthcoming
publication \cite{val}.

\newpage

%%%%%%%%%%%%%%%%%%%%%%%%%%%%%%%%%%%%%%%%%%%%%%%%%%%%%%%%%%%%%%%%%%%%%%%%%%%%%%%%%%%%%%%%%%%%%%%%%%%%%%%%%%%%%%%%%%%%%%%%%%%%%%%%%%%%%%%%%%%%%%%%

\end{document}